\documentclass[11pt,a4paper]{article}
\usepackage{epsfig}
 
\newlength{\dinwidth}
\newlength{\dinmargin}
\def\be{\begin{equation}}
\def\ee{\end{equation}}
\def\bear{\begin{eqnarray}}
\def\eear{\end{eqnarray}}
\def\bfi{\begin{figure}}
\def\efi{\end{figure}}
\def\btab{\begin{table}}
\def\etab{\end{table}}
\def\bc{\begin{center}}
\def\ec{\end{center}}
\def\bi{\begin{itemize}}
\def\ei{\end{itemize}}
\def\nn{\nonumber}

\def\text{\textstyle}
\def\mai{I}

\def\mathswitch#1{\relax\ifmmode#1\else$#1$\fi}

\newcommand{\lsim}
{\;\raisebox{-.3em}{$\stackrel{\displaystyle <}{\sim}$}\;}

\newcommand{\tosv}
{{\scriptscriptstyle \to}}

\newcommand{\qbar}{{\bar q}}
\newcommand{\gl}{\tilde{g}}
\newcommand{\sq}{\tilde{q}}
\newcommand{\sqb}{\bar{\tilde{q}}}
\newcommand{\al}{\alpha}
\newcommand{\as}{\alpha_{\mathrm{s}}}
\newcommand{\si}{\mathswitch \sigma}
\newcommand{\sih}{\mathswitch \hat \sigma}

\newcommand{\nf}{\mathswitch {n_{\mathrm{f}}}}

\newcommand{\CA}{\mathswitch {C_{\mathrm{A}}}}
\newcommand{\CF}{\mathswitch {C_{\mathrm{F}}}}
\newcommand{\shat}{{\hat s}}

\newcommand{\msq}{\mathswitch m_{\sq}}
\newcommand{\mgl}{\mathswitch m_{\gl}}
\newcommand{\rhohat}{\mathswitch \hat{\rho}}

\newcommand{\NLL}{\mathrm{NLL}}

\newcommand{\LO}{\mathrm{LO}}
\newcommand{\NLO}{\mathrm{NLO}}

\newcommand{\onebf}{{\bf 1}}
\newcommand{\eigbf}{{\bf 8}}
\newcommand{\eigbfa}{{\bf 8_{\mathrm \bf A}}}
\newcommand{\eigbfs}{{\bf 8_{\mathrm \bf S}}}
\newcommand{\tptbf}{{\bf 10}\oplus {\bf \overline{10}}}
\newcommand{\twsbf}{{\bf 27}}

\newcommand{\ord}{{\cal O}}

\newcommand{\GeV}{\unskip\,\mathrm{GeV}}

\newcommand{\TeV}{\unskip\,\mathrm{TeV}}
\newcommand{\fba}{\unskip\,\mathrm{fb}}

\newcommand{\BA}{\Lambda}
\newcommand{\C}{\Omega}

\newcommand{\Sgl}{\bar{S}}
\newcommand{\Tgl}{\bar{T}}
\newcommand{\Ugl}{\bar{U}}
\newcommand{\A}{\Sgl}

\begin{document}

\thispagestyle{empty}
\def\thefootnote{\fnsymbol{footnote}}
\setcounter{footnote}{1}
\null
\date
\strut\hfill  DESY 09-078\\
\strut\hfill  PITHA 09/11\\
\vskip 0cm
\vfill
\begin{center}

{\Large \bf Soft gluon resummation for the production of
   gluino-gluino and squark-antisquark pairs at the LHC}

\bigskip
\bigskip
\bigskip
\bigskip

{\large \sc
A.~Kulesza$^a$ and L.~Motyka$^{b,c}$}
~\footnote{%
{anna.kulesza@physik.rwth-aachen.de},
{leszek.motyka@desy.de}}

\bigskip
\bigskip

\begin{it}

$^a$Institut f\"ur Theoretische Physik E, RWTH Aachen, D--52056 Aachen, Germany \\
\bigskip

$^b$II Institute for  Theoretical Physics, 
University of Hamburg,\\ Luruper Chaussee 149, D-22761, Germany \\
\bigskip

$^c$ Institute of Physics, Jagellonian University, Reymonta 4,\\ 30-059
Krak\'{o}w, Poland

\end{it}

\bigskip
\bigskip

\bigskip
\bigskip
\bigskip

\end{center}

\vfill
\bc
{\bf Abstract}
\ec

We study the effect of soft gluon emission in the hadroproduction of
 gluino-gluino and squark-antisquark pairs at the next-to-leading
logarithmic (NLL) accuracy within the framework of the minimal
supersymmetric model. We present the calculation of 
the one-loop soft anomalous dimension matrices controlling the colour
evolution of the underlying hard-scattering processes. 
The numerical results for resummed cross sections for  
proton-proton collisions at the Large Hadron Collider are discussed in detail.

\par
\vskip 1cm
\noindent
\par
\null
\setcounter{page}{0}
\clearpage
\def\thefootnote{\arabic{footnote}}
\setcounter{footnote}{0}

\newpage

\section{Introduction}

Supersymmetry (SUSY) is one of the most promising candidates 
for the theory of physics beyond the Standard Model (SM). In the coming years,
experiments at the Large Hadron Collider (LHC) will undertake searches 
for new physical phenomena. A large part of this effort will be 
devoted to looking for signals of SUSY.

One of the most studied SUSY models is the Minimal Supersymmetric 
Standard Model (MSSM)~\cite{mssm}, characterized by the minimal content of 
supersymmetric particles and $R$-parity conservation.  
Within the MSSM, the dominant production processes of sparticles at the LHC 
are those involving pairs of coloured particles, i.e.\ squarks  
and gluinos, in the final state~\cite{susyrev}. The exact discovery reach of the LHC is model-dependent 
but it is expected that the discovery of squarks and gluinos should be possible 
for masses of up to around 2 TeV~\cite{tdrs}. 
Since the hadroproduction cross sections for strongly-interacting
sparticles depend only on the masses of squarks and gluinos~\cite{HLS,DEQ,BHSZ1,BHSZ2},
measurements of total cross sections for coloured sparticle production may be used 
to determine values of the fundamental MSSM parameters, e.g.\ the masses of
sparticles~\cite{HLS,BHSZ1,BHSZ2,Baer}, or to draw exclusion limits for the 
mass parameters~\cite{D0,CDF}. The precision of the mass
determination (exclusion limit) will crucially depend on the accuracy
of the corresponding theoretical predictions. With the large
production rates expected at the LHC it is thus of utter importance to
study the total cross sections for the hadroproduction of squarks and
gluinos with the highest possible theoretical accuracy.

The leading-order (LO) total cross sections of $\ord (\as^2)$ were calculated long
time ago~\cite{HLS,DEQ}. The corresponding next-to-leading 
order (NLO) SUSY-QCD corrections are known for all hadroproduction 
processes of pairs of squarks and gluinos~\cite{BHSZ1,BHSZ2}. 
The NLO electroweak (EW) corrections~\cite{HKMT} for the processes involving
squarks in the final state are also known, as well as the LO EW $\ord(\alpha^2)$ total
cross sections and the $\ord(\alpha \as)$ LO EW-QCD interference predictions~\cite{HKMT,LOEWsq}.

The NLO SUSY-QCD corrections have been found to be positive and large.  Among the pair-production 
processes of coloured sparticles at the LHC, the gluino-pair ($\gl\gl$) 
production receives the largest NLO SUSY-QCD correction  
that may reach 100\% for gluino mass $m_{\gl} = 1$~TeV and 
$1.2 \TeV \lsim m_{\sq} \lsim 2 \TeV$~\cite{BHSZ2}. 
The corrections to the squark-antisquark ($\sq \sqb$) total cross section 
can be also sizable, of order of 30\% for the squark mass 
$m_{\sq}=1$~TeV, and are the second largest in a certain range of mass 
parameters. The occurrence of large corrections indicates that computation 
of higher order SUSY-QCD corrections is necessary in order 
to achieve precise theoretical predictions.

A large part of the NLO SUSY-QCD corrections to the total cross section for 
$\gl \gl$ and $\sq \sqb$ processes comes from production close to 
threshold~\cite{BHSZ2}. The threshold region is reached when the square of the 
partonic center-of-mass (c.o.m.) energy, $\shat$, approaches $4\,m^2$, 
where $m$ is the average particle mass in the produced pair.
The velocity of the produced heavy particles in the partonic c.o.m.\ system 
$\beta \equiv \sqrt{1- 4 m^2/ \shat}$ is then small, $\beta \ll 1$.
In this region two types of corrections dominate: 
Coulomb corrections due to exchange of gluons between
slowly moving massive particles and soft gluon corrections due to
emission of low energy gluons off the coloured initial and
final states.  The soft gluon corrections are enhanced by powers of 
large logarithms of $\beta$, i.e. at the NLO one finds terms in the relative
corrections proportional to $\as \log^2 (\beta^2)$ that become sizeable when 
$\beta^2 \sim \exp(-1/\sqrt{\as})$.
At the $n$-th order of the perturbative expansion in the strong coupling $\as$ 
the total cross sections receive corrections
proportional to $\as^n \log^{k}(\beta^2)$ where $k=2n, \dots, 0$.
Sufficiently close to the partonic threshold fixed-order 
expressions for the cross sections are bound to fail. However, the 
logarithmic contributions can be taken
into account to all orders in $\as$ by means of threshold resummation.
Resummed predictions are particularly 
important for processes with large masses in the final states since then 
the bulk of production comes from the threshold region. This is exactly the
case for production of sparticles  which are expected to
be heavier than the SM particles. Additionally, if partonic
subprocesses involve gluons in the initial state, the soft-gluon effects, and
thus the impact of resummation, are
expected to be significant due to the high colour charge of the gluons.

Crucially, calculation of the soft gluon corrections provides a reliable
estimate of unknown higher order terms beyond the NLO, what results in 
reduction of the theoretical uncertainty  due to scale variation.
In a recent letter~\cite{KM} we have presented results for
threshold-resummed cross sections at the next-to-leading
logarithm (NLL) accuracy 
for hadroproduction of $\gl\gl$ and $\sq
\sqb$ pairs at the LHC. A dominant part of the
next-to-next-to-leading (NNLO) correction for the $\sq \sqb$ production 
consisting of terms coming from the expansion of the
resummed exponent at the next-to-next-to-leading logarithm (NNLL) level,
Coulomb terms and the universal scale terms, was later calculated
in~\cite{LM}. Moreover, threshold resummation for single colour-octet
scalar at the LHC was also studied~\cite{IKM}. Further work on resummation for 
production of coloured sparticle pairs is to be found in~\cite{BBKKLN,BFS}.

The hadronic production processes of $\gl\gl$ and $\sq \sqb$ pairs are scattering 
processes with a non-trivial colour flow structure. At the NLL level resummation requires including 
contributions from soft gluons emitted at wide angles. Such emission is 
sensitive to the colour flow of the underlying hard scattering and the 
evolution of the colour exchange is governed by the 
soft anomalous dimension matrix~\cite{BS,KS,KOSthr,KOScol,BCMNgen}. 
The one-loop soft anomalous dimension matrices were first calculated for heavy-quark and 
dijet production~\cite{BS,KS,KOScol,BCMNtop}. In~\cite{KM} we have presented the explicit form of 
the one-loop soft anomalous dimension matrices
for partonic subprocesses contributing to $\gl\gl$ hadroproduction. The general 
results for any $2\to n\,$ QCD process with massless particles
in the final state were derived at one-~\cite{BCMNgen}, 
and two-loops~\cite{MDS}. 
The two-loop anomalous dimension for 
the pair production of heavy quarks was also determined in the 
threshold limit~\cite{MU}, and employed later
in~\cite{LM}. Recently, the structure of the massive two-loop matrix
for any $2\to n$ process has been studied in~\cite{MSS}.

In this paper we discuss in detail the derivation of the analytical results
presented in~\cite{KM} and carry out a thorough study of the
numerical results as well as perform resummation of the leading Coulomb
corrections for the $\sq \sqb$ and $\gl \gl$ production processes.

\section{Leading order results}

The hadronic cross section for the process $h_1 h_2 \to k l$ reads
\be
\si_{h_1 h_2  \to k l}(S, \{m^2\}) = \sum_{i,j} \int d x_1 \;d x_2
\;f_{i/h_{1}}(x_1,\mu_F ) \;f_{j/h_{2}}(x_2,\mu_F ) \;\sih_{i j  \to k
  l }(\shat, \{m^2\},\mu_F^2,\mu_R^2 ) \;,
\label{eq:hadrxs1}
\ee
where $S$ ($\shat$) is the square of the hadronic (partonic)
center-of-mass energy and $ \{m^2\}$ stands for all masses
entering the calculations. The parton distribution functions (pdfs) $f_{i/h}$
are taken at the factorisation scale $\mu_F$. We set $\mu_F$ equal to 
the renormalisation scale $\mu_R$ in our calculations.

The expressions for the LO partonic cross sections $\sih_{ij  \to k l }(s, \{m^2\},\mu_F^2,\mu_R^2 )$  
for all squark and gluino hadroproduction processes can be found
in~\cite{BHSZ2}. Here we present the contributions to the LO cross
sections for the $\sq \sqb$
and $\gl \gl$ production coming from different colour channels. For each partonic process we define the corresponding colour basis in
the $s$-channel.

For the $\sq \sqb$ production we consider the processes
\be
q_i (p_i,\al_i) \; \bar q (p_j,\al_j) \;\to\; \sq (p_k, \al_k)
\; \sqb (p_l, \al_l)
\label{eq:proc:qq:sq}
\ee
and 
\be
g (p_i, a_i) g (p_j,a_j) \;\to\; \sq (p_k,\al_k)
\; \sqb(p_l,\al_l) \;,
\label{eq:proc:gg:sq}
\ee
where  $p$ are particle four-momenta and $\alpha$ and $a$ are 
colour indices in the fundamental and adjoint representation of SU(3),
correspondingly.
In the quark-channel~(\ref{eq:proc:qq:sq}) 
we have only two possible colour exchanges: the singlet and the octet,
$\{\onebf,\eigbf\}$, and the basis consists of two colour tensors
\bear
c^{q,\sq}_{\onebf} &=& \delta^{\al_{i}\al_{j}} \delta^{\al_{k}\al_{l}} \;, \nn \\
c^{q,\sq}_{\eigbf} &=& -{1\over 6} \delta^{\al_{i}\al_{j}} \delta^{\al_{k}\al_{l}}+ {1\over 2} \delta^{\al_{i}\al_{k}} \delta^{\al_{j}\al_{l}}  \;.
\label{eq:base:qq:sq}
\eear
In the gluon channel, the basis is built out of three tensors
corresponding to $\{\onebf,\eigbfs,\eigbfa \}$ representations
\bear
c^{g,\sq}_{\onebf} &=&  \delta^{a_i a_j} \, \delta^{\al_k
  \al_l}\;, \nn \\
c^{g,\sq}_{\eigbfs} &=&  T^b _{\alpha_l \alpha_k} d^{b a_i a_j} \;, \nn \\
c^{g,\sq}_{\eigbfa} &=& i T^b _{\alpha_l \alpha_k} f^{b a_i a_j} \;.
\label{eq:base:gg:sq}
\eear
where  $T^{b}$ matrices are the SU(3) generators.

At the leading order, two partonic channels contribute to the $\gl \gl$
production:
\be
q (p_i,\alpha_i) \,\qbar (p_j,\alpha_j) \to \gl(p_k,a_k)\,\gl(p_l,a_l) \;,
\label{eq:proc:qq:gl}
\ee
and
\be
g(p_i,a_i) \,g(p_j,a_j) \to \gl(p_k,a_k)\, \gl(p_l,a_l) \;.
\ee
For the process~(\ref{eq:proc:qq:gl}) the colour basis is the same as
for the $\sq\sqb$ production in the
gluon-channel~(\ref{eq:proc:gg:sq}) and is given
by~(\ref{eq:base:gg:sq}) after interchanging the indices
$(ij) \leftrightarrow (kl)$. For the  $gg$ channel there are eight independent 
colour tensors. Following~\cite{KOScol} we choose an orthogonal basis, 
\mbox{$\{c^{g,\gl}_I\}, \; I=1,2,\ldots,8$,}
consisting of five  tensors $c^{g,\gl} _1$, $c^{g,\gl} _2$, $c^{g,\gl} _3$, $c^{g,\gl} _4$ and 
$c^{g,\gl} _5$ corresponding to the $\onebf$, $\eigbfs$, $\eigbfa$, $\tptbf$ and $\twsbf$  
representations in the $s$-channel, and three additional tensors,
$c^{g,\gl} _6$, $c^{g,\gl} _7$, and $c^{g,\gl}_8$. The base tensors are 
\bear
c^{g,\gl} _1 &=& {1\over 8}\delta^{a_i a_j} \delta^{a_k a_l}\;, \nn \\
c^{g,\gl} _2 &=& {3\over 5}d^{a_i a_j b} d^{b a_k a_l}\;,\nn \\
c^{g,\gl} _3 &=& {1\over 3}f^{a_i a_j b} f^{b a_k a_l}\;,\nn \\
c^{g,\gl} _4 &=& {1\over 2} \left( \delta^{a_i a_k}\delta^{a_j a_l}
                     -  \delta^{a_i a_l}\delta^{a_j a_k} \right)
                       -{1\over 3} f^{a_i a_j b} f^{b a_k a_l}\;,\nn \\
c^{g,\gl} _5 &=&  {1\over 2} \left( \delta^{a_i a_k} \delta^{a_j a_l} 
       + \delta^{a_i a_l} \delta^{a_j a_k} \right)
        -{1\over 8}\delta^{a_i a_j} \delta^{a_k a_l} \nn \\
        &-&{3\over 5}d^{a_i a_j b} d^{b a_k a_l}\;,\nn \\
c^{g,\gl} _6 &=& {i\over 4} \left(
f^{a_i a_j b} d^{b a_k a_l} + d^{a_i a_j b} f^{b a_k a_l}
\right)\;,\nn \\
c^{g,\gl} _7 &=& {i\over 4} \left(
f^{a_i a_j b} d^{b a_k a_l} - d^{a_i a_j b} f^{b a_k a_l} \right)\;, \nn \\
c^{g,\gl} _8  &=& {i\over 4} \left(
d^{a_i a_k b} f^{b a_j a_l} + f^{a_i a_k b} d^{b a_j a_l} \right)\;. 
\label{eq:base:gg:gl}
\eear

In the set of basis defined above, we obtain the following
colour-channel contributions to the total cross section for $q_i \bar q _j \to \sq\sqb$: 
\begin{eqnarray} 
\sigma^{(0)} _{q_i \bar q_j \to \sq\sqb, \bf 1}  & = &  
{8\over 9}\, {\pi \hat\as^2 \over \shat}\,
\left[ \, \beta_{\sq}\, \left( -{4 \over 9} 
- {4 m_-^4\over 9(m_{\gl}^2 \shat + m_-^4)}
\right)\, - \, \left({4\over 9} + {8 m_-^2 \over 9\shat}\right)\,L_1\, \right]\,, 
\\
\sigma^{(0)} _{q_i \bar q_j \to \sq\sqb, \bf 8}  & = &  
\delta_{ij} \, {\nf \pi \as^2 \over \shat}\, 
{4 \over 27}\,\beta^3 _{\sq}\nn \\
&+&  \delta_{ij} \, {\pi \as \hat\as \over \shat}\,
\left[\,
\beta_{\sq}\, \left({4\over 27} + {8m_- ^2 \over 27 \shat}\right)
\, + \, \left( {8 m_{\gl} ^2 \over 27 \shat} \, + \, 
{8m_-^4 \over 27 \shat^2}\right) \,L_1 
\right] \nonumber \\
&+&   {1\over 9}\,{\pi \hat\as^2 \over \shat}\,
\left[ \, \beta_{\sq}\, \left( -{4 \over 9} 
- {4 m_-^4\over 9(m_{\gl}^2 \shat + m_-^4)}
\right)\, - \, \left({4\over 9} + {8 m_-^2 \over 9\shat}\right)\,L_1\, \right]\,, 
\end{eqnarray}
for $gg \to \sq\sqb$:
\begin{eqnarray}
\sigma^{(0)} _{g g \to \sq\sqb, \bf 1} & = &
{\nf \pi \as^2 \over \shat}\, \left[\,
\beta _{\sq} \left( {1\over 24} + {m_{\sq} ^2  \over 6 \shat}\right)
\, + \, \left( {m_{\sq} ^2 \over 6\shat} -  {m^4_{\sq} \over 3\shat^2} \right)\,
\log\left({1-\beta_{\sq} \over 1 + \beta_{\sq}}\right)\, \right] \, ,
\\
\sigma^{(0)} _{g g \to \sq\sqb, \eigbfs}+ \sigma^{(0)}_{g g \to
  \sq\sqb, \eigbfa} & = &
{\nf \pi \as^2 \over \shat}\, \left[\,
\beta _{\sq} \left( {1\over 6} + {29 m_{\sq} ^2  \over 12 \shat}\right)
\, + \, \left( {7m_{\sq} ^2 \over 6\shat} +  {2m^4_{\sq} \over 3\shat^2} \right)\,
\log\left({1-\beta_{\sq} \over 1 + \beta_{\sq}}\right)\, \right] \, ,
\end{eqnarray}
for $q\bar q \to\gl\gl$:
\begin{eqnarray}
&& \sigma^{(0)} _{q\bar q \to \gl\gl, \bf 1} =
{\pi \hat\as^2 \over \shat}\,
\left[ \, \beta_{\gl}\, \left( {4 \over 27} 
 +  {4 m_-^4\over 27 (m_{\sq}^2 \shat + m_-^4)} \right)\, - \, 
\left({8 m^2 _{-}\over 27\shat} 
- {8 m^2 _{\gl} \over 27(\shat-2m^2 _{-})}\right) \, L_2 \, \right],
\\
&& \sigma^{(0)} _{q\bar q \to \gl\gl, \eigbfs}+ \sigma^{(0)} _{q\bar q
  \to \gl\gl, \eigbfa} =
{\pi \as^2 \over \shat}\, 
\beta_{\gl} \, \left({8\over 9} + {16 m_{\gl}^2 \over 9\shat} \right)
\nn  \\
& +&   {\pi \as \hat\as \over \shat}\,
\left[\,
\beta_{\gl}\, \left(-{4\over 3} - {8m_- ^2 \over 3 \shat}\right)
\, + \, \left( {8 m_{\gl} ^2 \over 3\shat} \, + \, 
{8m_{-} ^4 \over 3 \shat^2}\right) \,L_2 \right] 
\nonumber \\
&+ &    {\pi \hat\as^2 \over \shat}\,
\left[ \, \beta_{\gl}\, \left( {28 \over 27} 
 +  {28 m_-^4\over 27 (m_{\sq}^2 \shat + m_-^4)} \right)\, - \, 
\left({56 m^2 _{-}\over 27\shat} 
+ {16 m^2 _{\gl} \over 27(\shat-2m^2 _{-})}\right) \, L_2 \, \right]\,, 
\end{eqnarray}
and for $gg \to\gl\gl$:
\begin{eqnarray}
\sigma^{(0)} _{g g\to\gl\gl,\bf 1}& = & 
   {1\over2}\, \sigma_{\mathrm{sym}}\;, \\
\sigma^{(0)} _{g g\to\gl\gl, \eigbfs}+\sigma^{(0)} _{g g\to\gl\gl, \eigbfa}   & = & 
 \sigma_{\mathrm{sym}}\,+\,\sigma_{\mathrm{asym}} \;,\\
\sigma^{(0)} _{g g\to\gl\gl, \bf 10}  & = & 0 \;,\\
\sigma^{(0)} _{g g\to\gl\gl, \bf 27}  & = &  
   {3\over2}\, \sigma_{\mathrm{sym}} \;,\\
\sigma^{(0)} _{g g\to\gl\gl, I}  & = & 0 \quad \mbox{for}\quad I=6 \dots 8\;,
\end{eqnarray}
with
\begin{eqnarray}
 \sigma_{\mathrm{asym}} & = &
{\pi \as^2 \over \shat }\, \left[\,
\beta _{\gl} \left( -{21 \over 16} - {6 m_{\gl} ^2 \over \shat}\right) \right. \\
 & & \left.
\, - \, \left( {9 \over 16} + {9m^2 _{\gl} \over 4\shat} 
 + {9m^4 _{\gl} \over 2 \shat^2}\right) \, 
\log\left( { 1 - \beta_{\gl} \over 1 + \beta_{\gl} }\right)\,\right], 
\nonumber \\
\sigma_{\mathrm{sym}} & = &
{\pi \as^2 \over \shat}\, \left[\,
\beta _{\gl} \left( -{9 \over 16} - {9 m_{\gl} ^2 \over 4\shat}\right)
\right. \\
& & 
\left. 
\, - \, \left( {9 \over 16} + {9m^2 _{\gl} \over 4\shat} -
{9m^4 _{\gl} \over 2\shat^2}\right) \, 
\log\left( { 1 - \beta_{\gl} \over 1 + \beta_{\gl} }\right)\,\right],
\nonumber
\end{eqnarray}
and
\[ 
\beta_{\sq} = \sqrt{1 -  {4m^2 _{\sq} \over \shat}}, \qquad 
\beta_{\gl} = \sqrt{1 -  {4m^2 _{\gl} \over \shat}}, \qquad 
m_{-} ^2 = m^2 _{\gl} - m^2 _{\sq},
\]
\[
L_1 = \log\left(
{\shat(1-\beta_{\sq}) + 2m^2 _{-} \over \shat(1+\beta_{\sq}) + 2m^2 _{-}}\right),
\qquad
L_2 = \log\left(
{\shat(1-\beta_{\gl}) - 2m^2 _{-} \over \shat(1+\beta_{\gl}) - 2m^2
  _{-}}\right) \;,
\]
where $\hat \as$ is the SUSY Yukawa coupling.

\section{Threshold  resummation -- general framework}

The resummation for $2\to 2$ processes with all four external legs
carrying colour was studied extensively in the literature.
The resummed cross section for the heavy-quark
production was constructed in~\cite{KS,BCMNtop}, and for the dijet
(multiple jet) production
in~\cite{KOSthr,KOScol,BCMNgen}.
Here we briefly review the derivation of the resummed cross sections 
for the production of 
two coloured final state particles of equal mass $m$, in the form
presented in~\cite{BCMNtop}. In our calculations me make use
of the framework of~\cite{KS,KOSthr,KOScol}.

Using the hadronic threshold variable $\rho \equiv 4 m^2/S$ we rewrite 
the cross section~(\ref{eq:hadrxs1}) 
\be
\si_{h_1 h_2  \to k l}(\rho, \{m^2\}) \!= \!\sum_{i,j} \sum_I \!\!\int\!\! d x_1 d x_2
\, d \rhohat \; \delta\left(\rhohat - \frac{\rho}{x_1 x_2}\right) 
f_{i/h_{1}}(x_1,\mu ) f_{j/h_{2}}(x_2,\mu ) \,\si_{ ij   \to k
  l,I }(\rhohat,\{ m^2\},\mu^2),
\ee
where the index $I$ sums over all possible colour states of the hard scattering.

At higher orders in perturbation theory, 
the partonic cross section $\sih$ contains terms of general structure
$\as^n \log^m \beta^2$, $m\leq 2n$, with $\beta = \sqrt{1- \rhohat}$.  
These terms are singular in the threshold limit $\rhohat \to 1$. 
The singularities can be systematically treated by taking Mellin moments
of the cross section   
\bear
\label{eq:Nspacexsec}
\tilde \si_{h_1 h_2 \to kl} (N, \{m^2\})&\equiv& \int_0^1 d \rho \; 
\rho^{N-1}\; \si_{h_1 h_2\to kl}(\rho,\{ m^2\}) \\ \nn
&=&  \; \sum_{i,j} \tilde f_{i/{h_1}} (N+1,\mu^2)\,
\tilde f_{j/{h_2}} (N+1, \mu^2) \, 
\tilde{\si}_{ij \to kl}(N,\{m^2\},\mu^2)  \;.
\eear
The moments of the parton distributions  $f_{i/h}(x_i,\mu^2)$ are defined
in the standard way,~\footnote{Note that from now on we will use the tilde sign to mark symbols for 
$N$-space quantities.}  
\be
\tilde f_{i/h} (N,\mu^2) \,\equiv\, \int_0^1 dx \, x^{N-1} \, 
f_{i/h}(x,\mu^2) \;,
\ee
and 
the moments of the partonic cross section $ab \to kl$ are given by 
\be
\tilde{\si}_{ij \to kl}(N, \{m^2\}, \mu^2)\,\equiv\, 
\int_0^1 d \rhohat \, \rhohat^{N-1}\,  
\si_{ij \to kl} (\rhohat, \{m^2\},\mu^2)\,.
\ee
Taking the Mellin moments transforms the logarithms in
$\beta^2$ into the logarithms of the Mellin variable $N$
which are then resummed to all orders in $\as$.

Following~\cite{CMNTmp} we define the differential distribution
\bear
&& \frac{d \si_{h_1 h_2 \to kl}}{d \xi}(\rho, \{m^2\}) \; = \\ \nn
&=&  \, \sum_{i,j} \sum_I 
\int d x_1 \;d x_2\;f_{i/{h_1}}(x_1,\mu ) \;f_{j/{h_2}}(x_2,\mu ) \,
\delta( \xi - x_1 x_2)\, \si_{ij  \to kl, I }(\rho /\xi, \{m^2\}, \mu^2 )\\ \nn
&=& \, \sum_{i,j} \sum_I \, \si_{ij  \to kl, I }(\rho /\xi, \{m^2\}, \mu^2 )\;
\int_{C - i \infty}^{C + i \infty}
dN \, \xi^{-N} \, \tilde f_{i/{h_1}}(N,\mu^2 ) \, \tilde f_{j/{h_2}}(N,\mu^2 ) \;.
\eear
Since soft radiation carries colour charge, it can change the colour
state of the underlying hard scattering for hadronic processes with
two or more coloured partons in the final state. This has to be taken
into account while writing the form of the cross section with
long-distance and short-distance effects factorised, and leads to~\cite{KOSthr,BCMNtop} 
\bear
\frac{d \si^{\rm (res)}_{h_1 h_2 \to kl}}{d \xi}(\rho, \{m^2\}) &=& 
\sum_{i,j} \sum_{I,J} \,
h_{ij\to kl,I}^* (\rho /\xi, \{m^2\}, \mu^2) \,  
h_{ij\to kl,J} (\rho /\xi,\{m^2\}, \mu^2)  \\ \nn 
&\times&  \frac{1}{2\pi i} \, \int_{C - i \infty}^{C + i \infty}
dN \xi^{-N}  \tilde f_{i/{h_1}}(N,\mu^2 ) \, \tilde f_{j/{h_2}}(N,\mu^2 ) \;
\tilde \omega^{\rm (res)}_{ij \to kl, IJ}(N, Q, \mu) \;,
\label{eq:res0}
\eear
with $Q^2 = 4m^2$. The function $h_{ij\to kl,I}$ ( $h_{ij\to kl,J}^*$)
is a  colour-dependent hard-scattering amplitude (conjugate of)
absorbing the far, i.e. of the order of the scale of the process $Q$,
off-shell effects. All the logarithmic dependence on $N$,
originating from soft and collinear radiation, is
contained in the function  $\tilde \omega^{\rm (res)}_{ij \to kl, IJ}(N, Q, \mu)$.

In the approach of~\cite{KS,KOSthr,KOScol,Sdy}, resummation follows from
refactorisation of partonic cross sections. In the case of threshold
resummation, the cross sections are factorised w.r.t.\ fixed fractions of
energy as opposed to fractions of momenta in the standard collinear
factorisation. Using the  refactorised form of  the cross
section for the  production of two massive coloured particles in the final state~\cite{KS},
we have, up to
corrections of $\ord (1/N)$,
\bear
\label{eq:fact0}
 \frac{d \si^{\rm (res)}_{h_1 h_2 \to kl}}{d \xi}(\rho, \{m^2\}) &=& 
\sum_{i,j} \sum_{I,J} 
h_{ij \to kl,I}^* (\rho /\xi, \{m^2\}, \mu^2)  \,
h_{ij \to kl,J} (\rho /\xi,\{m^2\}, \mu^2)  \\ 
&\times & \frac{1}{2\pi i}\, \int_{C - i \infty}^{C + i \infty}
dN \xi^{-N}  \tilde f_{i/{h_1}}(N,\mu^2 ) \, \tilde f_{j/{h_2}}(N,\mu^2) 
\nn \\ 
&\times & 
\frac{\tilde \psi_{i/i} (N,Q/\mu,\as(\mu^2))\,\tilde \psi_{j/j} (N,Q/\mu,\as(\mu^2))}
{\tilde f_{i/i} (N,\mu^2)\, \tilde f_{j/j} (N,\mu^2)}\,
\tilde S_{ij \to kl,IJ} (Q/(N\mu),\as(\mu^2))\;. \nn 
\eear
Following~\cite{KS,KOSthr} the soft
eikonal function $\tilde S$ represents coupling of the soft gluons to
the initial and final state particles. Consequently, the soft function carries dependence on the possible colour exchanges and we sum over all possible colour structures $I,J$ at hard
vertices.  As an object of purely eikonal
character~\cite{KS,Sdy,DMS} the soft function can depend on the scales only 
through their ratio.
The parton-in-parton distributions $\psi_{i/i}$ are defined at fixed
fraction of energy of parton $i$ in the partonic
center-of-mass frame as opposed to the light-cone distributions $f_{i/i}$
which are defined at fixed momentum fraction. The parton distribution functions
and the soft function can be defined explicitly in terms of operator
matrix elements~\cite{KOSthr}.

As a consequence of refactorisation, the soft function $\tilde S_{ij, IJ}$ and the distributions
functions $\tilde \psi_{i/i},\ \tilde f_{i/i}$
obey the corresponding renormalisation group
equations (RGEs)~\cite{KS,KOSthr,KOScol,Sdy}. Solutions of these RGEs give functions
which resum the large logarithms in question. In fact for the $2 \to
2$ production process involving coloured massive particles only the solution
of the RGE for the soft function 
\be
\left(\mu \frac{\partial}{ \partial \mu} +\beta(g) \frac{\partial}{ \partial
  g} \right) \tilde S_{ij \to kl,IJ} = - \Gamma_{ij \to kl,IK}^\dagger \tilde
S_{ij \to kl,KJ} - \tilde S_{ij \to kl,IL}\Gamma_{ij \to kl,LJ} 
\label{eq:rgesol}
\ee
is needed. As shown in~\cite{KS}, the resummed initial-state jet factors
for the ratio of $\tilde \psi_{i/i}$ to $\tilde f_{i/i}$ functions in Eq.~(\ref{eq:fact0}) can be obtained directly from
the resummed cross section for production of a colour singlet state
through the Drell-Yan mechanism, for which results are known~\cite{Sdy,CTdy}. We write this ratio in the form
\be 
\frac{\tilde \psi_{i/i} (N,Q/\mu,\as(\mu^2))}
{\tilde f_{i/i} (N,\mu^2)} 
= R_i(\as(\mu^2)) \Delta_i (N,Q^2,\mu^2)  \left[ \tilde U_{i \bar i} (Q/(N\mu),\as(\mu^2))\right]^{-1/2} \;,
\label{eq:psitof}
\ee
where the function $R_i(\as(\mu^2))$ is an $N$-independent and
infrared-safe function of the coupling.  The soft
eikonal function $\tilde U_{i \bar i} $ describes soft gluon emission and
exchange by the annihilating initial state partons in the Drell-Yan
process. The radiative factor $ \Delta_i$ represents both the soft and
collinear radiation from an incoming parton.
Inserting the expression for the ratio of $\psi$ to $f$ functions,
Eq.~(\ref{eq:psitof}), into Eq.~(\ref{eq:fact0}) leads to
\bear
 \frac{d \si^{\rm (res)}_{h_1 h_2 \to kl}}{d \xi}(\rho, \{m^2\})  &=&
\sum_{i,j} \sum_{I,J} 
h_{ij \to kl,I}^* (\rho /\xi, \{m^2\}, \mu^2)  \,
h_{ij \to kl,J} (\rho /\xi, \{m^2\}, \mu^2 )\nn \\ \nn
&\times& \frac{1}{2\pi i} \int_{C - i \infty}^{C + i \infty}
dN \xi^{-N}  \tilde f_{i/{h_1}}(N,\mu^2 ) \tilde f_{j/{h_2}}(N,\mu^2 ) 
 R_i(\as(\mu^2)) R_j(\as(\mu^2)) \\
&\times&  \Delta_i (N,Q^2,\mu^2) \Delta_j (N,Q^2,\mu^2)
\bar {\tilde S}_{ij \to kl,IJ} (Q/(N\mu),\as(\mu^2))\;,
\label{eq:fact}
\eear
where we introduce $\bar {\tilde S}_{ij \to kl,IJ} \equiv \tilde
S_{ij \to kl,IJ} / \tilde
  U_{i \bar i} $. Consequently, the soft anomalous dimension matrix
  $\bar {\Gamma}_{ij \to kl}$ corresponding to the  function $\bar
  {\tilde S}_{ij \to kl,IJ}$ is given by 
\be
 \bar{\Gamma}_{ij \to kl, IJ}(\as) = \Gamma_{ij \to  kl, IJ}(\as) 
- \delta_{IJ} \Gamma_{i \bar i} (\as)\;,
\label{eq:gammabar}
\ee
where $\Gamma_{i \bar i}$ is the anomalous dimension associated with the Drell-Yan soft
function  $\tilde U_{i \bar i} $. The results for $\Gamma_{i \bar i}$ can be found in
the literature~\cite{KOScol,Sdy}.

In general, for a given colour basis, the soft anomalous dimension matrix $\Gamma_{IJ}$
is not diagonal, leading to resummed expressions in terms of
path-ordered exponentials~\cite{KOSthr}. Through the diagonalisation
of the soft anomalous dimension matrix a simpler form for the resummed cross sections,
involving a sum of exponentials, can be obtained. 
However, the diagonalisation procedure can be avoided if from the
beginning the calculations are performed in the colour basis in which the
the soft anomalous dimension matrix is diagonal, i.e.
\be
\Gamma_{ij \to kl}(\as) = {\rm diag} (\dots, \lambda_{ij \to kl,I}(\as), \dots) \;.
\ee
The soft function then reads, up to NLL,
\bear
\tilde S_{ij \to kl,IJ}(Q/(N\mu), \as(\mu^2))&=&  \tilde S_{ij \to kl,IJ}^{(0)} \nn \\
&\times&
\exp \left[ \int_\mu^{Q/N} \frac{d
    q}{q} \left[\lambda_{ij \to kl,I}^*(\as(q^2)) +\lambda_{ij \to kl,J}(\as(q^2))\right]
\right] \;,
\eear
with the $0$-th order term in the perturbative expansion of $
\tilde S_{ij\to kl,IJ}(1,\as(Q^2/N^2))$ straightforwardly related to
the colour structure, i.e.
\be
\tilde S_{ij\to kl,IJ}^{(0)} = \mathrm{Tr} (c_I^\dagger c_J ) \;,
\ee
where $\{c_I\}$ is the corresponding colour basis for the process $ij
\to kl$, see Section 2.  
If additionally the colour basis is orthogonal then the soft function
matrix becomes diagonal 
\be
\tilde S_{ij\to kl,IJ}(Q/(N\mu),\as(\mu^2))= \delta_{IJ}\tilde S_{ij\to kl,II}^{(0)} \exp \left[ \int_\mu^{Q/N} \frac{d
    q}{q} 2 {\rm Re} (\lambda_{ij\to kl,I} (\as(q^2))) \right] \;.
\label{eq:softevol}
\ee
In this case inserting the solution~(\ref{eq:softevol}) into
Eq.~(\ref{eq:fact}) leads to the following NLL expression
\bear
\label{eq:factsigC}
 \frac{d \si^{\rm (res)}_{h_1 h_2 \to kl}}{d \xi}(\rho, \{m^2\})  &=&
\sum_{i,j} \sum_{I} \frac{1}{2\pi i} \int_{C - i \infty}^{C + i
  \infty}dN \xi^{-N}  \\ 
&\times& \hat \sigma_{ij \to kl, I}^{(0)}(\rho /\xi, \{m^2\}, \mu^2)  \;
{\cal C}_{ij \to kl, I}(\rho /\xi, N,\{m^2\}, \mu^2) \nn \\ 
&\times& \tilde f_{i/{h_1}}(N,\mu^2 ) \tilde f_{j/{h_2}}(N,\mu^2 )  \Delta_i (N,Q^2,\mu^2) \Delta_j (N,Q^2,\mu^2)
 \Delta^{(s)}_{ij \to kl,I} (N,Q^2,\mu^2)\;,\nn 
\eear
where we identify
\be
\left| h_{ij \to kl,I}\right|^2
R_i R_j \tilde S_{ij\to kl,II}^{(0)} \equiv \sigma_{ij \to kl,I
}^{(0)}  {\cal C}_{ij \to kl, I} \;.
\label{eq:hrrssc}
\ee
The functions $ {\cal C}_{ij \to kl, I}$ are of perturbative nature and 
contain information about higher-order corrections which are
non-logarithmic in $N$. 
All the information on the soft non-collinear logarithmic corrections
is included in the radiative factor   $\Delta^{(s)}_{ij \to kl,I}$.
After performing integration over $\xi$ we obtain
\bear
\si^{\rm (res)}_{h_1 h_2 \to kl}(\rho, \{m^2\}) &=&
\sum_{i,j} \sum_{I} \frac{1}{2\pi i} \int_{C - i \infty}^{C + i \infty}
dN \rho^{-N+1} \tilde f_{i/{h_1}}(N,\mu^2 ) \tilde f_{j/{h_2}}(N,\mu^2 )\nn \\ 
&\times& \tilde {\si}^{(0)}_{ij\to kl,I}(N-1, \{m^2\},\mu^2)
\; 
\bar {\cal C}_{ij\tosv kl,I}(N-1, \{m^2\},\mu^2)\nn \\ 
&\times& \, \Delta_i (N,Q^2,\mu^2) \, \Delta_j (N,Q^2,\mu^2) \, 
\Delta^{(s)}_{ij\to kl,I} (N,Q^2,\mu^2) \,.
\label{eq:resummed}
\eear
The functions $\bar {\cal C}_{ij\tosv kl,I}$ are related to the 
$ {\cal C}_{ij \to kl, I}$ functions in Eq.~(\ref{eq:factsigC}) and have a
perturbative expansion of the form  $\bar{\cal C}_{ij\tosv kl,I} = 1+ \sum_{n=1}
\as^n  \bar {\cal C}^{(n)}_{ij\tosv kl,I}$. In general, the values of the
coefficients $\bar {\cal C}^{(n)}_{ij\tosv kl,I}$ are obtained by
comparing the resummed cross section expanded in $\as$ with the expression
for the full higher-order cross section in $N$ space.
The expression for the resummed hadronic cross section in the Mellin-moment
space can be easily derived from Eq.~(\ref{eq:resummed}) and reads
\bear
&&\si^{\rm (res)}_{h_1 h_2 \to kl}(N, \{m^2\}) =  
\sum_{i,j} \sum_{I}   \tilde f_{i/{h_1}}(N+1,\mu^2 )\, 
\tilde f_{j/{h_2}}(N+1,\mu^2 )\,
\tilde{\si}^{(0)}_{ij\to kl,I}(N,\{m^2 \},\mu^2)\nn \\ 
&&\times\;
 \bar{\cal C}_{ij\tosv kl,I}(N, \{m^2\},\mu^2) \,
\Delta_i (N+1,Q^2,\mu^2) \, \Delta_j (N+1,Q^2,\mu^2) \,
\Delta^{(s)}_{ij\to kl,I} (N+1,Q^2,\mu^2).\nn \\ 
&&
\label{eq:res:fact}
\eear
The expressions for the radiative factors $\Delta_i,\
\Delta^{(s)}_{ij\to kl,I}$ are presented below. Since  
after expansion of the exponentials the non-trivial terms
contained in $\bar{\cal C}_{ij \to kl,I}$ generate contributions
of the NNLL and higher orders, we keep $\bar {\cal C}_{ij \to kl,I}=1$ for
the rest of the calculations. Additionally, in Appendix A we 
list the results for the Mellin moments of the
colour-channel contributions
$\tilde{\si}^{(0)}_{ij\to kl,I}$ to the leading-order partonic cross sections.

\subsection{Soft radiation factors}

The expressions for the radiative factors in the $\overline{\mathrm{MS}}$ 
factorisation scheme read, up to the NLL level~\cite{KS,KOSthr,BCMNtop},              
\bear
&& \log  \Delta_{i}(N,Q^2,\mu^2)  =  \int_0^1 dz \,\frac{z^{N-1}-1}{1-z}
\int_{\mu^2}^{ Q^2(1-z)^2} \frac{dq^2}{q^2} A_i(\as(q^2))\, , \nn \\
&& \log \Delta^{(s)}_{ij\tosv kl,I}(N,Q^2,\mu^2) = 
 \int_0^1  dz \frac{z^{N-1}-1}{1-z} \frac{\as((1-z)^2 Q^2)}{\pi} D_{ ij\tosv
   kl,I}^{(1)}. 
\label{eq:radfact}
\eear
As already noted, the radiative factor $\Delta_{i}$ represents the soft and collinear gluon
radiation from the incoming partons, whereas the function $\Delta^{(s)}_{ij\to kl,I}$ takes into
account soft and large-angle gluon radiation.
The coefficient $A_i$ is a 
power series in the coupling constant $\as$,
$
A_i(\as)= \frac{\as}{\pi}{A_i}^{(1)}+ \left(\frac{\as}{\pi}\right)^2
{A_i}^{(2)}+ \dots \;
$
The universal leading logarithm (LL) and NLL coefficients $A_i^{(1)}$, $A_i^{(2)}$ are
well known~\cite{KT,CET} and given by 
\be
A_i^{(1)}= C_i, \qquad A_i^{(2)}=\frac{1}{2} \; C_i \left(\left( 
\frac{67}{18} - \frac{\pi^2}{6}\right) \CA - \frac{5}{9} \nf \right) \;,
\ee  
with \mbox{$C_g=\CA=3$} for radiation off gluon lines and \mbox{$C_q=\CF=4/3$} for
radiation off quark lines. In the case the
soft anomalous dimension matrix $\Gamma_{ij \to kl}$ is diagonal in the orthogonal
colour basis, the relation between the $D_{ ij\to
   kl,I}^{(1)}$ coefficients and the $\bar \Gamma_{ij \to kl}$
 eigenvalues,  $\bar \lambda_{ij \to kl,I}$, reads~\cite{KS,KOSthr,KOScol} 
\be
\frac{\as}{\pi} D^{(1)}_{ij \to kl,\, I} = 2{\rm Re}\,(\bar \lambda_{ij \to
  kl, I}(\as)) \;.
\label{eq:D1coeff}
\ee
The customary NLL expansions of the radiative factors~(\ref{eq:radfact}) are presented in
Appendix B.

\section{Soft anomalous dimension matrices}

Due to the same colour
structure the soft anomalous dimension matrices for the $\sq \sqb$
production and for the heavy quark
production are the same. The results for the heavy quark
production are available in the
literature~\cite{KS}. 
For the $\gl\gl$ production, however, a separate calculation
is needed.

In order to obtain resummed
cross sections up to the NLL accuracy we need the one-loop $\ord
(\as)$ result for the soft  anomalous dimension matrix, $\Gamma^{(1)}_{ij \to kl,IJ}$ .
The $\Gamma_{ij \to kl,IJ}$ matrices are 
given in terms of matrices of renormalisation constants $Z_{ij \to kl,
  IJ}$ for the soft function~\cite{BS}-\cite{KOScol}, \cite{DMS}. 
In the $\overline{\rm MS}$ scheme in the
$d=4-\epsilon$ dimensions we have 
\be
\Gamma^{(1)}_{ij \to kl, IJ}(g_s) = -\frac{g_s}{2} \frac{\partial}{\partial g_s} {\rm Res}_{\epsilon
  \to 0} Z^{(1)}_{ij \to kl,IJ}(g_s, \epsilon) \,.
\ee 
The calculation of  $\Gamma^{(1)}_{ij \to kl,IJ}$ reduces then to
evaluating the UV-divergent part of the one-loop correction to
the $S_{ij \to kl,IJ}$ function. 

\begin{figure}[t]
\begin{center}
\epsfig{file=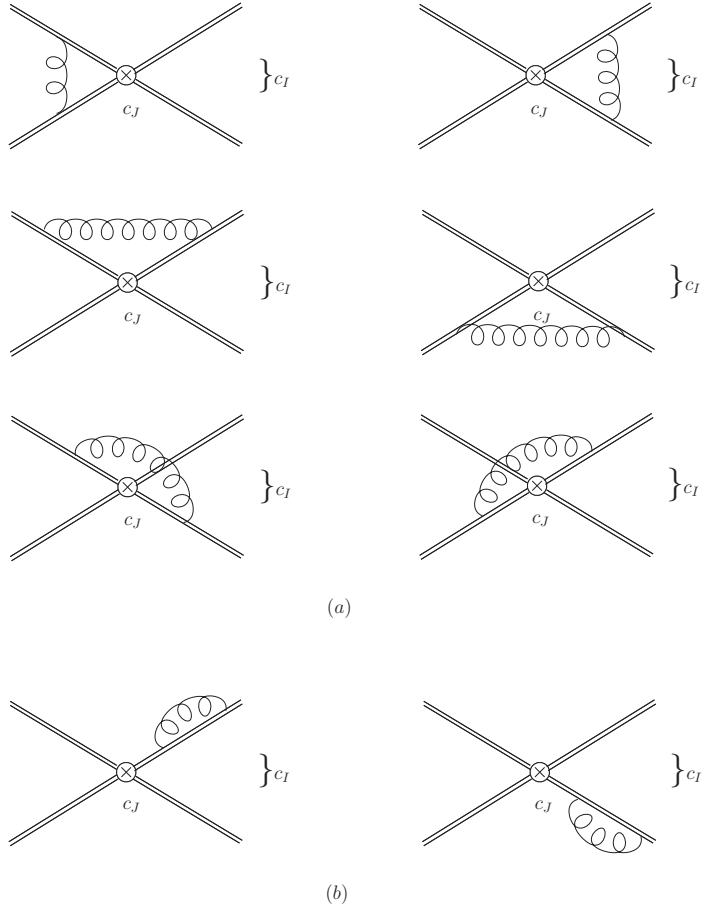,width=0.6\columnwidth} \hspace{0.05\columnwidth}
\end{center}
\caption{\it
Vertex (a) and self-energy (b) one-loop contributions to the soft
function $S_{ij \to kl,IJ}$ with  $i,j$ massless
and $k,l$ massive particles.
\label{eikonalgraphs}
}
\end{figure}

In Fig.~\ref{eikonalgraphs}, following~\cite{KS}, we show the set of eikonal 
one-loop diagrams contributing to the soft function $S_{ij \to kl,IJ}$.
All one-loop integrals needed for
the calculation of the anomalous dimension 
matrices for $q\bar q \to \gl \gl$ and $gg \to \gl \gl$ are known.
In our calculation we use
the results for the eikonal one-loop integrals from Ref.~\cite{BS,KS} which 
have been obtained in the axial gauge.
Although the integrals are the same as in the heavy-quark
production case, their contributions to the soft anomalous dimension
matrices come weighted with different colour factors. We compute 
these colour factors in the set of basis presented in Section 2. The
calculations are performed in
two independent ways: using the {\tt FeynCalc} package~\cite{Mertig:1990an} for 
{\tt Mathematica}, and using the Group Theory (Colour) Factors of Feynman diagrams 
package~\cite{colform} for {\tt FORM}~\cite{form}.
The resulting expressions for the soft anomalous dimension matrices for the 
$\gl\gl$ production are presented below.

\subsection{Gluino-pair production}

We introduce the notation
\bear
\Tgl &\;\equiv\;& 
\log\left({ m^2-\hat t \over \sqrt{m^2 \hat s }}\right)\,
\; - \; {1 - i\pi \over 2}\; ,
\nn \\ \nn
\Ugl &\;\equiv\;& 
\log\left({ m^2-\hat u \over \sqrt{m^2 \hat s }}\right)\,
\; - \; {1 - i\pi \over 2}\; ,
\nn \\ 
\Sgl &\equiv & -\frac{L_{\beta} + 1}{2}\;,
\eear
where the Mandelstam variables are given by
$$
\hat s = (p_1 + p_2)^2, \qquad \hat t = (p_1 - p_3)^2, \qquad 
\hat u = (p_1 - p_4)^2,
$$
and $\;L_{\beta} = \frac{1}{\beta}(1 - 2m^2 / \hat s)
\left( \log{1-\beta \over 1+\beta} + i\pi \right)$. 
We also define 
\mbox{$\;\BA  \!\equiv \! \Tgl +\Ugl\,$,} \mbox{$\;\C  \!\equiv \! \Tgl - \Ugl\,$.}

In the basis~(\ref{eq:base:gg:sq}) we obtain the one-loop soft anomalous dimension matrix of
the form
\be
\bar \Gamma^{(1)}_{q\bar q\to \gl\gl} (\as) \; =\;
{
{\as \over \pi} \,\left[\,
\left(
\begin{array}{ccc}
\rule{0em}{3.5ex} \;6\A\; & 0 & -\C \\
\rule{0em}{3.5ex} 0 &\, 3\A + {3\over 2}\BA \,& -{3\over 2}\C \\
\rule{0em}{3.5ex} -2\C & -{5\over 6}\C & \, 3\A + {3\over 2}\BA \,\\
\end{array}
\right) \, - \, \frac{4}{3}i\pi\,{\mathbf{\hat{I}}} \;\, \right]. 
}
\label{eq:qqbargamma}
\ee 
The one-loop soft anomalous dimension matrix for the $gg$ channel, calculated in the basis~(\ref{eq:base:gg:gl}) has
the block form
\be
\bar \Gamma^{(1)}_{gg \to \gl\gl} (\as)  \;=\;
{\as \over \pi} \; \left[\,
\left(
\begin{array}{cc}
\rule{0ex}{2ex}\; \bar \Gamma_5\; & {\mathbf{\hat 0}} \\
\rule{0ex}{2ex}{\mathbf{\hat 0}}  & \; \bar \Gamma_3\;\\
\end{array}
\right)
\, - \, 3i\pi\,{\mathbf{\hat{I}}}\;
\right]\, ,
\label{eq:gggamma}
\ee
where the five-dimensional matrix $\bar \Gamma_5$  reads
\be
\bar \Gamma_5 \; = \; 
{
\left(
\begin{array}{ccccc}
\rule{0em}{3.5ex}\;\;6\A\;\;   & 0     & 6\C    & 0   &  0  \\
\rule{0em}{3.5ex}0 & \;3\A + {3\over 2}\BA\; &    {3\over 2}\C  & 3\C   &  0  \\
\rule{0em}{3.5ex}\;{3\over 4}\C\;  & {3\over 2}\C     & \;3\A + {3\over2}\BA \;&
 0  & \;{9\over 4}\C\;   \\
\rule{0em}{3.5ex}0  & {6\over 5}\C     & 0     &\;\; 3\BA \;\;& {9\over 5}\C   \\
\rule{0em}{3.5ex}0  & 0     & {2\over 3}\C     & \;{4\over 3}\C   & 4\BA-2\A\; \\
\end{array}
\right)}\;
\ee
and the three-dimensional matrix $\bar \Gamma_3$ is diagonal, 
\be
\bar \Gamma_3 = 
\mathrm{diag}\,(\,3(\Sgl+\Ugl)\,,
                \,3(\Sgl+\Tgl)\,,
                \,3(\Tgl+\Ugl)\;).
\ee

Although both  
$\tilde S_{ij \to kl, IJ}$  and $\tilde U_{i \bar i}$ are gauge-dependent functions,
results for the one-loop $\bar{\Gamma}^{(1)}_{ij \to kl, IJ}$ matrices presented here  are gauge-invariant.
The gauge dependence in the sum of vertex and
self-energy corrections cancels against the gauge dependence of the Drell-Yan
anomalous dimension~\cite{KS, LSV}. The $\tilde U_{i \bar i}$ soft function, in turn, carries the same gauge dependence 
as the ratio of the incoming jet functions $(\tilde \psi_{i/i}/\tilde
f_{i/i})^2$~\cite{KOScol, DMS}.  In practice, we fix the gauge $A^0=0$
to calculate ${\Gamma}^{(1)}_{ij \to kl, IJ}$ and  use $\Gamma^{(1)}_{i \bar i} (\as) = C_i \as/\pi$ in Eq.~(\ref{eq:gammabar}).

At the production threshold $\beta_{\gl} \to 0$ 
the soft anomalous dimension matrices $\bar \Gamma_{q\bar
  q\to \gl\gl}$ and $\bar \Gamma_{q\bar q\to \gl\gl}$ approach the diagonal form.
The off-diagonal terms, proportional to $\C$, vanish like
$\beta_{\gl}$ for $\beta_{\gl} \to 0$ and thus may be neglected.
Using Eq.~(\ref{eq:D1coeff}) we obtain 
\bear
\{D_{q\bar q \to \gl\gl,\,I} ^{(1)}\}&=& \{0,-3,-3\}\\
\{D_{gg\to \gl\gl,\,I} ^{(1)}\} &=& \{0,-3,-3,-6,-8;-3,-3,-6\} \;,
\eear
where the index $I$ indicates a colour channel, defined by the base
tensor $c^{q,\gl}_I$ and $c^{g,\gl}_I$ as in (\ref{eq:base:gg:sq})
and (\ref{eq:base:gg:gl}), correspondingly.
Note that the values of the  $D^{(1)}$-coefficients are the negative
values of the quadratic Casimir operators for the SU(3) 
representations for the outgoing state. 
This agrees with the physical picture of the soft gluon 
radiation from the total colour charge of the heavy-particle pair 
produced at threshold~\cite{BCMNtop}.

\subsection{Squark-antisquark production}
For completeness, we also list here the NLL coefficients  $D^{(1)}_{
  ij\tosv \sq \sqb,\,I}$ which we need for our numerical calculations.
They have been first obtained in the calculation of resummed cross
sections
for the heavy quark production and read~\cite{KS,BCMNtop} 
 \bear
\{D^{(1)}_{q \qbar \to \sq \sqb,I}\} &=& \{0, -3\} \\
\{D^{(1)}_{gg \to \sq \sqb,I}\} &=& \{0, -3, -3\}
\eear
with the index $I$ indicating the corresponding tensors in the
$s$-channel colour basis $c^{q,\sq}_I$ and $c^{g,\sq}_I$, given in (\ref{eq:base:qq:sq}) 
and (\ref{eq:base:gg:sq}).

\subsection{Checks of analytical results}

The computational framework applied here has been tested by re-deriving the 
known results for the one-loop soft anomalous matrices for the pair-production of massive
particles. We have reproduced the
results for the heavy quark production~\cite{KS} both in the
$q\bar q$ and in the $gg$ channel.

Another test of our calculations has been based on a comparison between 
expansion of the resummed cross section, Eq.~(\ref{eq:res:fact}), and the
Mellin moments of the NLO corrections taken in the threshold limit. The analytic form of the NLO corrections in the
threshold limit in momentum space is known for all squark and gluino
production processes~\cite{BHSZ1}.    
More precisely, for each of the partonic processes, 
$q_i \bar q_j \to \sq\sqb$, $g g \to \sq\sqb$, $q\bar q \to \gl\gl$, and  
$gg \to \gl\gl$, we have extracted terms with $\log^2 N$ 
and  $\log N$ from the ${\cal O (\as)}$ expansion of the
corresponding resummed formula. 
In each case the result has been compared with the collection of terms
logarithmic in $N$ in the Mellin transform of the $\ord (\as)$
correction taken in the large $N$ limit. The contributions enhanced by double logarithms, 
i.e.\ ${\cal O}(\as\log^2 N)$, depend only on the colour charges of the
incoming partons and thus do not provide a check of the soft anomalous
dimension matrices. A non-trivial cross-check is, however, provided
by the contributions to the NLO correction with single logarithms, ${\cal
  O}(\alpha_s\log N)$, that are sensitive to the eigenvalues of the
$\bar \Gamma$ matrices. Our result for the 
resummed cross section, Eq.~(\ref{eq:res:fact}), agree in this way with results of  Ref.~\cite{BHSZ1}.

\section{Resummation of leading Coulomb corrections}

Important higher order corrections to cross sections for production of
coloured particles come from multiple exchanges of Coulomb gluons
between the produced particles.  This type of
corrections should be then taken into account to all orders~\cite{BFS,CoulLO,Coultop}. In the threshold limit, the Coulomb
corrections 
are enhanced by the inverse powers of $\beta$.   
At $n$-th order in perturbation theory the leading corrections are of the 
form $C_{\rm Coul}^{(n)} \as^n/\beta^n$. These leading contributions can be
summed to all orders using~\cite{CoulLO}
\be
\hat\sigma ^{(C)} _{ij \to kl,\, \mai} \; = \; 
\sigma ^{(0)} _{ij \to kl,\, \mai} \; 
\Delta^{(C)}\! 
\left( {\pi\,\alpha_s \over \beta}\,  \kappa_{ij \to kl,\, \mai}\,\right),
\ee
where
\be
\Delta^{(C)} (z) \; = \,  {z \over \exp(z) -1}\,,
\ee
and $\beta=\beta_{\sq}$ for $\sq\sqb$ production, 
$\beta=\beta_{\gl}$ for $\gl\gl$ production.
For the processes of interest, the $\kappa$ coefficients, calculated in the
set of colour basis introduced in Section 2, are given by
\bear
\kappa_{q\bar q \to \sq\sqb,\, I} & = & 
\left(\,-{4 \over 3}\,,\,{1\over 6} \,\right), \nn \\
\kappa_{g g \to \sq\sqb,\, I} & = & 
\left(\,-{4 \over 3}\,,\,{1\over 6}\,,\,{1\over 6}\, \right), \nn \\
\kappa_{q\bar q \to \gl\gl,\, I} & = & 
\left(\,-3\,,\,-{3\over 2}\,,\,-{3\over 2}\,\right), \nn \\
\kappa_{g g \to \gl \gl,\, I} & = & 
\left(\,
-3\,,\,-{3\over 2}\,,\,-{3\over 2}\,,\,0\,,\,1\,;\,-{3\over 2}\,,\,
-{3\over 2}\,,\,0\,\right)\, . \nn 
\eear

The ${\cal O}(\alpha_s)$ Coulomb correction is a part of the full NLO 
result. Since we are interested in corrections above NLO, we subtract
it from $\hat\sigma ^{(C)} _{ij \to kl,\, \mai}$. In this way we
define the correction $\delta\hat\sigma ^{(C)} _{ij \to kl,\, \mai}$
due to leading (in terms of powers of $1 / \beta$) Coulomb
contributions above NLO
\be
\delta\hat\sigma ^{(C)} _{ij \to kl,\, \mai} \; = \; 
\hat\sigma ^{(C)} _{ij \to kl,\, \mai} \; - \;
\left. \hat\sigma ^{(C)} _{ij \to kl,\, \mai} \right|_{(\NLO)}\;.
\ee
The corresponding Coulomb correction at the hadronic level then reads
\be
\delta\si^{(C)}_{h_1 h_2  \to k l}(\rho,\{m^2\}) \,=\,
\sum_{i,j; \;\mai}\; \int d x_1 \, d x_2\; d \rhohat
\,\delta\left(\rhohat-\frac{\rho}{x_1 x_2}\right)\,
f_i(x_1,\mu ) \;f_j(x_2,\mu ) \, 
\delta\sih^{(C)} _{ ij \to kl, {\mai}}(\rhohat, \{m^2\}, \mu^2) \;.
\label{eq:chadr}
\ee

\section{Predictions for squarks and gluino production at the LHC}

We investigate in detail the effect of the soft gluon corrections on
the cross sections for two sparticle production processes at the LHC,
$pp \to \gl \gl$ and  $pp \to \sq \sqb$,  at  $\sqrt{S} = 14
\TeV$. The main results obtained in this section are the
resummation-improved 
total cross sections. We also study
the effect of the resummed leading Coulomb corrections to the total
cross sections. Moreover, we present a detailed analysis of the
soft gluon corrections in partonic channels of hadronic processes,
including also the dependence of soft gluon effects on the colour 
structure of the hard matrix element. 
All numerical calculations were performed using two independent computer codes.

\subsection{Inversion and matching}

The resummation-improved cross sections are obtained through matching the NLL resummed expressions with the full NLO cross sections,
\bear
\label{hires}
\si^{\rm (match)}_{h_1 h_2 \to kl}(\rho, \{m^2\}) \; & = & \;
\si^{\rm (NLO)}_{h_1 h_2 \to kl}(\rho, \{m^2\})  \\
& & \; + \;  \sum_{i,j=q,\qbar,g}\,
\int_\mathrm{CT} 
\; \rho^{-N}\; \tilde f_{i/h_1} (N+1, \mu^2)  \; 
\tilde f_{j/h_{2}} (N+1,\mu^2) 
\nn \\
&&\qquad \qquad \qquad \times\;
\left[ 
\tilde \si^{\rm (res)}_{ij\to kl} (N, \{m^2\})
\; - \; \tilde \si^{\rm (res)}_{ij\to kl} (N, \{m^2\})
{ \left. \right|}_{\scriptscriptstyle({\NLO})}\, \right], \nn
\eear
where $\tilde \si^{\rm (res)}_{ij\to kl}$ is given through
Eq.~(\ref{eq:res:fact}) together with Eq.~(\ref{eq:Nspacexsec}), and  $
\tilde \si^{\rm
  (res)}_{ij\to kl} \left. \right|_{\scriptscriptstyle({\NLO})}$
represents its perturbative expansion truncated at the order of $\as$
associated with the NLO correction. 

The inverse Mellin transform (\ref{hires}) is evaluated numerically using 
a contour $ \mathrm {CT}$ in the complex-$N$ space according to the ``Minimal Prescription'' 
method developed in Ref.~\cite{CMNTmp}. More specifically, we use a 
contour parameterised by a parameter $\chi$,  $N= C_0+\chi \exp(\pm i
\phi)$, as described in~\cite{KSVew,KSVh}. In order to be able to use
available parameterisations of parton distribution functions in
$x$-space we apply the method introduced in~\cite{KSVew}. The NLO cross sections are
evaluated 
using \mbox{\sc Prospino}~\cite{prospino}, the numerical package based 
on calculations employing the $\overline{\mathrm{MS}}$ renormalisation and factorisation schemes.

\subsection{Numerical results}

In the phenomenological analysis we consider a wide range of gluino and squark masses. Left- and right-handed squarks of all flavours are assumed to be mass degenerate. For the $\gl\gl$ production we vary the gluino mass, $\mgl$, 
between 200~GeV and 2~TeV. Similarly, for the $\sq\sqb$ production we take~\footnote{For the highest masses considered here, the experimental exploration at the LHC will require luminosities of $\ord(100 \fba^{-1}$).} \mbox{200~GeV$\,<\msq <\,$2~TeV}. We present the results for a fixed ratio of gluino and squark masses, $r=\frac{\mgl}{\msq}$, and choose the following values 
$r=0.5, \; 0.8\; 1.2\; 1.6,\;2.0$. The $\sq \sqb$ cross section accounts 
for production of all $\sq \sqb$ flavour combinations apart from the 
ones with scalar top particles.

 For most of the phenomenological results, we use the
 CTEQ6M~\cite{cteq6m} parameterisation of parton distribution
 functions (pdfs). In addition, we give the total cross sections and
 study their scale dependence for the MSTW
 parameterisation~\cite{MSTW} of the pdfs. Unless explicitly specified
 otherwise, the CTEQ6 pdfs are applied. In both sets of pdfs the
 usual assumption of five massless quark flavours active at large
 scales is made. Consequently, in the NLO and NLL calculations we use
 the two-loop $\overline{\mathrm{MS}}$ QCD running coupling constant
 $\as$ with $\nf=5$. We also show some results obtained at the LO
 accuracy using CTEQ6L1 parameterisation of the pdfs and the one-loop
 running coupling constant with 5~flavors. For all 
parameterisations we consistently use  the corresponding default
 values of $\Lambda^{(5)}$. The effects due to virtual top quarks and virtual sparticles in the running 
of $\as$ and in the evolution of pdfs are thus not included in our predictions.
However, the value of the top mass $m_t=175 \GeV$ enters the matched
NLL cross sections through the NLO corrections.

\begin{figure}[h]
\begin{center}
\begin{tabular}{ll}
\epsfig{file=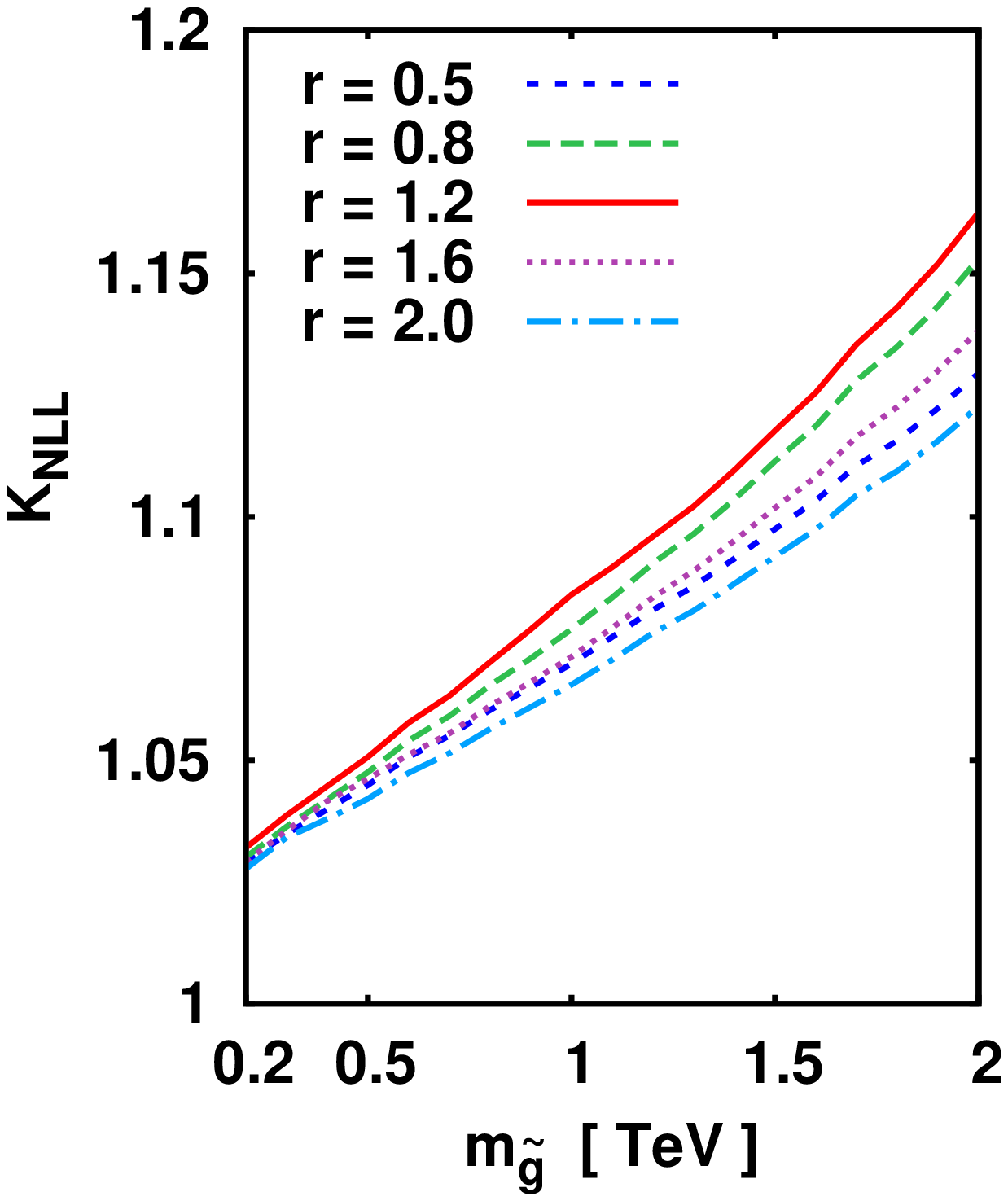,width=0.42\columnwidth} \hspace{0.05\columnwidth} &
\epsfig{file=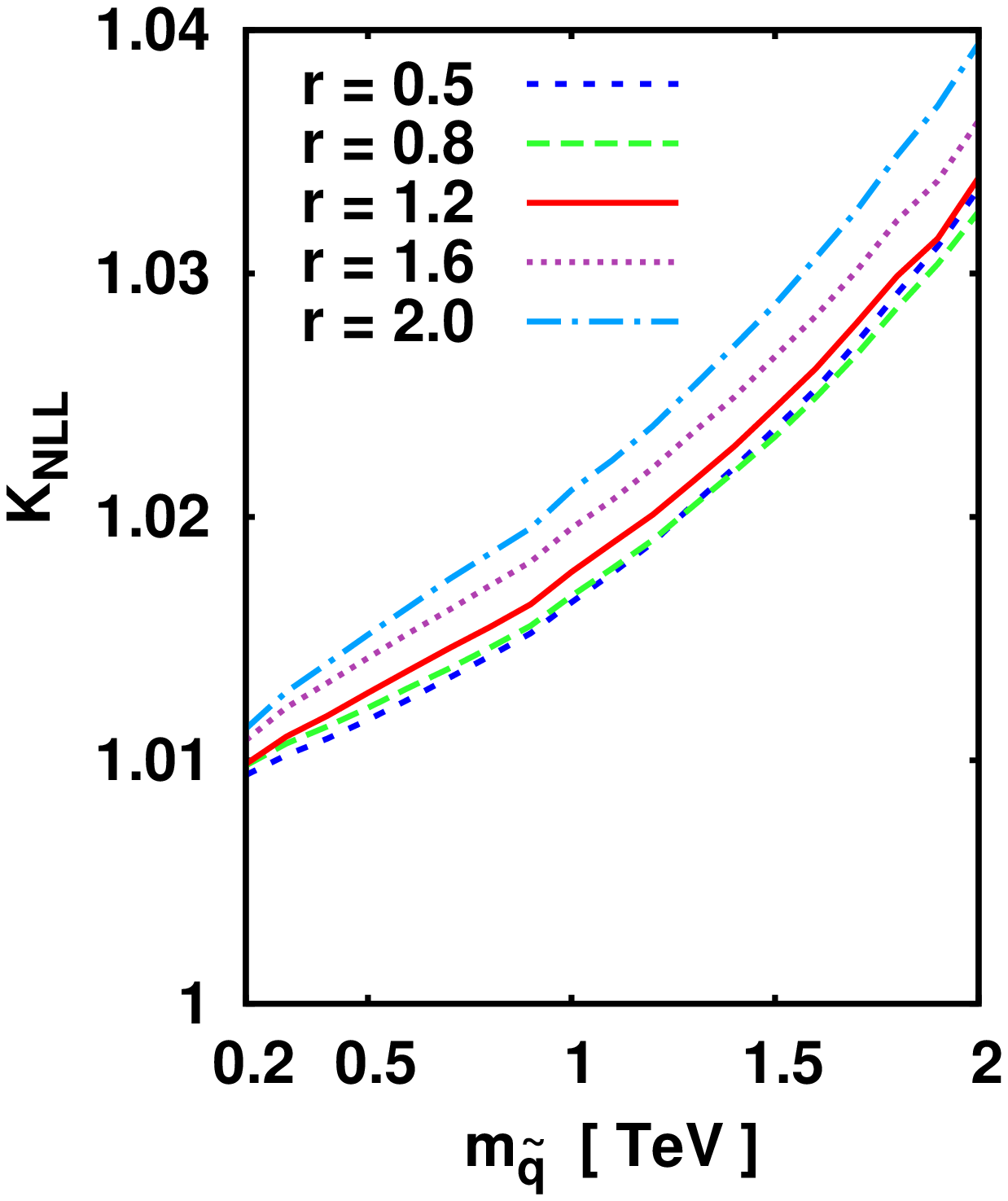,width=0.42\columnwidth}\\
{\large\bf a)} & {\large\bf b)} \end{tabular}
\end{center}
\caption{\it 
The NLL $K$-factor,
\mbox{$K_{\NLL}$},
for the $\gl\gl$ (a) and the $\sq\sqb$ (b) total production cross section 
at the LHC as a function of gluino and squark mass, respectively;
$r=\mgl / \msq$.
\label{fig:total}}
\end{figure} 

In Fig.\ \ref{fig:total} we present the relative enhancement of the NLO total cross sections due to soft gluon resummation, \mbox{$K_{\NLL}\,\equiv\, \si^{(\mathrm{match})}/\si^{\NLO}$}. The NLL $K$-factors are shown for $\gl\gl$ and $\sq\sqb$ production cross sections at the LHC, in Fig.~\ref{fig:total}a and Fig.~\ref{fig:total}b, respectively.
In the plots we set the scales $\mu_F=\mu_R=\mu_0$, where $\mu_0 = m_{\gl}$
($\mu_0 = m_{\sq}$) for the $\gl\gl$ production (the $\sq \sqb$ production).
The NLO CTEQ6M pdfs are used. $K_{\NLL}$ grows with the final-state
mass and depends on the mass ratio~$r$ in a moderate way. The relative
correction $K_{\NLL}-1$ reaches 16\% (8\%) for the
$\gl\gl$~production with $r=1.2$ and $\mgl= 2$~TeV (1~TeV), and 4\%
(2\%) for the $\sq\sqb$~production with $r=2$ and $\msq= 2$~TeV
(1~TeV). The stronger effect found in the $\gl\gl$ production follows
from the dominance of  the $gg\to \gl\gl$ channel and hence larger
colour factors. It comes from the fact that the size of the soft-collinear
radiative factor $\Delta_i$ increases with  higher colour charge of
the incoming parton. Similarly, the size of the soft non-collinear gluon corrections increases with the total colour charge of the final state, which may be the highest in the   
$gg\to \gl\gl$ case. 

\begin{figure}[h]
\begin{center}
\begin{tabular}{ll}
\epsfig{file=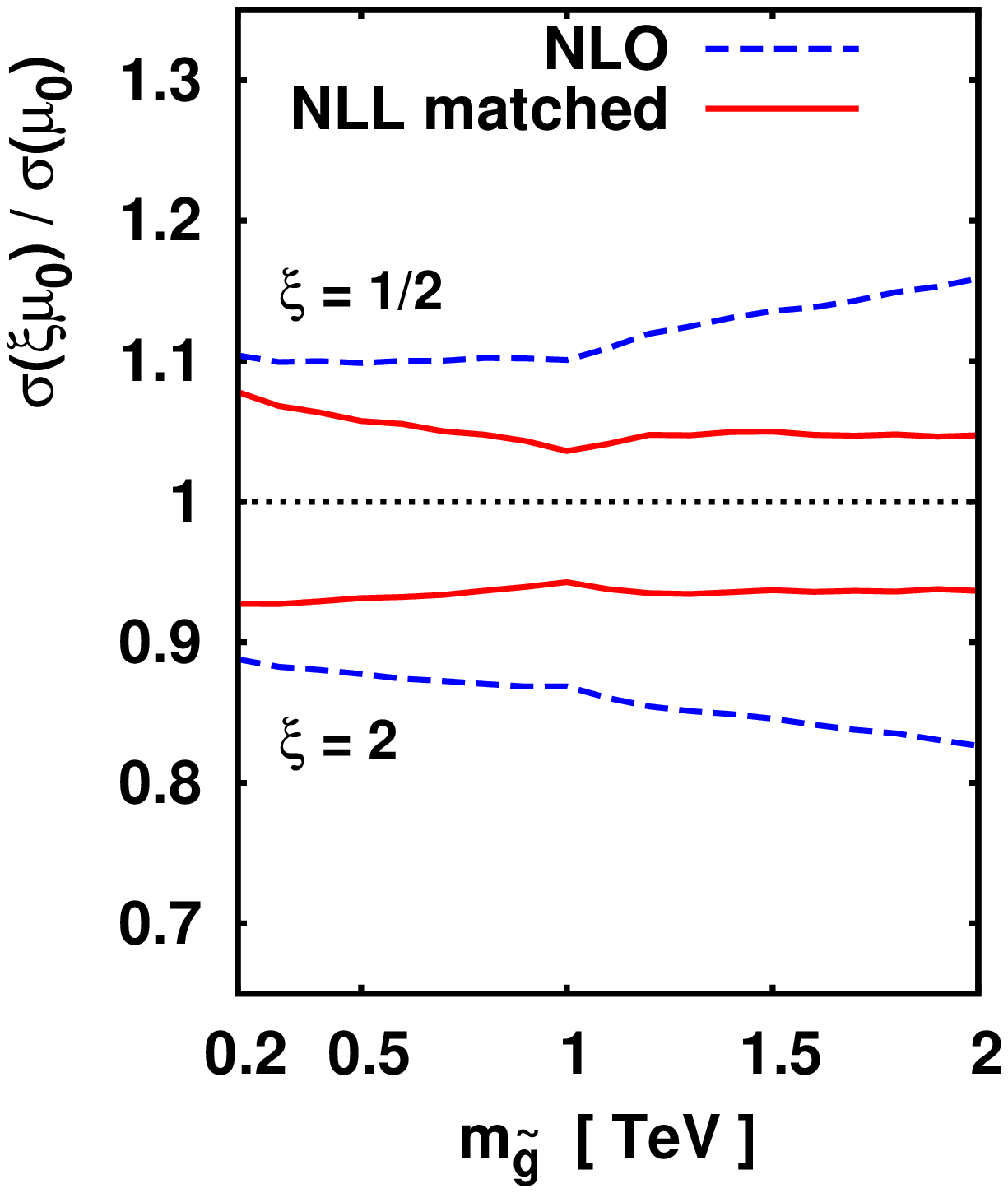,width=0.42\columnwidth} \hspace{0.05\columnwidth} &
\epsfig{file=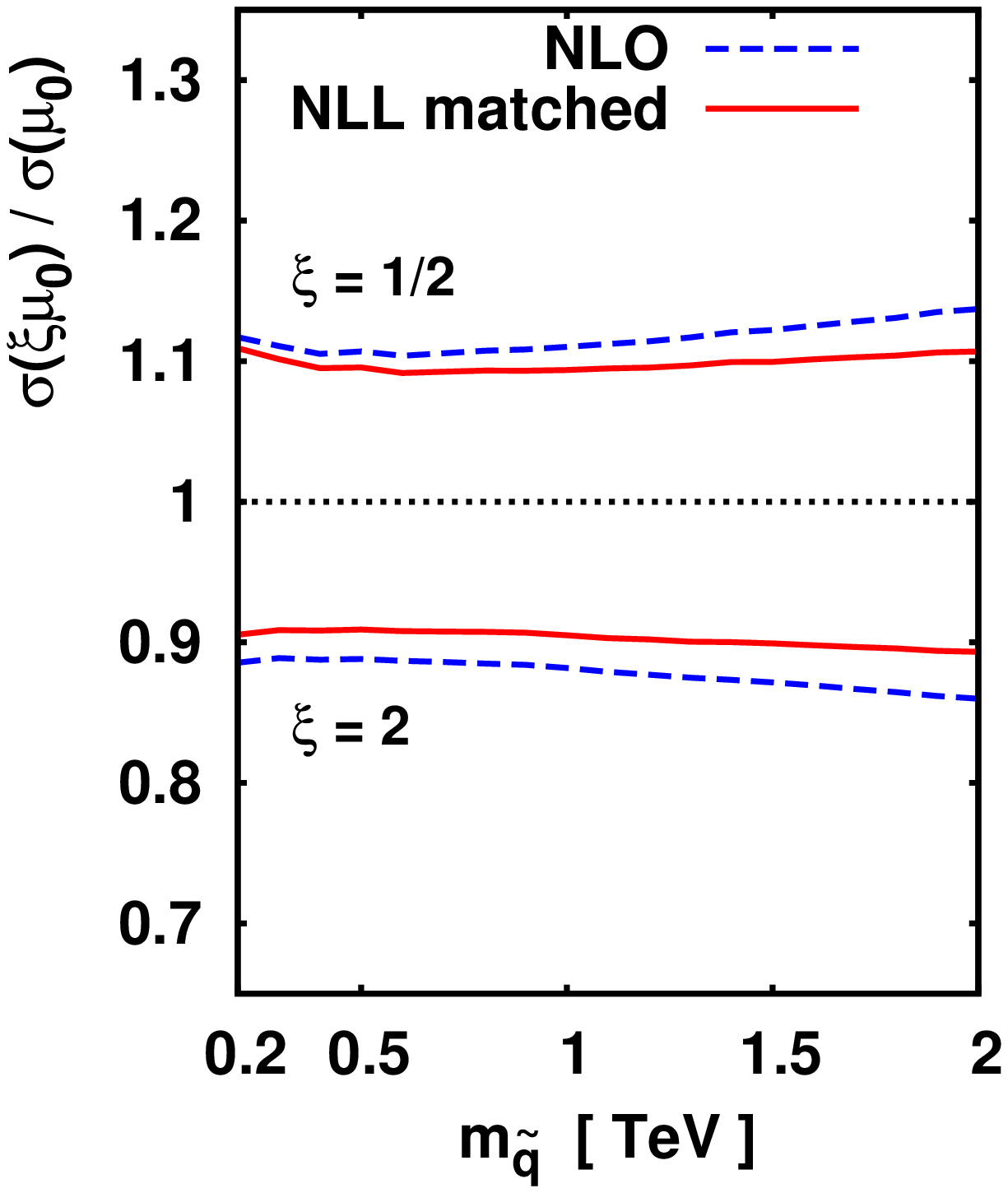,width=0.42\columnwidth} \\
{\large\bf a)} & {\large\bf b)} \end{tabular}
\end{center}
\caption{\it Scale dependence of the total $\gl\gl$ (a)  
and $\sq\sqb$ (b) production cross section at the LHC 
(see the text for explanation).
\label{fig:scale}
}
\end{figure} 

We also investigate the dependence of the matched NLL cross section
on the values of factorisation and renormalisation scales, in comparison 
to the NLO cross section. To illustrate our results we choose 
\mbox{$\mu=\mu_F=\mu_R$} and $r=1.2$. In Fig.~\ref{fig:scale}a and  Fig.~\ref{fig:scale}b we plot the ratios  \mbox{$\sigma^{\NLO}(\mu=\xi \mu_0) /
\sigma^{\NLO}(\mu=\mu_0)$} and 
\mbox{$\sigma^{\mathrm{(match)}}(\mu=\xi \mu_0)
/ \sigma^{\mathrm{(match)}}(\mu=\mu_0)$}, obtained by varying
$\xi$ between $\xi=1/2$ and $\xi=2$.
Due to resummation the scale sensitivity
of the $\gl\gl$ production cross section reduces significantly, by a
factor of $\sim 3$ ($\sim 2$) at $\mgl = 2$~TeV ($\mgl =1$~TeV). 
At  $\mgl>1$~TeV the theoretical error of the matched NLL $\gl\gl$ 
cross section, defined by changing the scale \mbox{$\mu=\mu_F=\mu_R$} 
around $\mu_0 = m_{\gl}$ by a factor of~2, is around 5\%. 
In the case of the $\sq \sqb$ production the reduction of the scale 
dependence due to including soft gluon corrections in the theoretical 
predictions is moderate.

\begin{figure}[h]
\begin{center}
\begin{tabular}{ll}
\epsfig{file=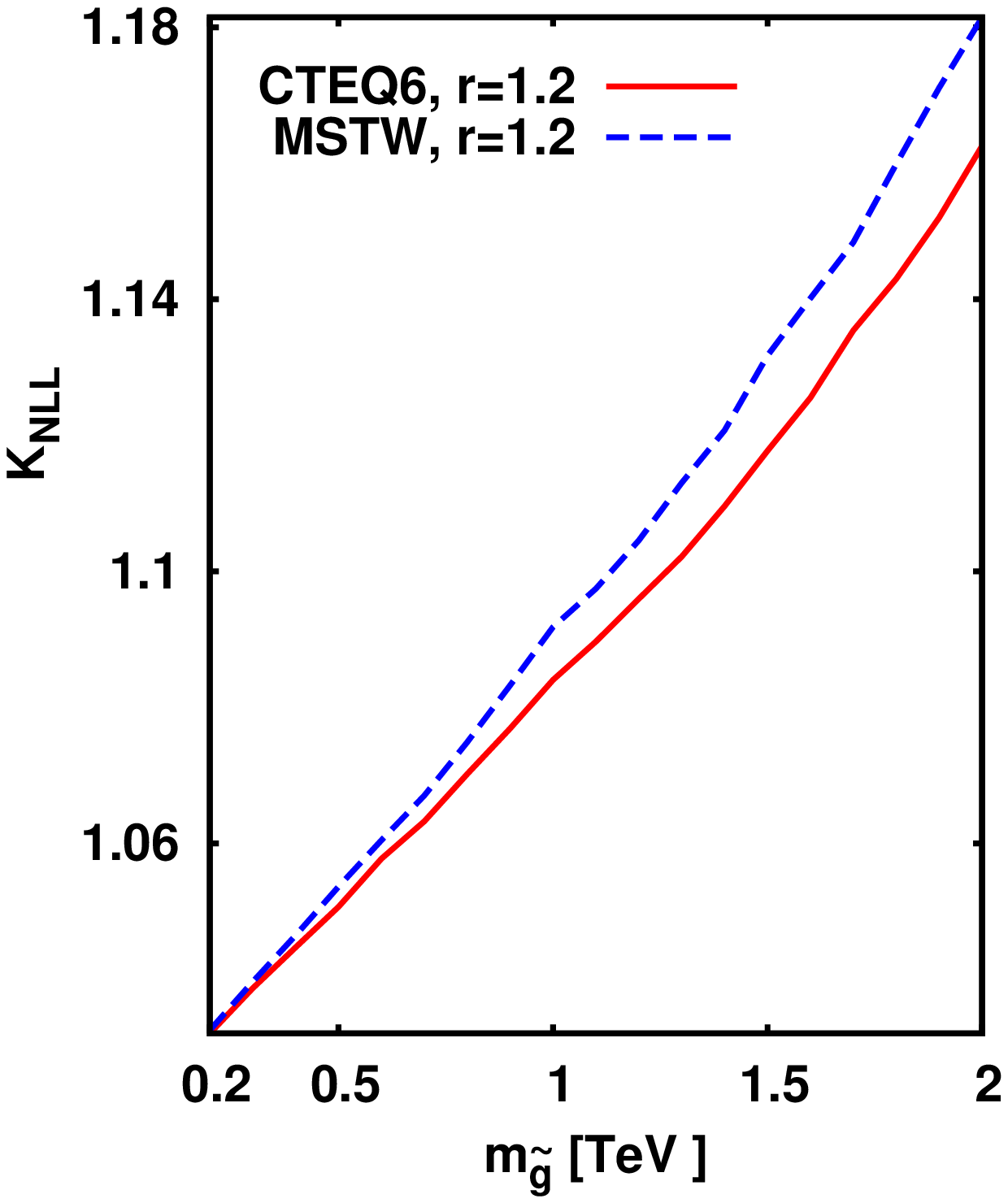,width=0.42\columnwidth} \hspace{0.05\columnwidth} &
\epsfig{file=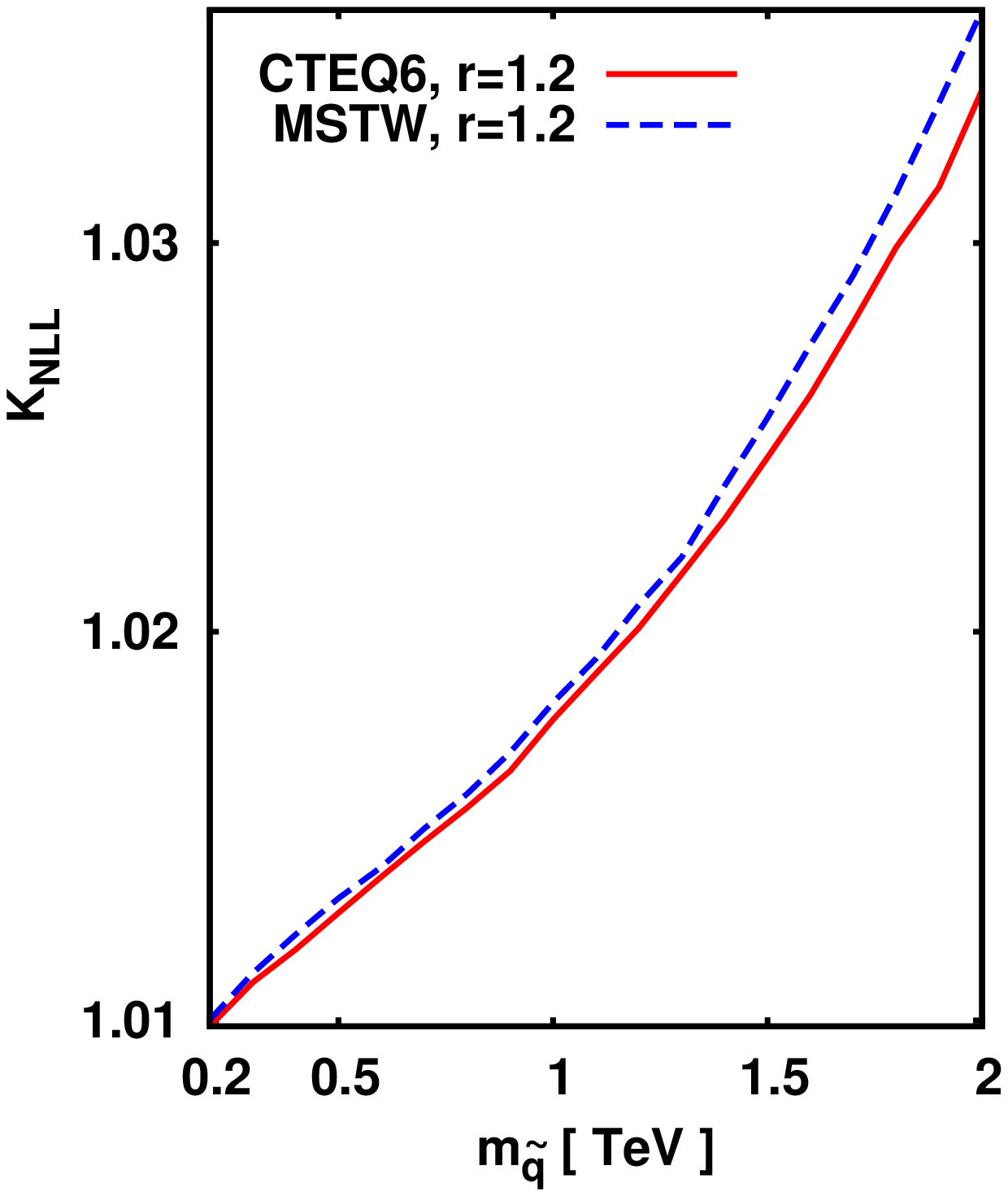,width=0.42\columnwidth} \\
{\large\bf a)} & {\large\bf b)} \end{tabular}
\end{center}
\caption{\it 
The relative NLL correction
\mbox{$K_{\NLL}$}
for the $\gl\gl$ (a) and the $\sq\sqb$ (b) total production cross section 
at the LHC as a function of gluino and squark mass, respectively;
$r=\mgl / \msq$. The continuous lines correspond to the CTEQ6M pdfs and 
the dashed lines to the MSTW pdfs.}
\label{fig:pdfs}
\end{figure} 

The dependence on the pdf parameterisation of the NLL
$K$-factors for $\gl\gl$ and $\sq\sqb$ production at the LHC is
presented in Fig.\ \ref{fig:pdfs}a and Fig.\ \ref{fig:pdfs}b,
respectively. The MSTW pdfs, shown with the dashed lines, lead to
slightly larger NLL~$K$-factors than the CTEQ6M pdfs. In this figure
we choose $r = m_{\gl} / m_{\sq} = 1.2$. The difference between the
$K$-factors for the two parameterisations is moderate but it grows
with increasing mass of the produced sparticles. It is expected since
for larger masses of sparticles larger values of the factorisation
scale $\mu_F$ and of the momentum fraction $x$ are probed. 
For larger values of $\mu_F$ and $x$ the pdfs are currently not so well
constrained, 
see e.g.\ the discussion of the uncertainties of the pdfs given in \cite{MSTW}.

\begin{figure}[p]
\begin{center}
\begin{tabular}{lll}
\epsfig{file=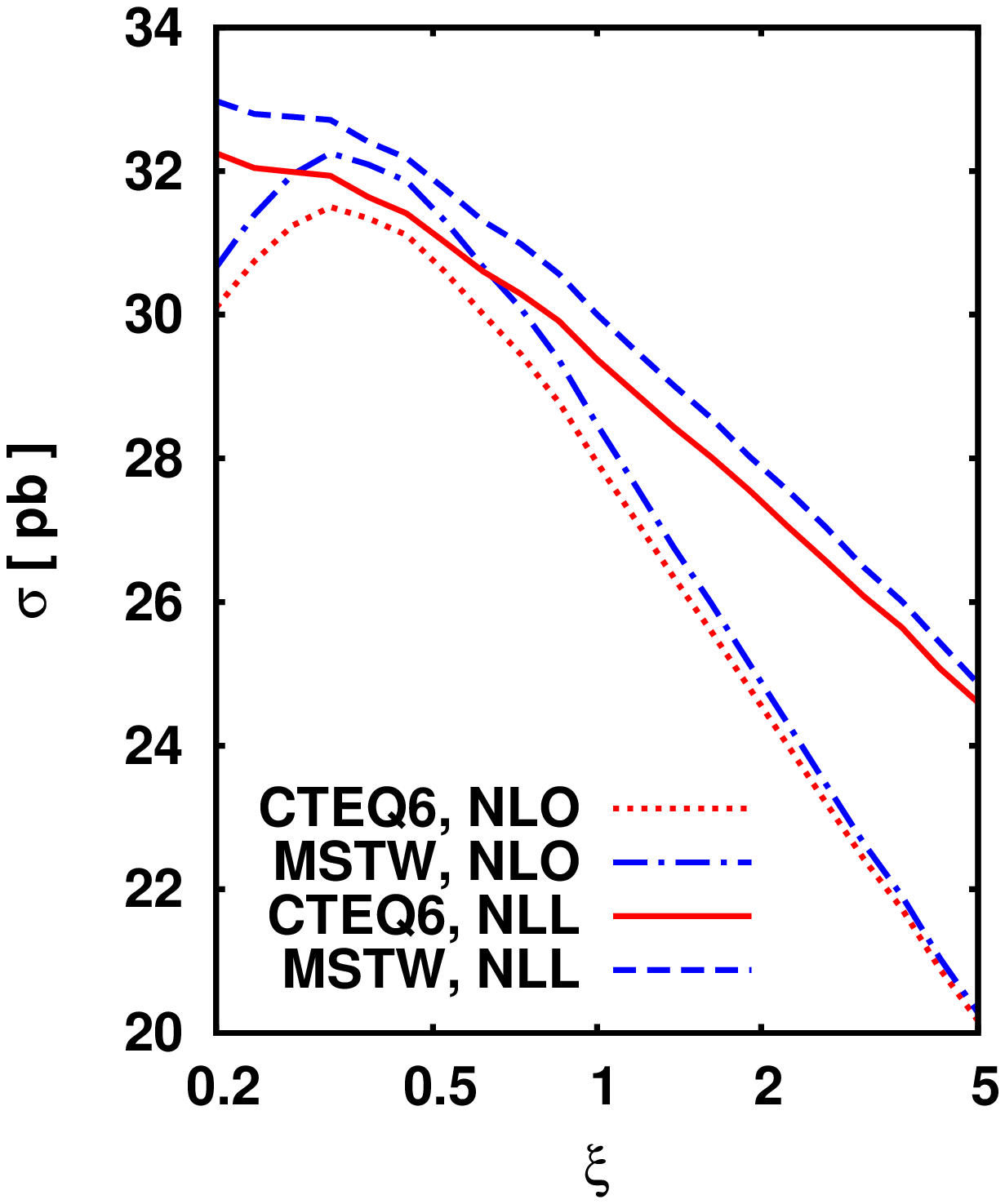,width=0.32\columnwidth} &
\epsfig{file=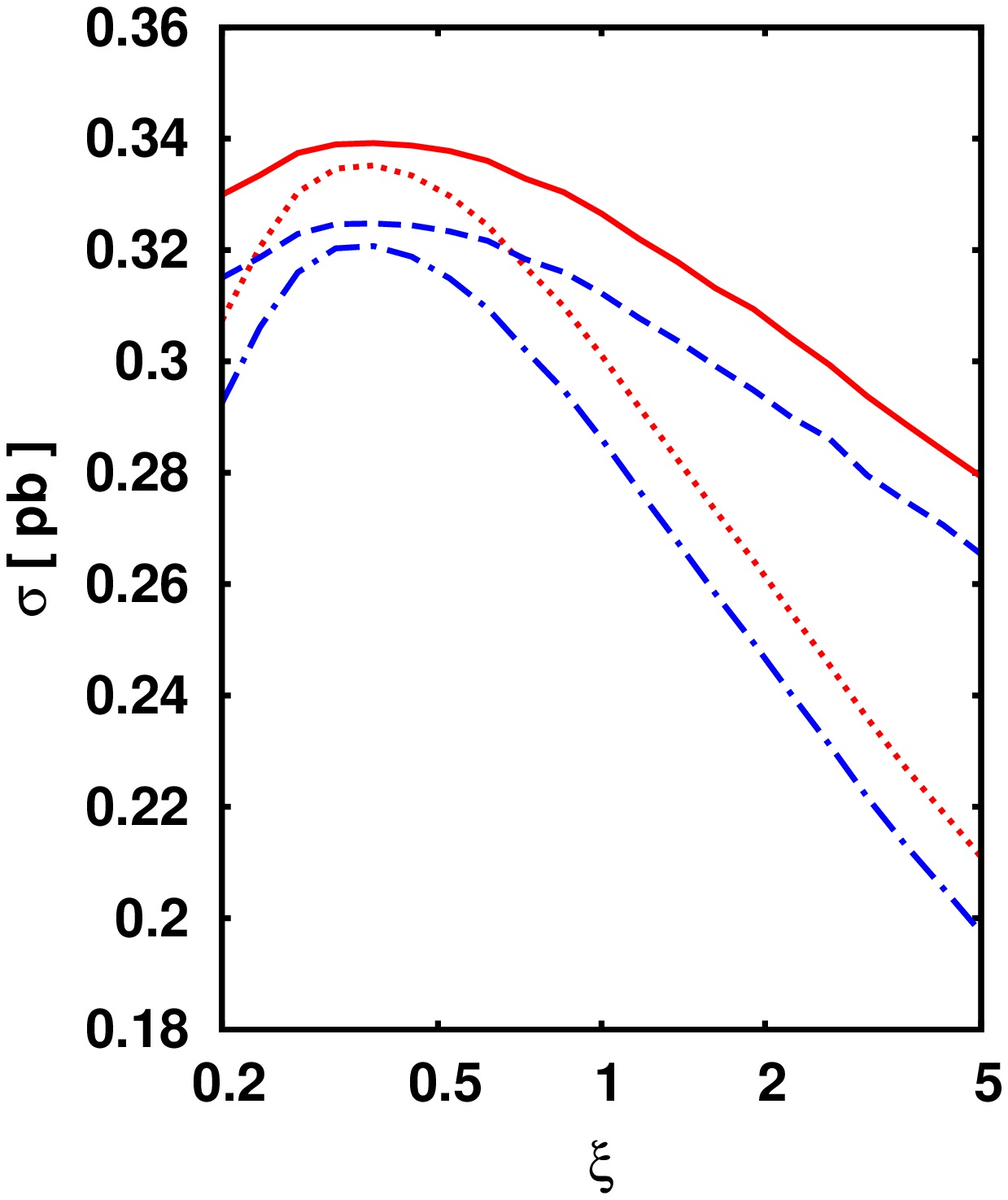,width=0.32\columnwidth}  &
\epsfig{file=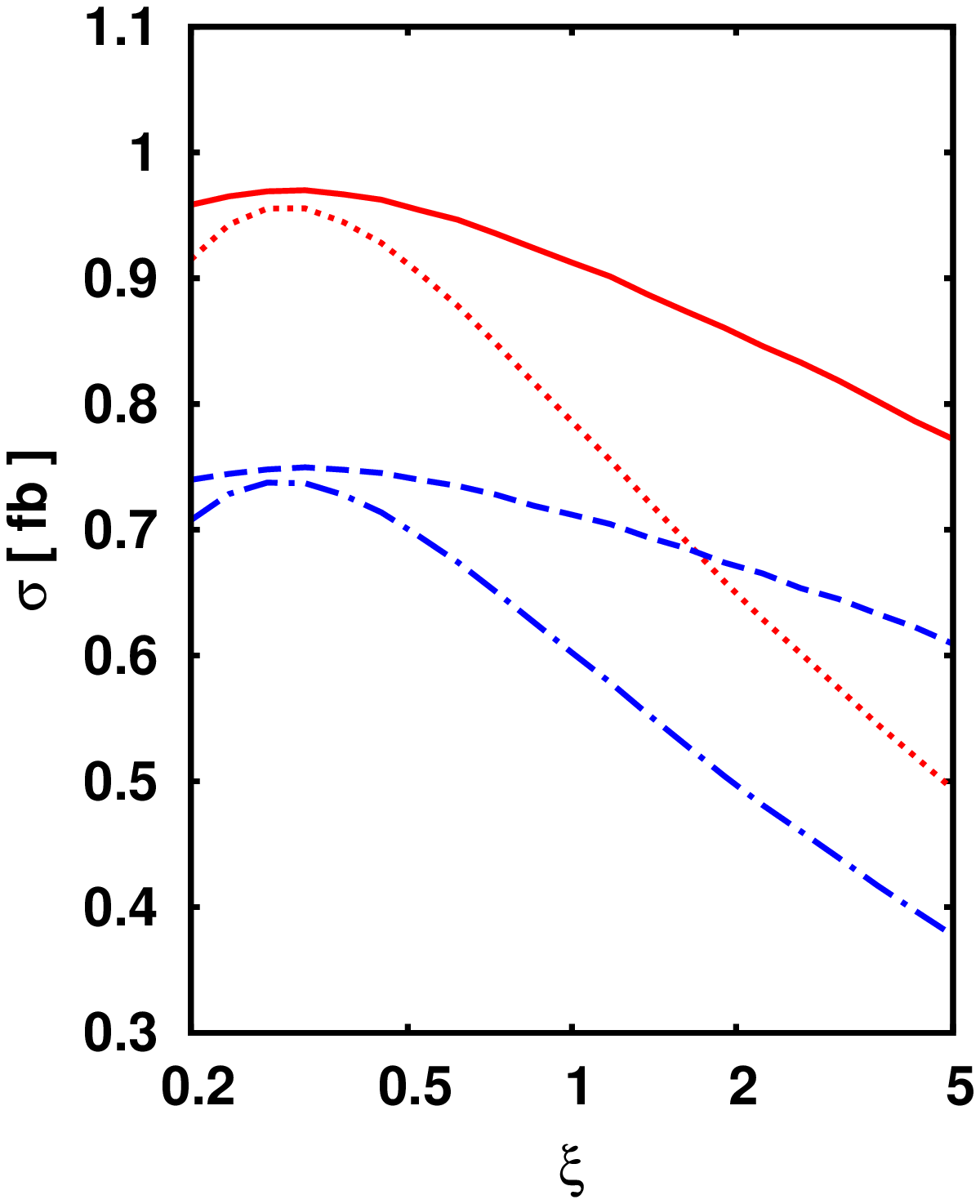,width=0.32\columnwidth}  \\
{\large\bf a)} & {\large\bf b)} & {\large\bf c)} \end{tabular}
\end{center}
\caption{\it 
The dependence of the total cross section $pp \to \gl\gl$ on the renormalisation and factorisation scale, $\mu = \mu_R = \mu_F = \xi m_{\gl}$, for $r = m_{\gl} / m_{\sq} = 1.2$ and: (a) $m_{\gl} = 0.5$~TeV,  (b) $m_{\gl} = 1$~TeV and (c) $m_{\gl} = 2$~TeV. The four lines in each plot correspond to the NLO and NLL cross section, each one evaluated with CTEQ6M and MSTW pdfs, see the legend. 
 \label{fig:pscaleg} 
}
\end{figure}

\begin{figure}[p]
\begin{center}
\begin{tabular}{lll}
\epsfig{file=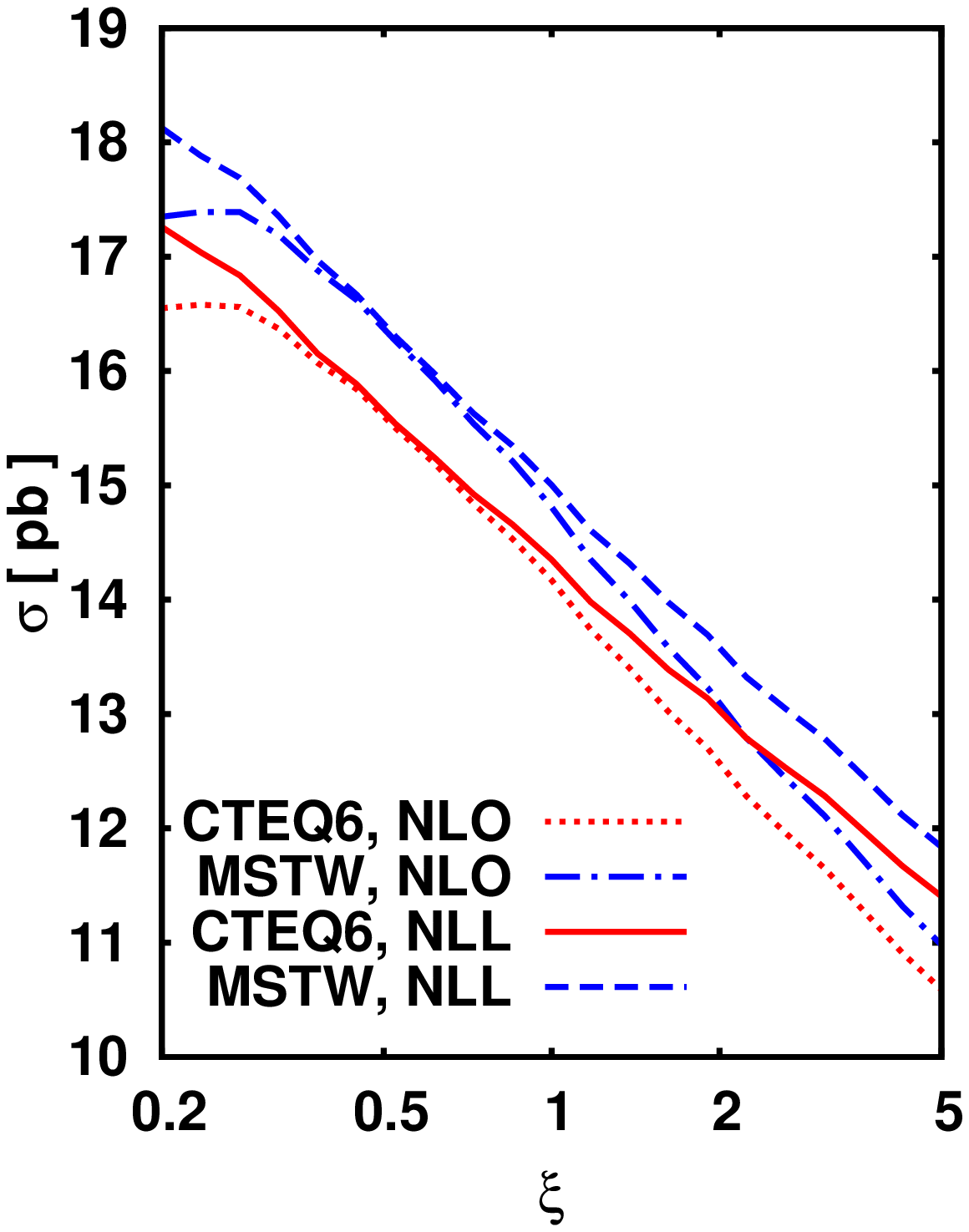,width=0.32\columnwidth} & 
\epsfig{file=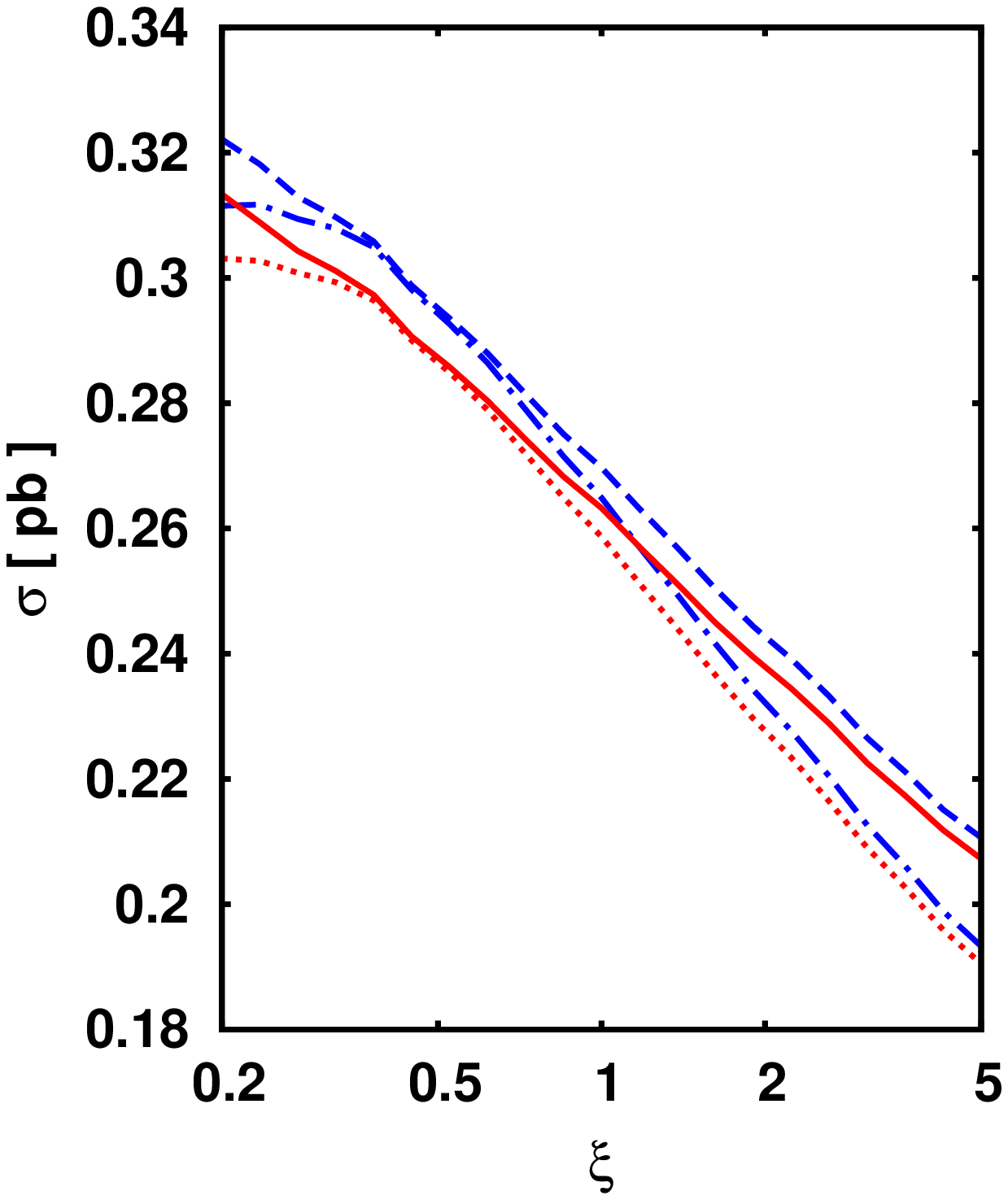,width=0.32\columnwidth}  & 
\epsfig{file=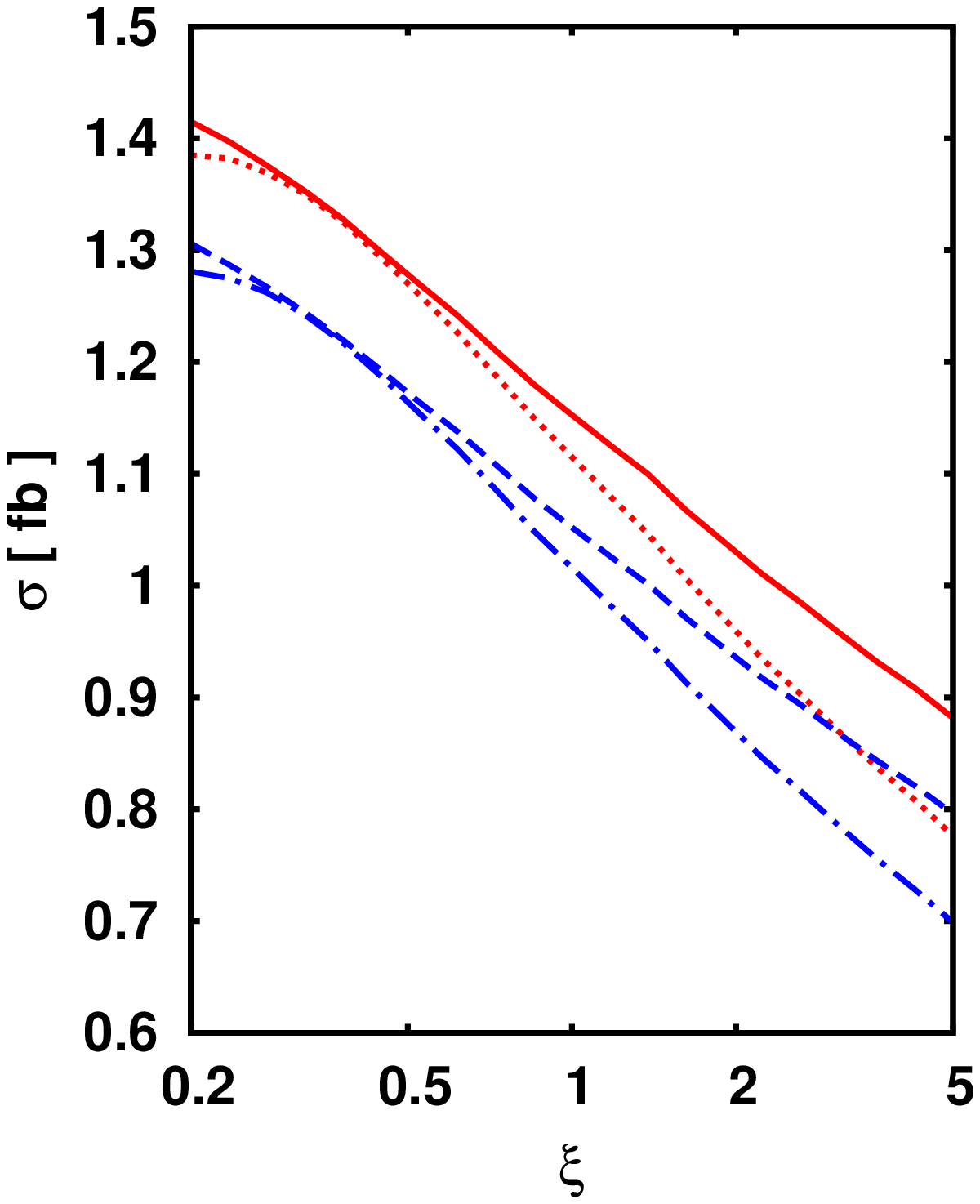,width=0.32\columnwidth}  \\
{\large\bf a)} & {\large\bf b)} & {\large\bf c)} \end{tabular}
\end{center}
\caption{\it 
The dependence of the total cross section $pp \to \sq\sqb$ on the renormalisation and factorisation scale, $\mu = \mu_R = \mu_F = \xi m_{\sq}$, for $r = m_{\gl} / m_{\sq} = 1.2$ and: (a) $m_{\sq} = 0.5$~TeV,  (b) $m_{\sq} = 1$~TeV and (c) $m_{\sq} = 2$~TeV. The four lines in each plot correspond to the NLO and NLL cross section, each one evaluated with CTEQ6M and MSTW pdfs, see the legend. 
 \label{fig:pscaleq} 
} 
\end{figure}

Next, we analyse the combined dependence of the NLO
and the matched NLL total cross sections on the factorisation and the
renormalisation scale and on the type of the pdfs. 
The results are shown in Fig.\ \ref{fig:pscaleg} and Fig.\ \ref{fig:pscaleq} for the $\gl\gl$ and $\sq\sqb$ production, respectively.  Again, the renormalisation and factorisation scales are assumed to be equal, 
$\mu_F = \mu_R = \mu  = \xi \mu_0$, where the scale $\mu_0$ is set
equal to the average mass of the produced particles. The parameter $\xi$ is varied between 0.2 and 5. In Fig.\ \ref{fig:pscaleg}a, Fig.\ \ref{fig:pscaleg}b and Fig.\ \ref{fig:pscaleg}c (Fig.\ \ref{fig:pscaleq}a, Fig.\ \ref{fig:pscaleq}b and Fig.\ \ref{fig:pscaleq}c) the gluino (squark) mass, $m_{\gl}$ ($m_{\sq}$), takes values of 0.5, 1 and 2~TeV, correspondingly, and $r=m_{\gl} / m_{\sq} = 1.2$.

The theoretical uncertainty of the $pp \to \gl\gl$ cross section 
due to differences in the parameterisations of the pdfs may be read out from 
Fig.\ \ref{fig:pscaleg}. At the NLO the uncertainty is given by the 
difference between the dotted (CTEQ6M) and the dash-dotted (MSTW) lines
and at the NLL by the difference between the continuous (CTEQ6M) and the
dashed (MSTW) lines. The absolute difference between the CTEQ6 curves and 
the MSTW curves is rather similar at the NLO and at the NLL accuracy.
It is clearly visible in Fig.\ \ref{fig:pscaleg} that the uncertainty introduced by the pdfs grows with the increasing gluino mass. In particular, for 
$m_{\gl}=0.5$~TeV this uncertainty is smaller than 3\%, for  
$m_{\gl}=1$~TeV it is smaller than 5\%, and for $m_{\gl}=2$~TeV it reaches already about $25\%$. We rephrase that the probable reason for the strong dependence of the cross sections on the pdfs at large gluino mass is the fact that the currently available pdfs are poorly constrained at large scales and at large parton~$x$. This uncertainty should be, however, substantially reduced after high $p_T$ jet measurements are perfomed at the LHC.
As seen in Fig.\ \ref{fig:pscaleg}, the variation of the $pp \to
\gl\gl$ cross section with the scale $\mu = \xi m_{\gl}$ is
substantially reduced after inclusion of the soft gluon resummation over the whole $\xi$ range.

The case of squark-antisquark production is illustrated in Fig.\ \ref{fig:pscaleq}. Clearly, for $\sq\sqb$ production, the soft gluon resummation introduces much smaller reduction in the scale dependence over the whole $\xi$ range than it was in the case of the $\gl\gl$ production. 
The relative uncertainty of the $pp\to \sq\sqb$ cross section due to the choice of the pdfs varies from about 3\% at $m_{\sq} = 1$~TeV by about 5\% at $m_{\sq} = 0.5$~TeV to about 8\% at $m_{\sq} = 2$~TeV. Again, this uncertainty should be reduced after suitable measurements at the LHC are performed.

\begin{figure}[p]
\begin{center}
\begin{tabular}{ll}
\epsfig{file=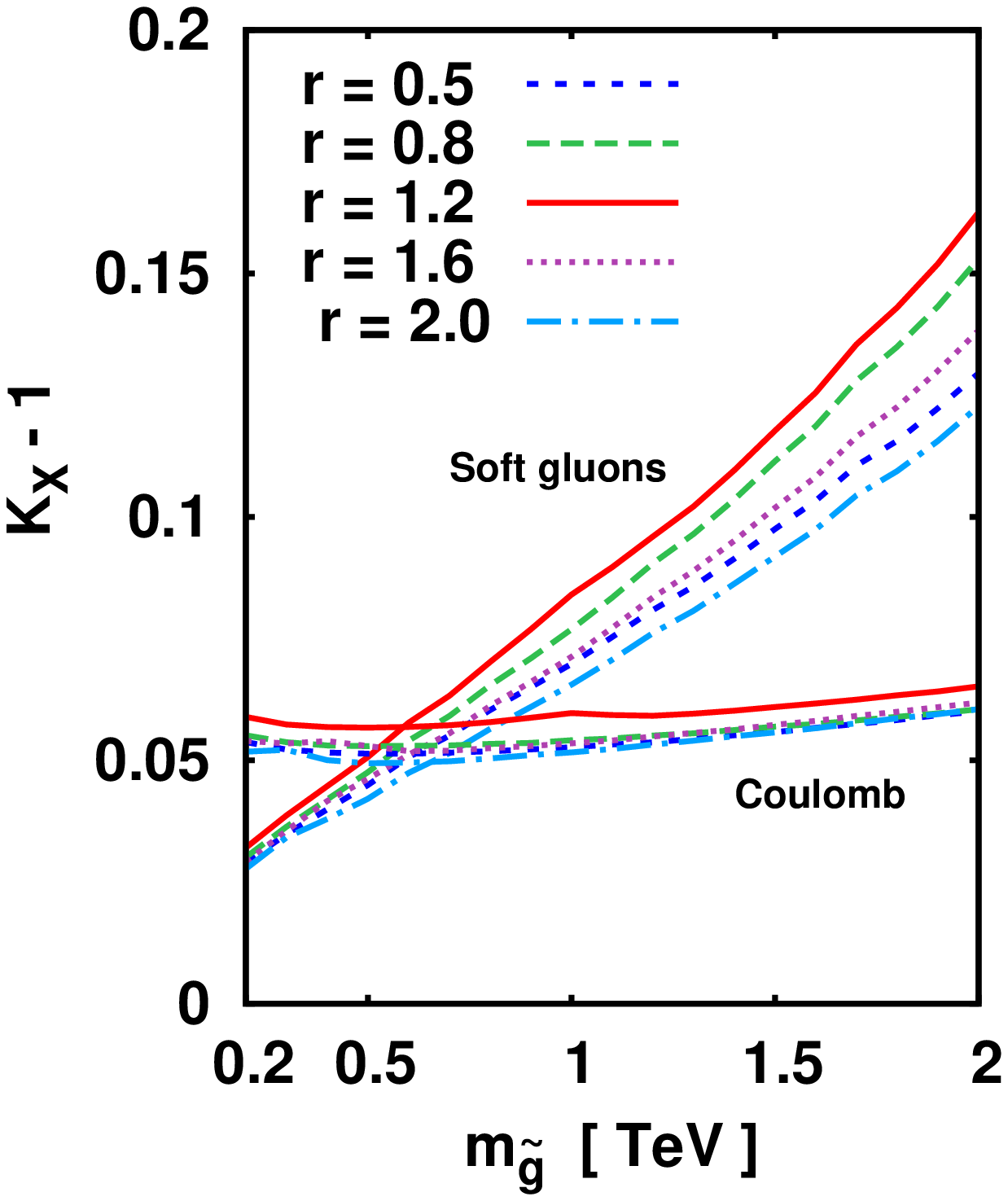,width=0.42\columnwidth} \hspace{0.05\columnwidth} &
\epsfig{file=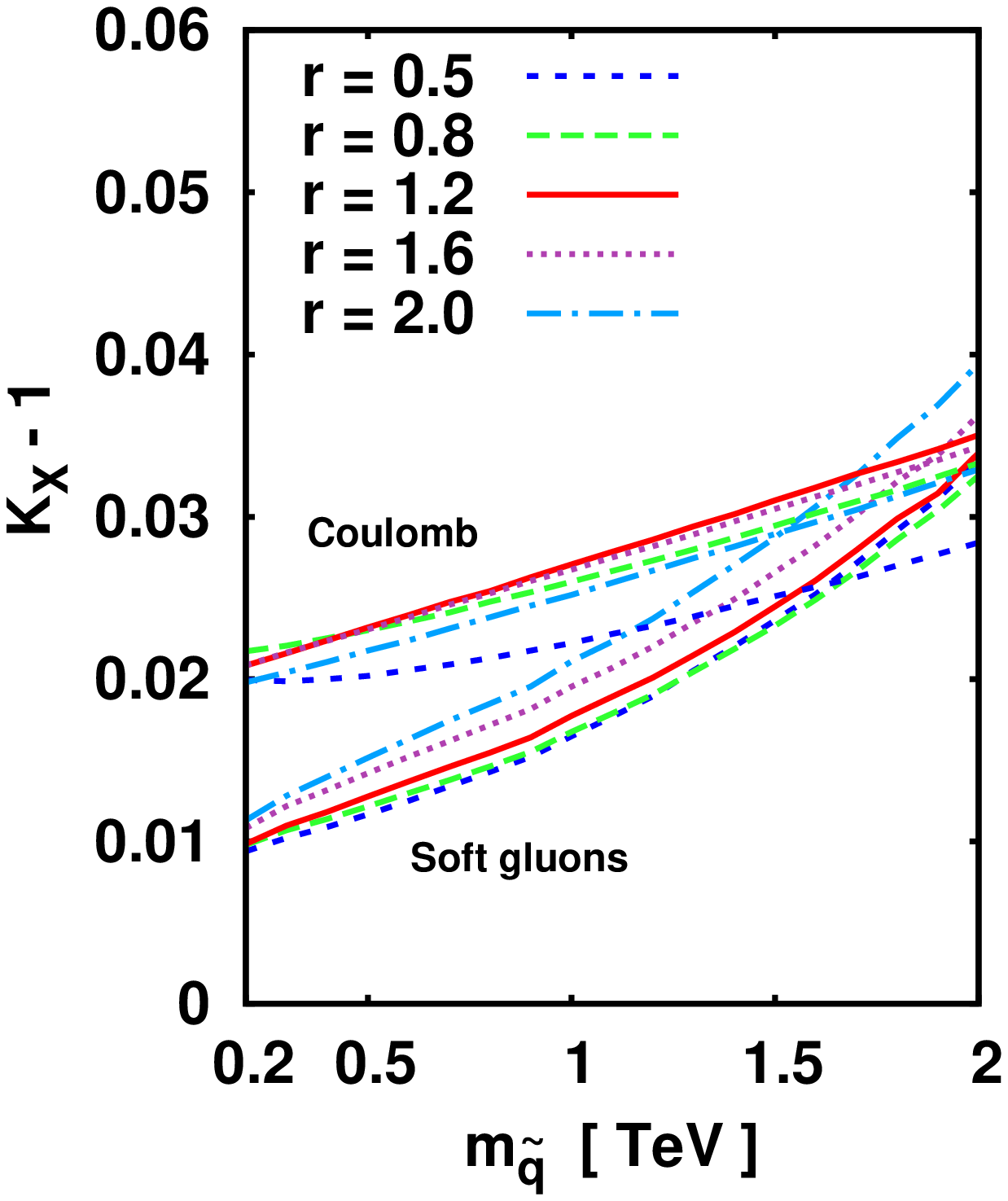,width=0.42\columnwidth} \\
{\large\bf a)} & {\large\bf b)} \end{tabular}
\end{center}
\caption{\it 
\label{fig:coul1}
The relative corrections, 
\mbox{$K_{\NLL}-1$} and \mbox{$K_{\mathrm{Coul}}-1$},
to the NLO cross sections for the $\gl\gl$ (a) 
and the $\sq\sqb$ (b) production at the LHC 
as a function of gluino and squark mass, respectively; 
$r=\mgl / \msq$.}
\end{figure} 

\begin{figure}[p]
\begin{center}
\begin{tabular}{ll}
\epsfig{file=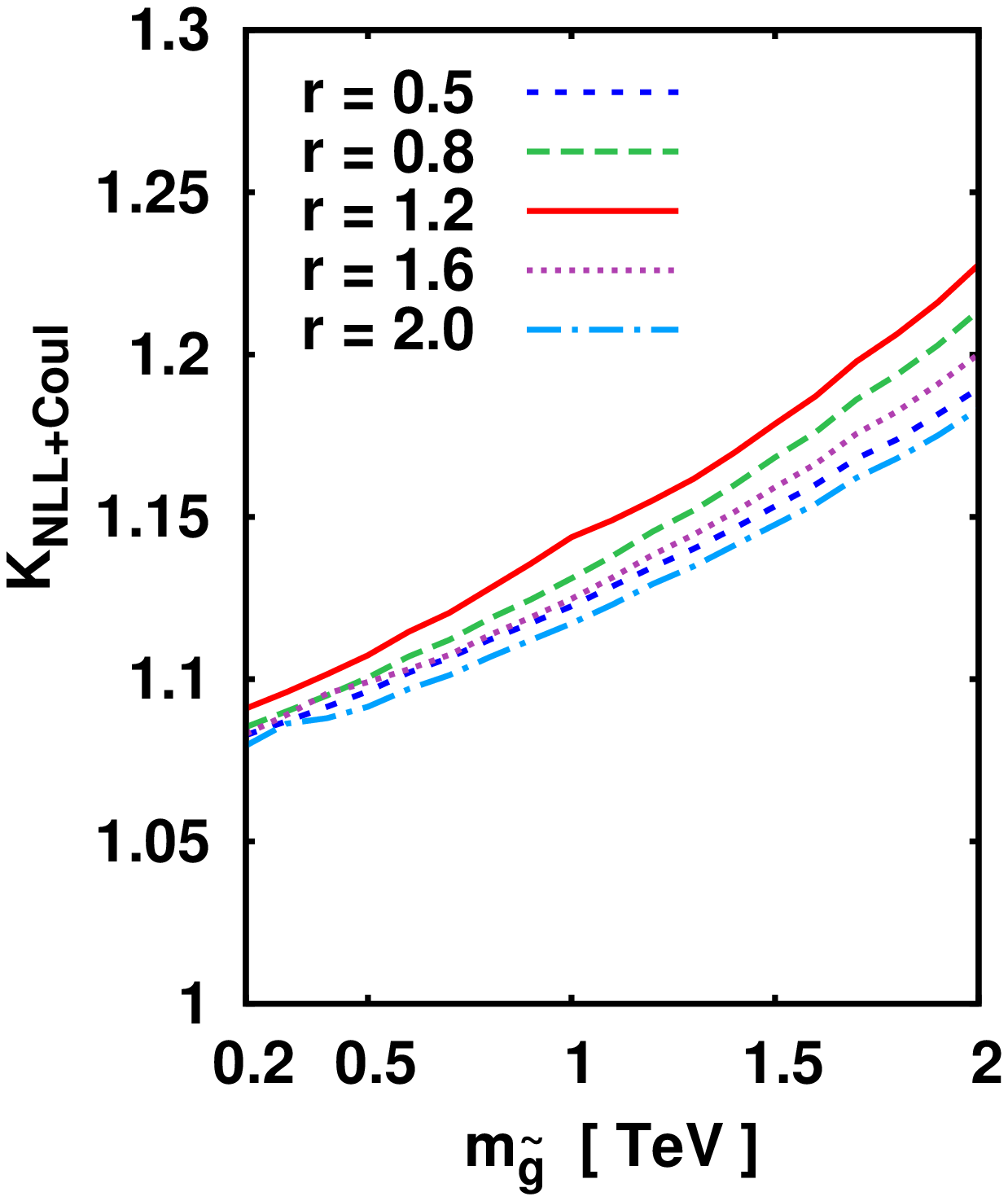,width=0.42\columnwidth} \hspace{0.05\columnwidth} &
\epsfig{file=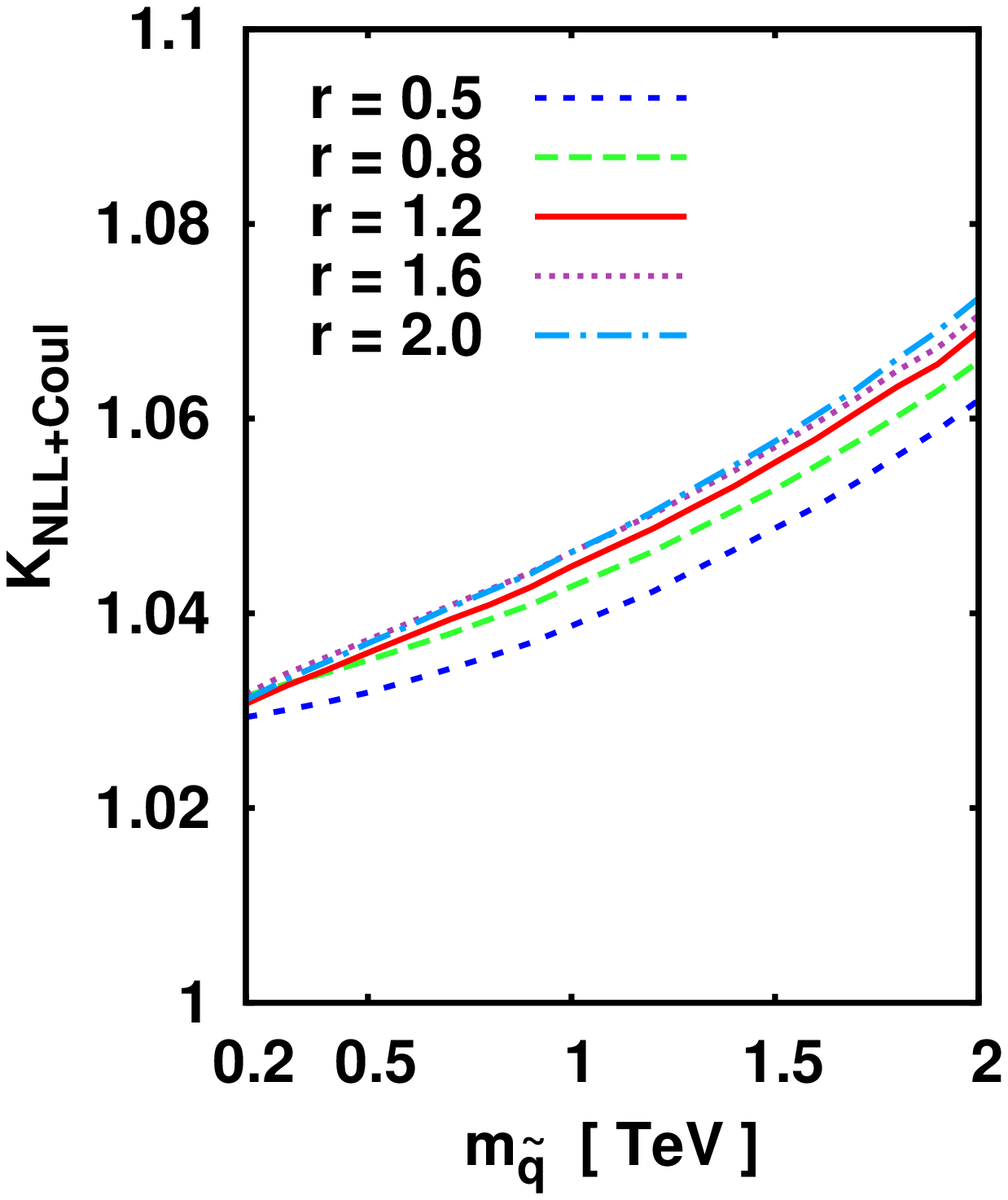,width=0.42\columnwidth} \\
{\large\bf a)} & {\large\bf b)} \end{tabular}
\end{center}
\caption{\it 
\label{fig:coul2}
The $K$-factors, \mbox{$K_{\mathrm{NLL+Coul}}$}
for the $\gl\gl$ (a) and the $\sq\sqb$ (b) total production cross sections 
at the LHC as a function of gluino and squark mass, respectively; 
$r=\mgl / \msq$.}
\end{figure} 

We also study the corrections to the cross sections for $pp\to \gl\gl$
and $pp\to \sq\sqb$ coming from the leading Coulomb exchanges beyond
the NLO accuracy. We define the corresponding Coulomb $K$-factor
\be
K_{\mathrm{Coul}} \;=\;  
{ 
\si^{\mathrm{(NLO)}} _{h_1 h_2  \to k l}(\rho,\{m^2\})
\,+\, \delta\si^{(C)}_{h_1 h_2  \to k l}(\rho,\{m^2\}) \over
\si^{\mathrm{(NLO)}} _{h_1 h_2  \to k l}(\rho,\{m^2\})}\,,
\ee
where we use Eq.~(\ref{eq:chadr}) to calculate $\delta\si^{(C)}_{h_1
  h_2  \to k l}(\rho,\{m^2\})$
with the two-loop $\as$ taken at the scale $\mu^2$.
The obtained numerical results for the relative Coulomb correction, 
$K_{\mathrm{Coul}}-1$, to the  $\gl\gl$ and $\sq\sqb$ cross sections 
are shown in Fig.\ \ref{fig:coul1}a and Fig.\ \ref{fig:coul1}b, respectively,
and compared to the corresponding relative soft gluon correction,
$K_{\mathrm{\NLL}}-1$. For the gluino pair-production we find that the
Coulomb 
$K$-factor exhibits only weak dependence on the gluino and squark masses. 
In more detail, $K_{\mathrm{Coul}}-1$ takes values between  5\% and 6\%. 
In fact, the Coulomb corrections exceed the soft gluon corrections for 
$m_{\gl} < 0.5$~TeV and are relatively important up to $m_{\gl} = 2$~TeV. 
For the $\sq\sqb$ production  $K_{\mathrm{Coul}}-1$ takes the values  
between 2\% and 3.5\%, depending on $m_{\sq}$ and $m_{\gl}$. 
The importance of the leading Coulomb corrections in $\sq\sqb$
production can be seen in
Fig.\ \ref{fig:coul1}b.

The NLL soft gluon corrections and the Coulomb corrections beyond the
NLO approximation~(\ref{eq:chadr})
can be combined additively. We define the $K$-factor that
accounts for both types of corrections in the following way:
\be
K_{\mathrm{NLL+Coul}} \; = \; 
{ \si^{(\mathrm{match})} _{h_1 h_2  \to k l}(\rho,m^2,\{\mu^2\}) \, + \,
  \delta\si^{(C)}_{h_1 h_2  \to k l}(\rho,m^2,\{\mu^2\}) \over
\si^{\mathrm{(NLO)}} _{h_1 h_2  \to k l}(\rho,m^2,\{\mu^2\})}\; .
\ee
The obtained $K$-factors, $K_{\mathrm{NLL+Coul}}$,
in  the  $\gl\gl$ and $\sq\sqb$ production at the LHC
are shown in Fig.\ \ref{fig:coul2}a and Fig.\ \ref{fig:coul2}b, respectively.
The results shown in Fig.\ \ref{fig:coul2} are our most complete estimates 
of the higher order QCD corrections in these processes.

\begin{figure}[h]
\begin{center}
\begin{tabular}{ll}
\epsfig{file=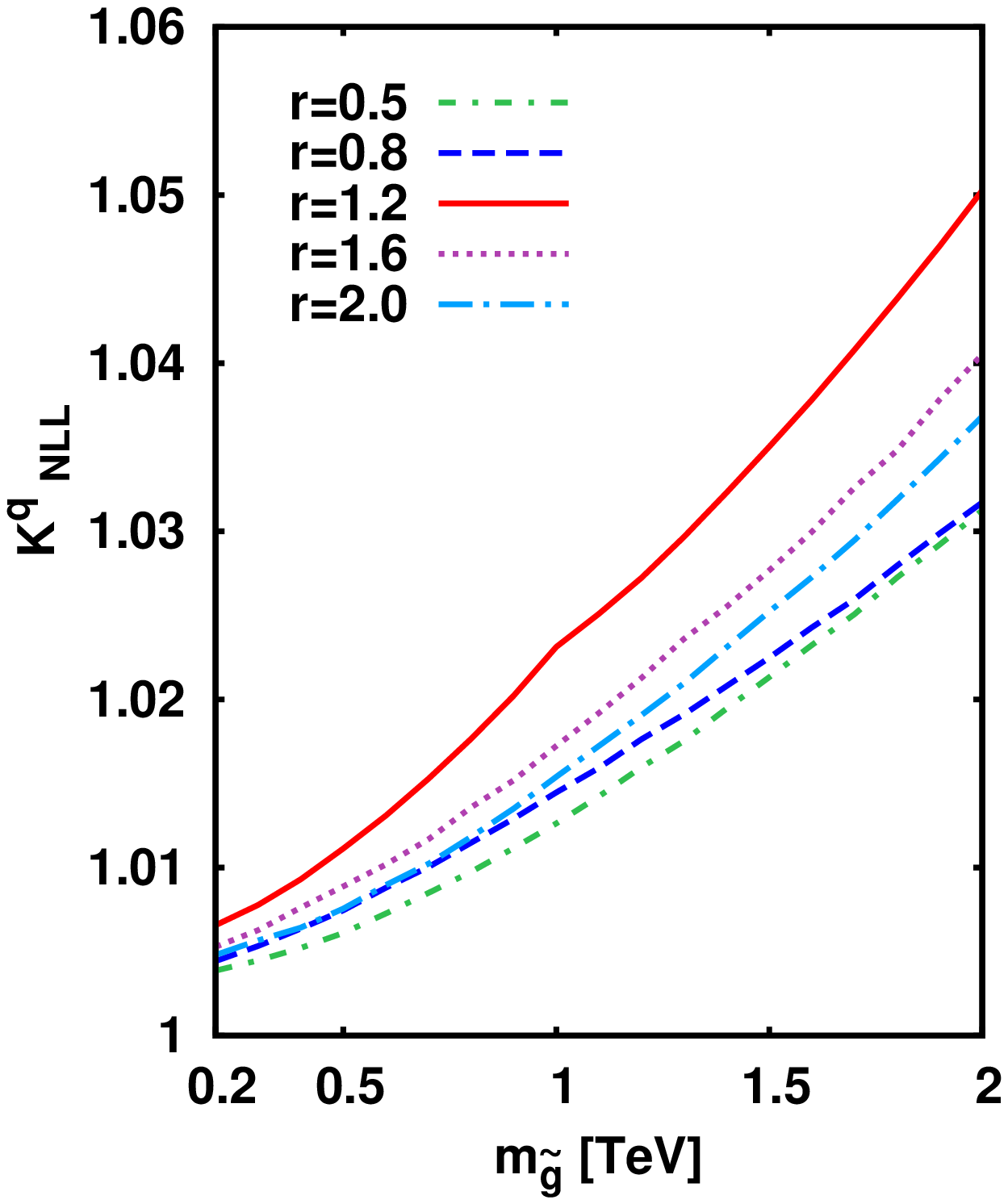,width=0.42\columnwidth}  \hspace{0.05\columnwidth} &
\epsfig{file=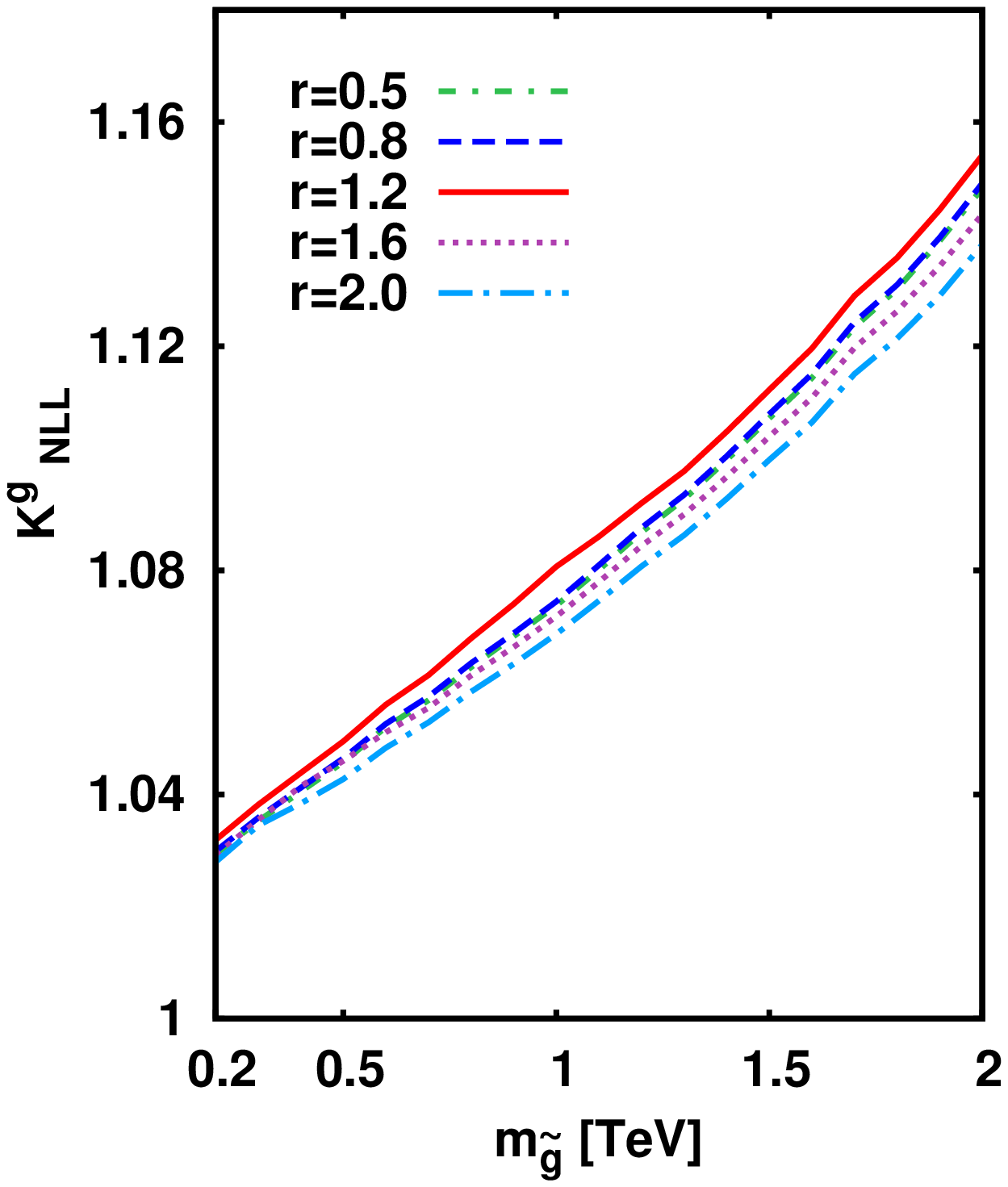,width=0.42\columnwidth} \\
{\large\bf a)} & {\large\bf b)} \end{tabular}
\end{center}
\caption{\it 
The $K$-factor, $K^i _{\NLL}$, for the partonic channels in the hadronic process
$pp \to \gl\gl$: (a) $q\bar q \to \gl\gl$ and (b)  $gg \to \gl\gl$. 
\label{fig:gpart}
} 
\end{figure} 
\begin{figure}[h]
\begin{center}
\begin{tabular}{ll}
\epsfig{file=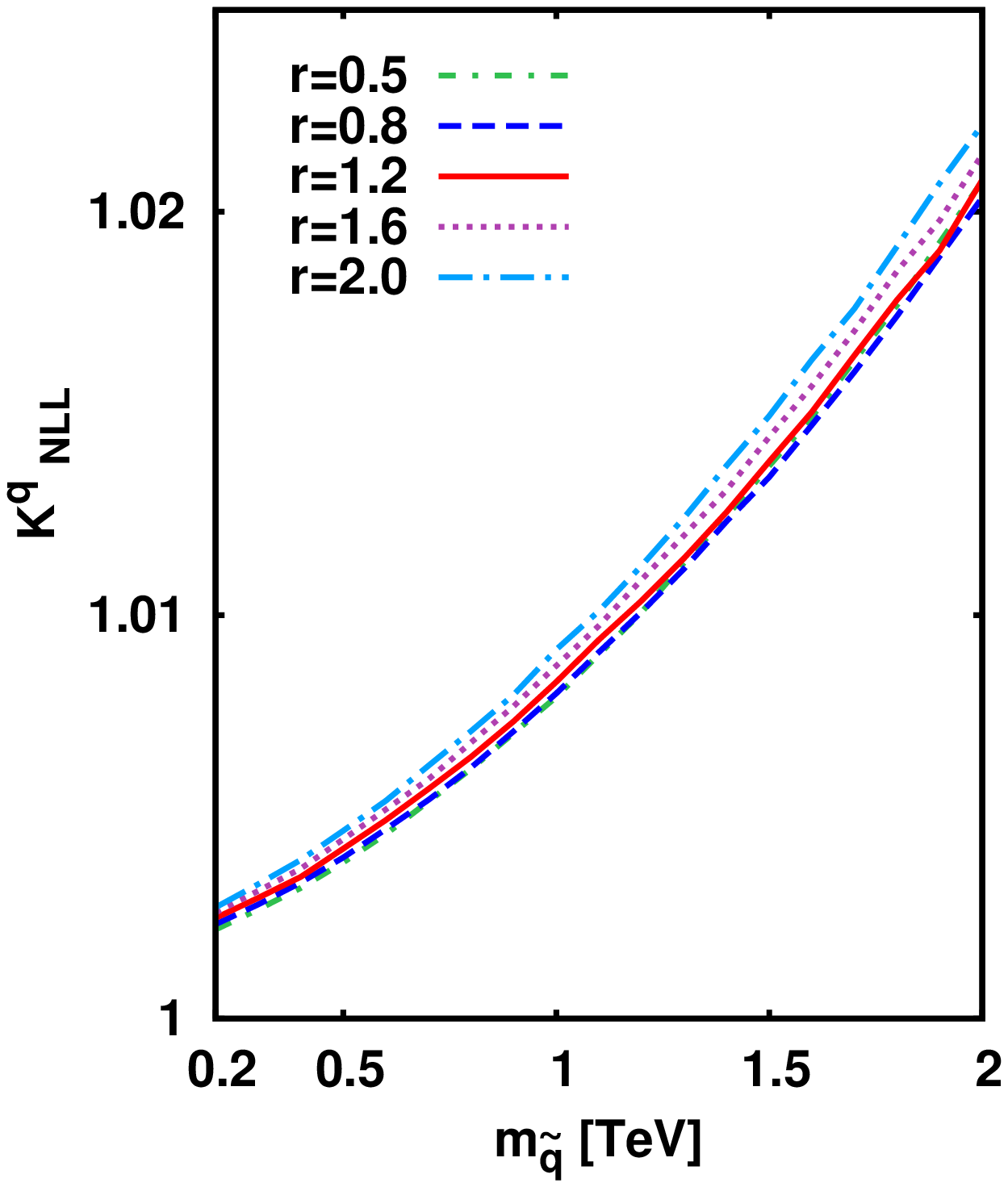,width=0.42\columnwidth}  \hspace{0.05\columnwidth} &
\epsfig{file=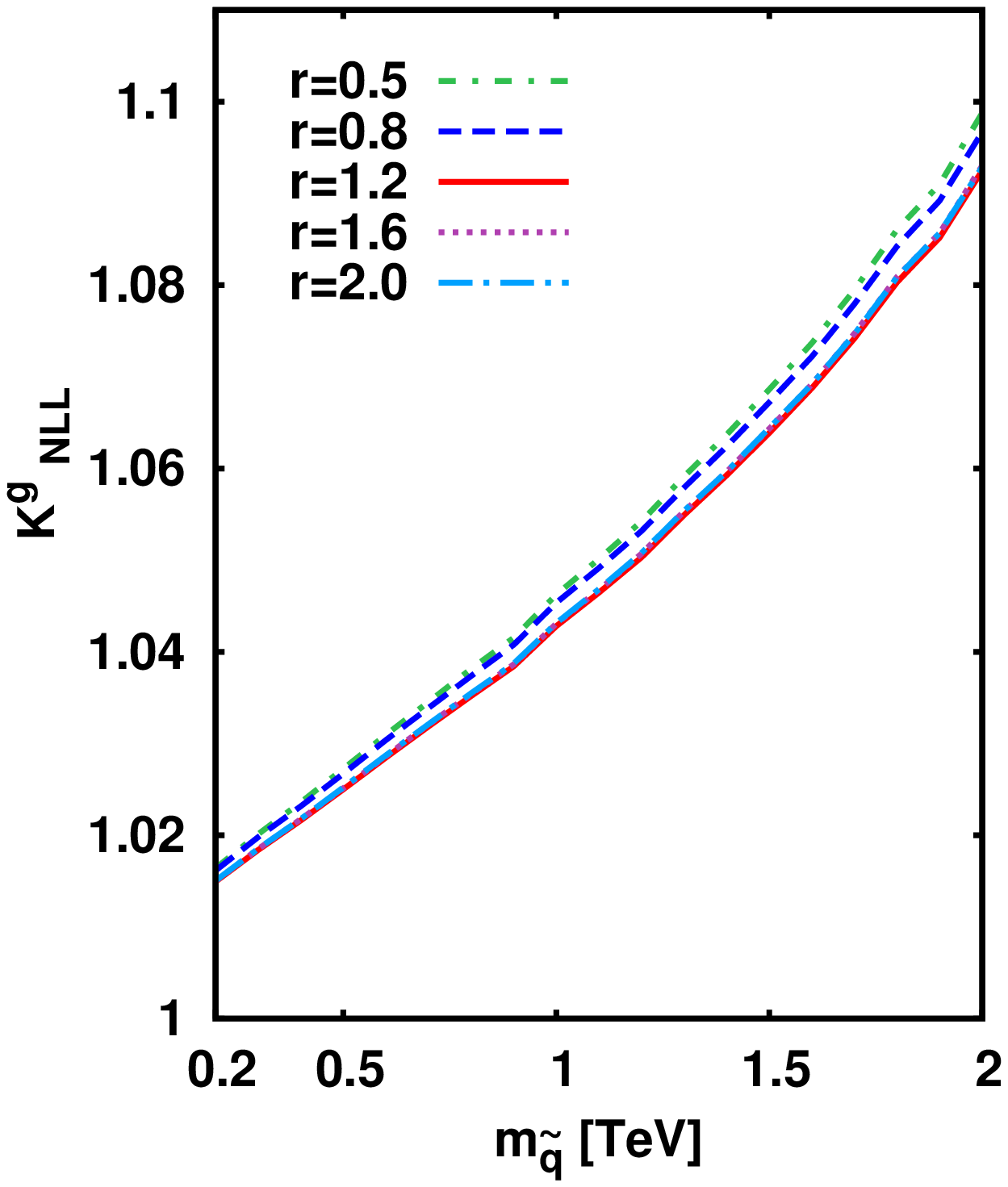,width=0.42\columnwidth} \\
{\large\bf a)} & {\large\bf b)} \end{tabular}
\end{center}
\caption{\it 
The $K$-factor, $K^i _{\NLL}$, for the partonic channels in the hadronic
process 
$pp \to \sq\sqb$: (a) $q\bar q \to \sq\sqb$ and (b)  $gg \to \sq\sqb$. 
\label{fig:qpart}
} 
\end{figure}

It is interesting to investigate in more detail the soft gluon corrections 
in partonic subchannels of the $\gl\gl$ and $\sq\sqb$ hadroproduction at the
LHC. For this purpose we define the NLL $K$-factors 
restricted to the subprocesses occurring through the $q\bar q$ 
and the $gg$ partonic collision, i.e. we define
$K^q _{\NLL} \, = \, \si^{(\mathrm{match})} _{pp \to q\bar q \to kl} /
 \si^{\NLO} _{pp \to q\bar q \to kl}$ 
and
$K^g _{\NLL} \, = \, \si^{(\mathrm{match})} _{pp \to gg \to kl} /
 \si^{\NLO} _{pp \to gg \to kl}$, where, as usual, 
$kl = \gl\gl$ or $kl = \sq\sqb$. The results for $K^i _{\NLL}$ in the $\gl\gl$
and in the $\sq\sqb$ production are given in Fig.\ \ref{fig:gpart} and  
Fig.\ \ref{fig:qpart}, respectively. The NLL~$K$-factors grow
with increasing masses of the produced particles in all cases. This 
is expected since the importance of the threshold logarithms strengthens 
with the higher masses of the produced particles.

It is clearly visible from  Fig.\ \ref{fig:gpart} and Fig.\ \ref{fig:qpart}
that the size of the soft gluon corrections increases with higher colour charges of all  
particles involved in the partonic reaction. The NLL correction
for the $gg$ mediated processes (see Fig.\ \ref{fig:gpart}a and 
Fig.\ \ref{fig:qpart}a) is much larger than the NLL correction for 
the $q\bar q$ mediated processes (Fig.\ \ref{fig:gpart}b and 
Fig.\ \ref{fig:qpart}b), $K^g _{\NLL}-1 > K^q _{\NLL}-1$,
both for the $\gl \gl$ and in $\sq\sqb$ production. More specifically, 
$K^g _{\NLL}-1$ is 3~---~5 times larger than $K^q _{\NLL}-1$, depending on
the final state particles and their masses. By comparing the $K$-factors
with the same partonic initial state in the production of $\gl\gl$ and 
$\sq\sqb$ (i.e.\ by comparing Fig.\ \ref{fig:gpart}a with Fig.\ \ref{fig:qpart}a and Fig.\ \ref{fig:gpart}b with Fig.\ \ref{fig:qpart}b) one finds that 
the soft gluon effects are larger for $\gl\gl$ production than for 
 $\sq\sqb$ in both partonic subchannels. Such results are expected since
due to larger colour factors the soft gluon corrections should be more 
pronounced when the particles with higher colour charges take part 
in the hard scattering.

\begin{figure}[p]
\begin{center}
\begin{tabular}{lll}
\epsfig{file=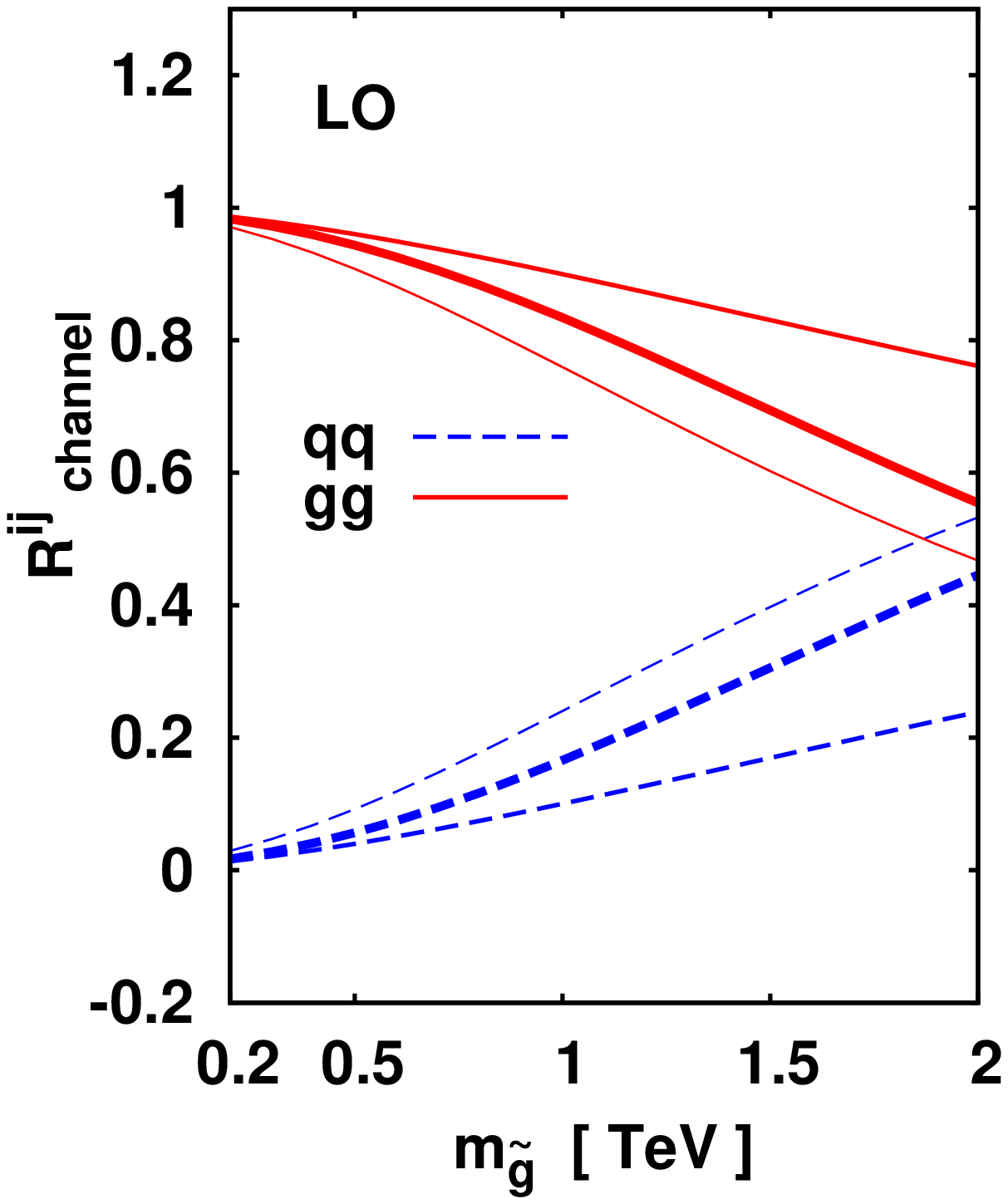,width=0.32\columnwidth}  &
\epsfig{file=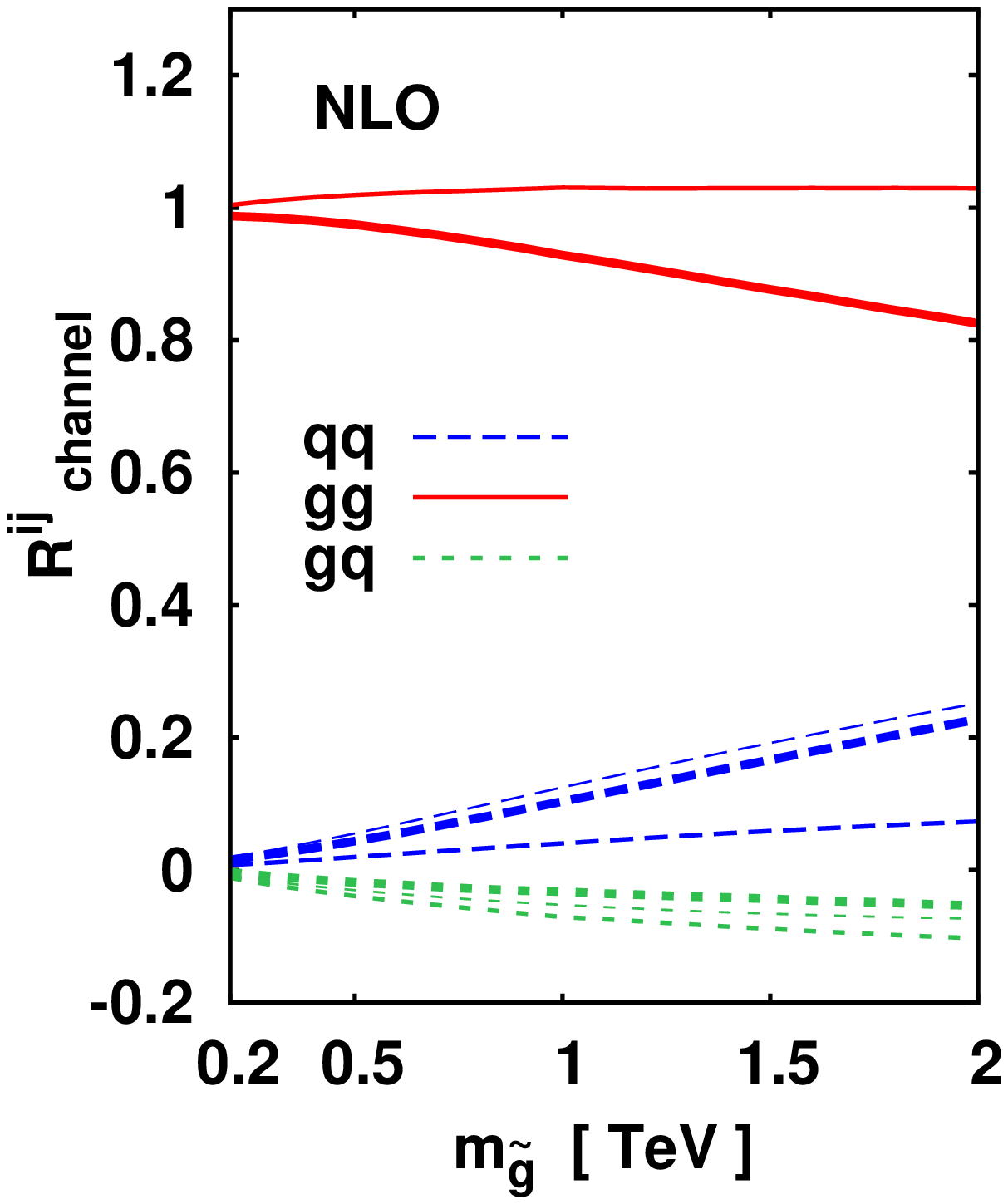,width=0.32\columnwidth} &
\epsfig{file=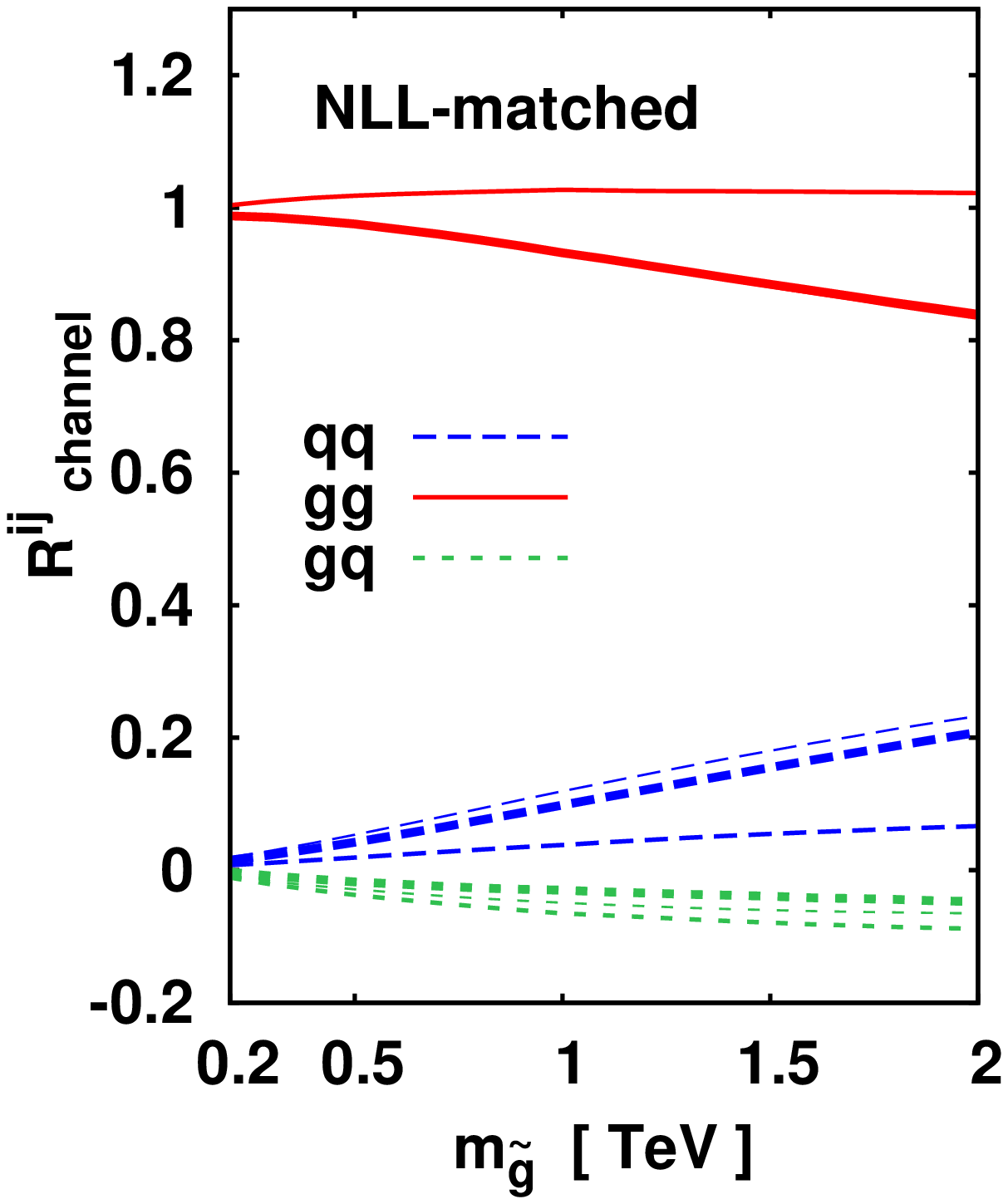,width=0.32\columnwidth} \\
{\large\bf a)} & {\large\bf b)} & {\large\bf c)} \end{tabular}
\end{center}
\caption{\it 
Relative contributions of the partonic channels to the $pp \to \gl\gl$ cross section
at the LO (a), NLO (b) and NLL-matched accuracy. 
The thick lines correspond to $r=0.5$, the medium lines to  $r=1.2$,
and the thin ones to $r=2.0$. The $gg$ lines in (b) and (c)
incidentally overlap for $r=0.5$ and $r=2.0$. 
\label{fig:gbreak}
} 
\end{figure} 
\begin{figure}[p]
\begin{center}
\begin{tabular}{lll}
\epsfig{file=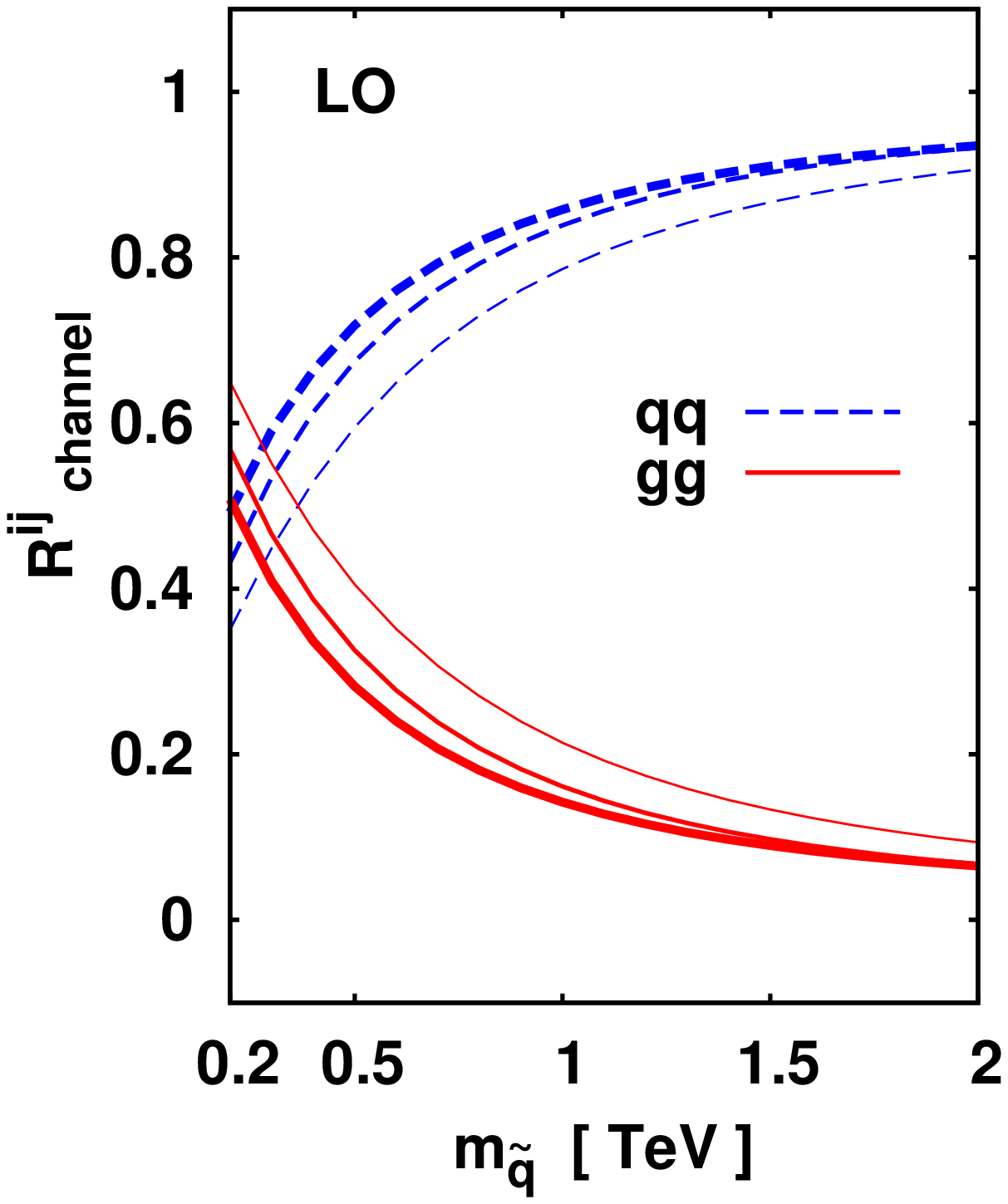,width=0.32\columnwidth}  &
\epsfig{file=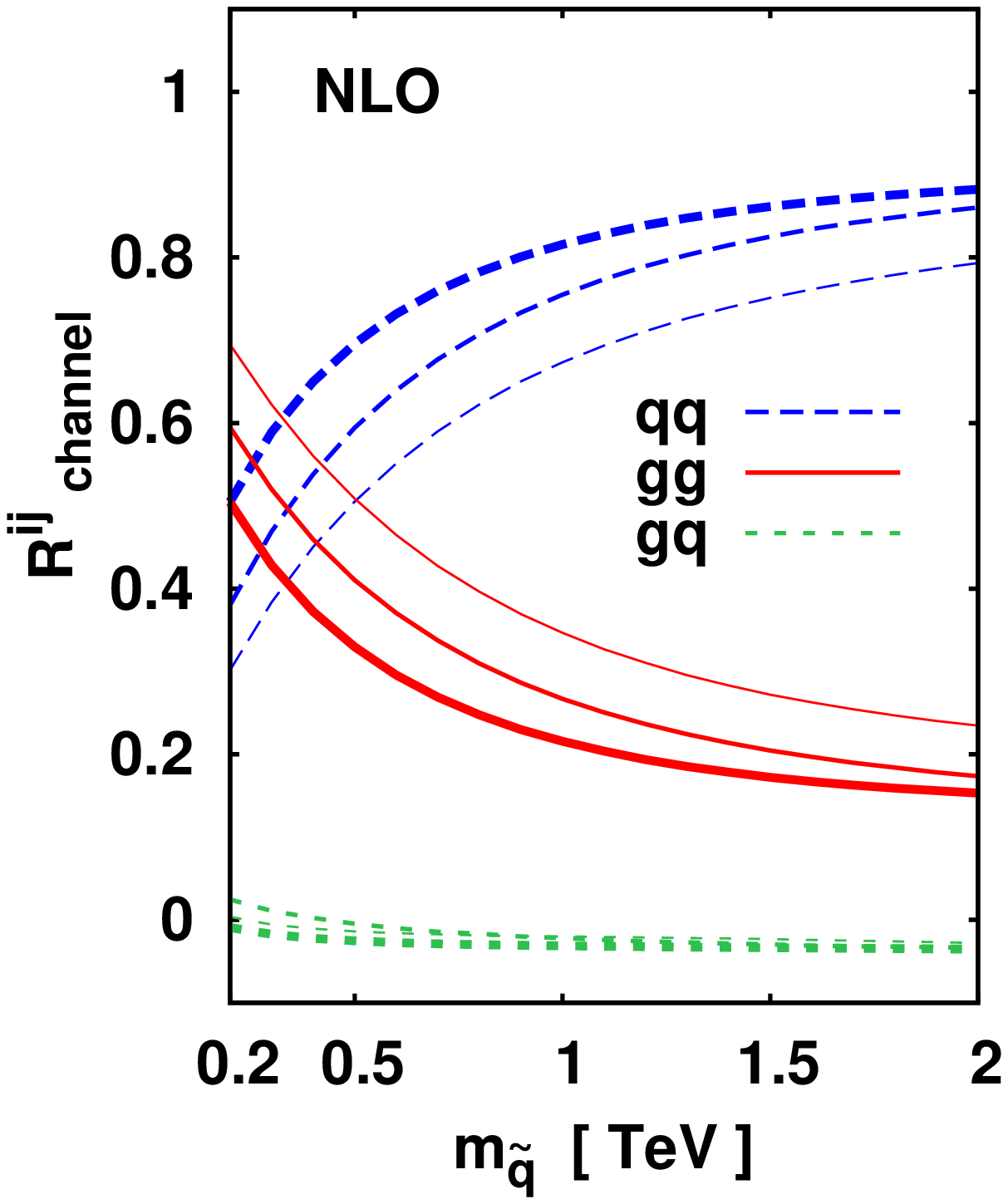,width=0.32\columnwidth} &
\epsfig{file=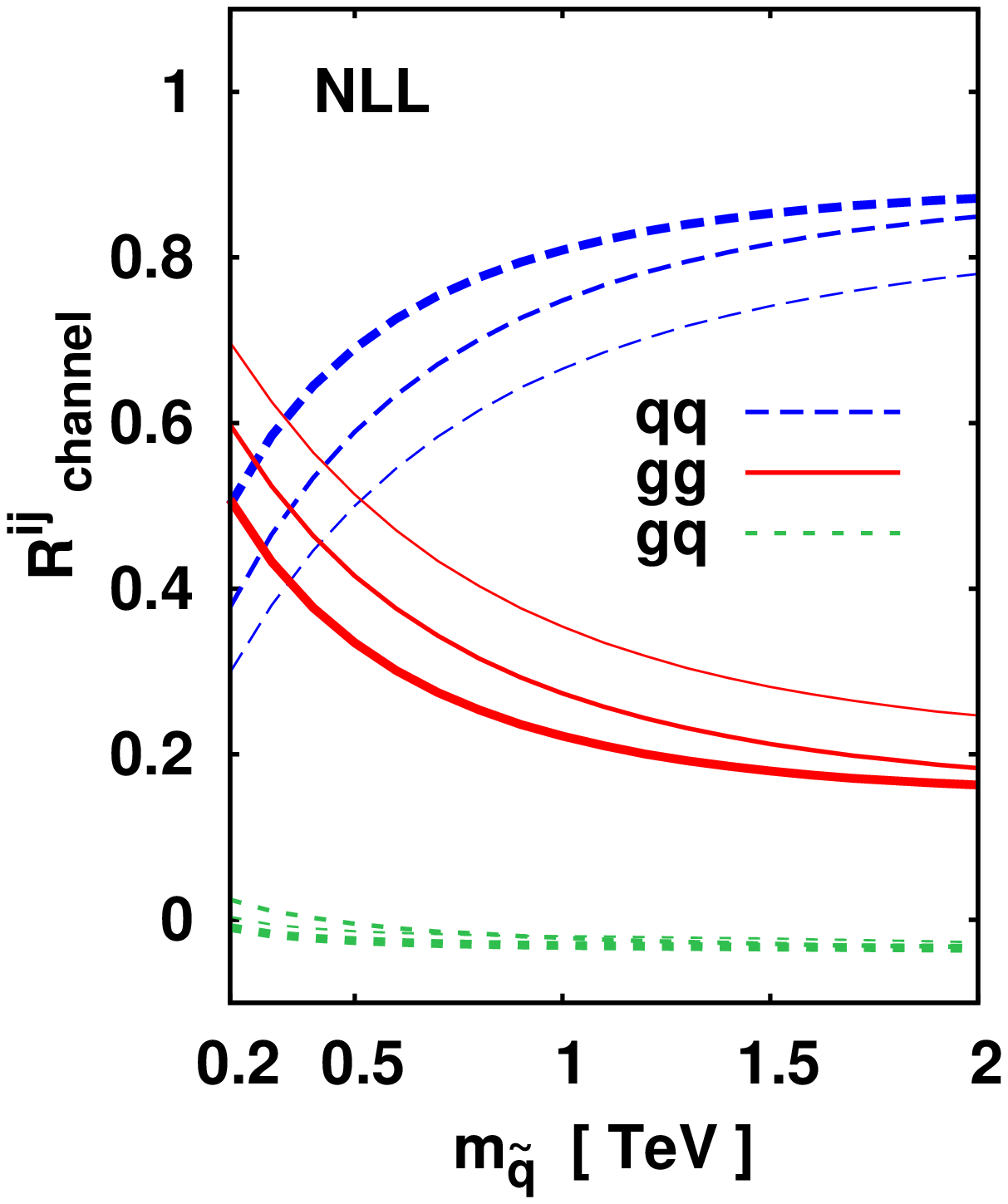,width=0.32\columnwidth} \\
{\large\bf a)} & {\large\bf b)} & {\large\bf c)} \end{tabular}
\end{center}
\caption{\it 
Relative contributions of the partonic channels to the $pp \to \sq\sqb$ cross section
at the LO (a), NLO (b) and NLL-matched (c) accuracy. The thick lines correspond to $r=0.5$, the medium lines to  $r=1.2$, and the thin ones to $r=2.0$.
\label{fig:qbreak}
} 
\end{figure} 

We also study the relative importance of the partonic subchannels 
in the considered sparticle production processes. For this purpose we
define the fractions $R^{ij} _{\mathrm{channel}} = {\sigma_{pp \to ij \to kl} \; / \; \sigma_{pp \to kl}}$
for each total hadronic cross section, $\sigma_{pp\to kl}$, where $kl = \gl\gl$ or $kl = \sq\sqb$.
The components $\sigma_{pp \to ij \to kl}$ of the hadronic cross sections 
come from the contributions with intermediate partons $ij$. We shall denote: 
$ij = q\bar q$  ($qq$ in the figures) for the combined 
$q\bar q$ and $\bar q q$ contributions,  $ij = gq$ for the combined
$g q$, $q g$, $g \bar q$ and $\bar q g$ contributions, and 
$ij=gg$ for the $gg$ contribution. For the considered processes the 
$ij = gq$ contribution does not appear at the LO.  
In Fig.\ \ref{fig:gbreak} and Fig.\ \ref{fig:qbreak} we show the obtained
values of $R^{ij} _{\mathrm{channel}}$ for the $\gl\gl$ and $\sq\sqb$ 
production at the LHC. The results are given at the LO 
(Fig.\ \ref{fig:gbreak}a and Fig.\ \ref{fig:qbreak}a), at the NLO
(Fig.\ \ref{fig:gbreak}b and Fig.\ \ref{fig:qbreak}b), and at the
matched NLL accuracy  
(Fig.\ \ref{fig:gbreak}c and Fig.\ \ref{fig:qbreak}c).
Clearly, the dominant contribution to the $\gl\gl$ hadroproduction comes 
from the $gg$ subprocess and the $\sq\sqb$ hadroproduction is dominated 
by the $q\bar q$ subprocesses except for the lower range of the squark masses
where $q\bar q$ and $gg$ contributions are similar. The inclusion of the NLO
effects leads to a relative enhancement of the $gg$ channel, and the values 
of $R^{ij} _{\mathrm{channel}}$ do not change visibly between the NLO
and the NLL results. The $gq$ channel appears beyond the LO and plays
a minor r\^{o}le. Note that $R^{gq} _{\mathrm{channel}}$ is negative
for the gluino-pair production (see Fig.\ \ref{fig:gbreak}b and Fig.\ \ref{fig:gbreak}c), which is explained by the fact that this contribution enters only beyond the LO where the simple probabilistic interpretation of the pdfs is lost.

\begin{figure}[p]
\begin{center}
\begin{tabular}{ll}
\epsfig{file=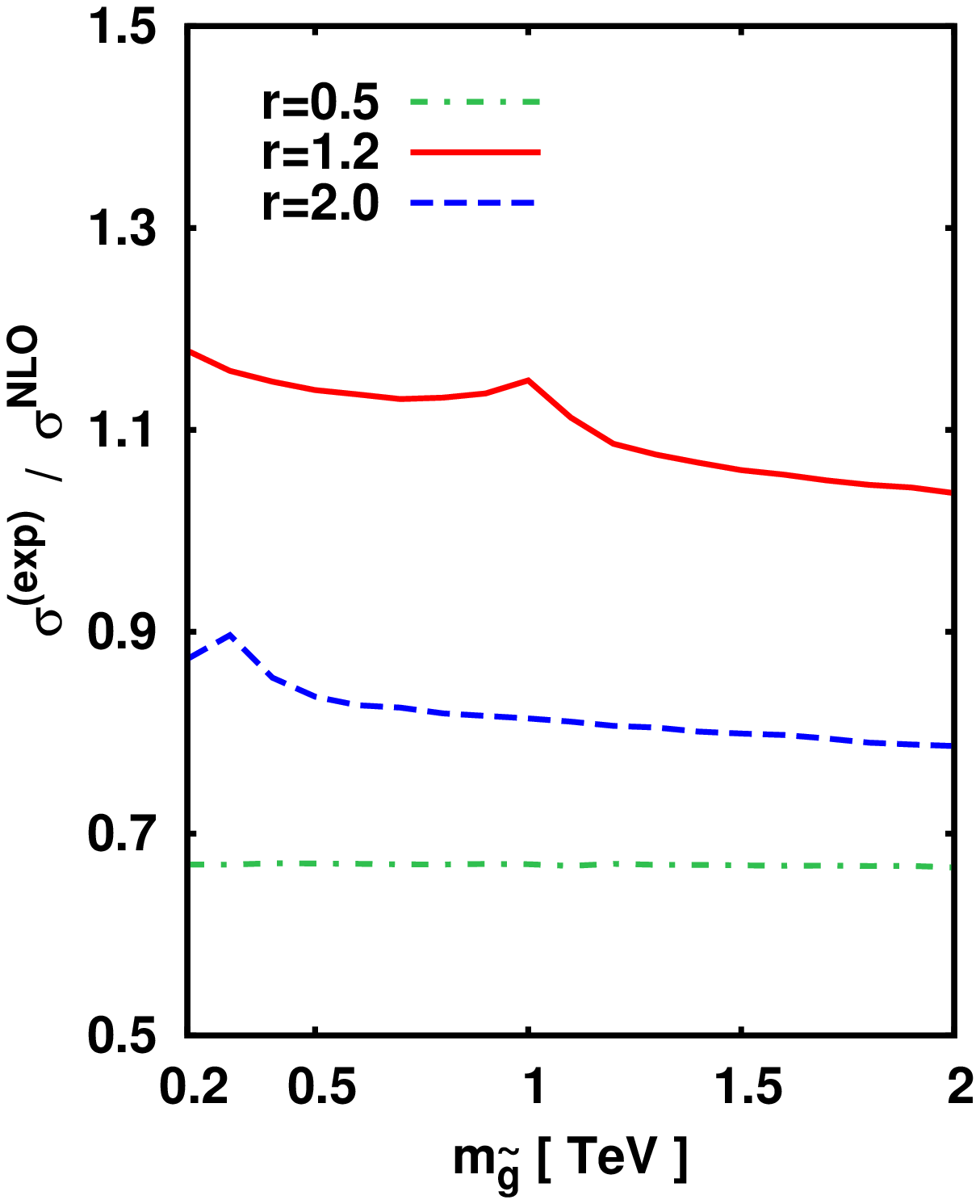,width=0.42\columnwidth} \hspace{0.05\columnwidth} &
\epsfig{file=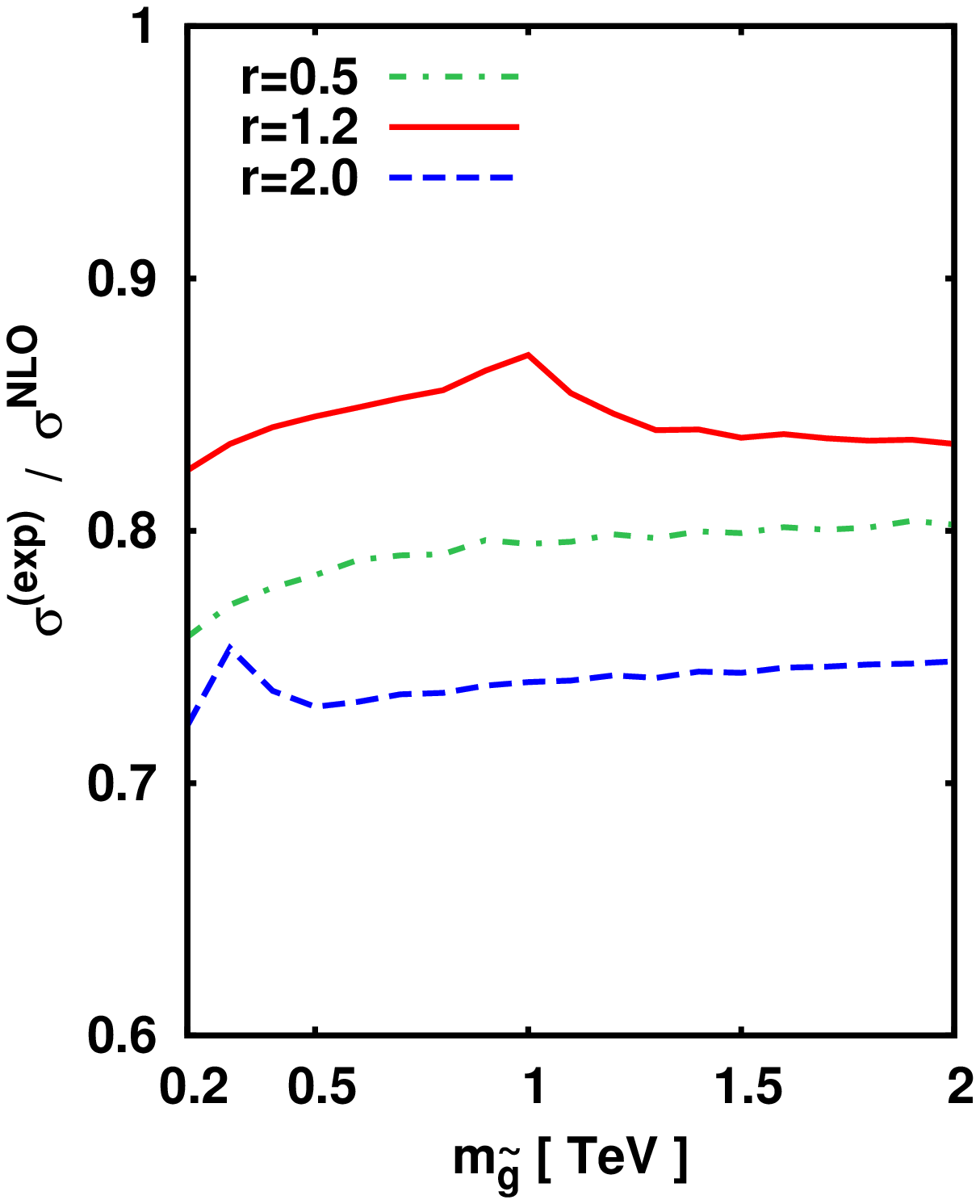,width=0.42\columnwidth}\\
{\large\bf a)} & {\large\bf b)}  \end{tabular}
\end{center}
\caption{\it 
The fraction of the NLO correction exhausted by the soft gluon 
contribution, $\sigma^{\mathrm{(exp)}} / \sigma_{(\NLO)}$, for
(a) $q\bar q \to \gl\gl$ and (b)  $gg \to \gl\gl$; $r=m_{\gl} / m_{\sq}$.
\label{fig:gfrac}
} 
\end{figure}

\begin{figure}[p]
\begin{center}
\begin{tabular}{ll}
\epsfig{file=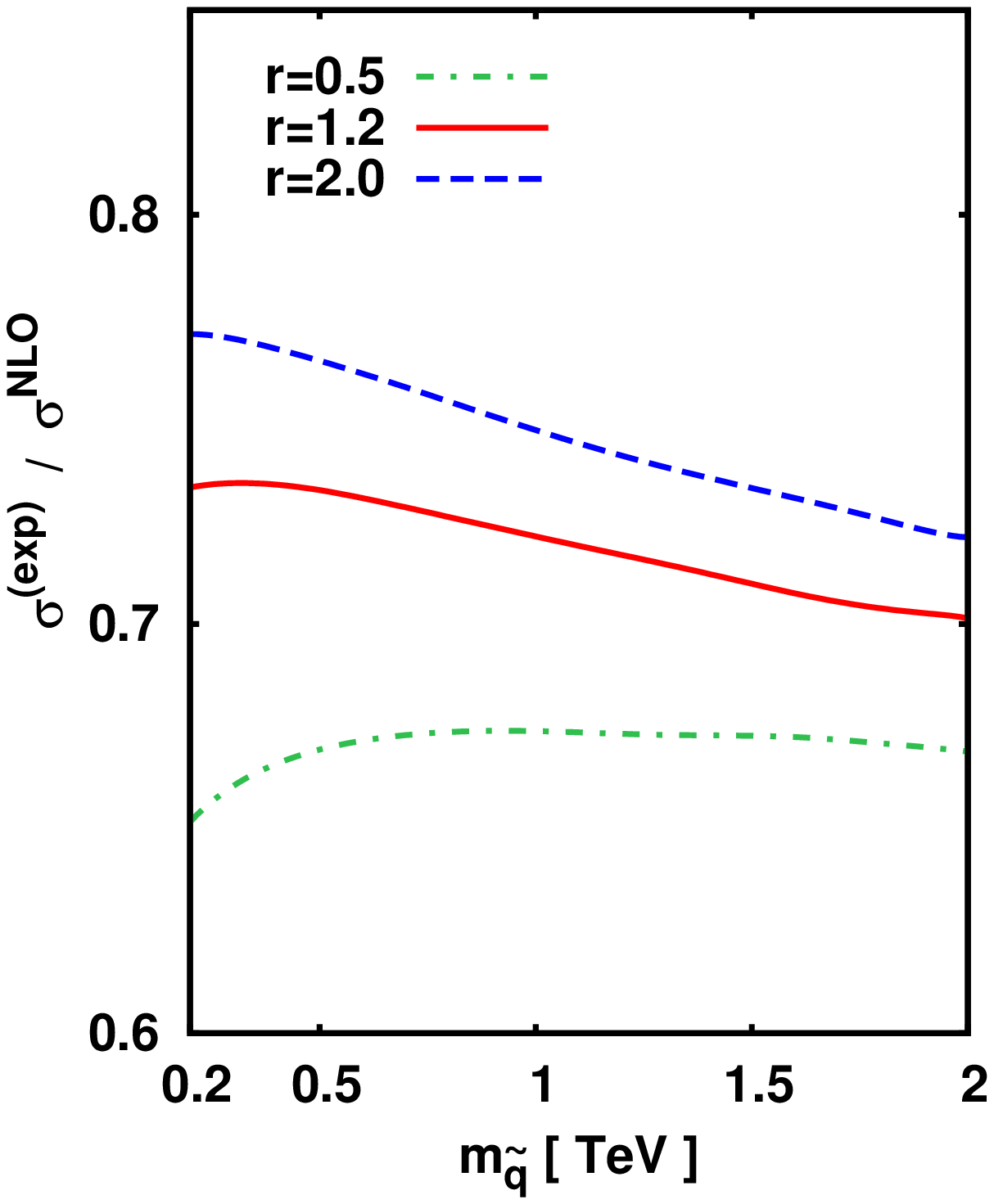,width=0.42\columnwidth} \hspace{0.05\columnwidth} &
\epsfig{file=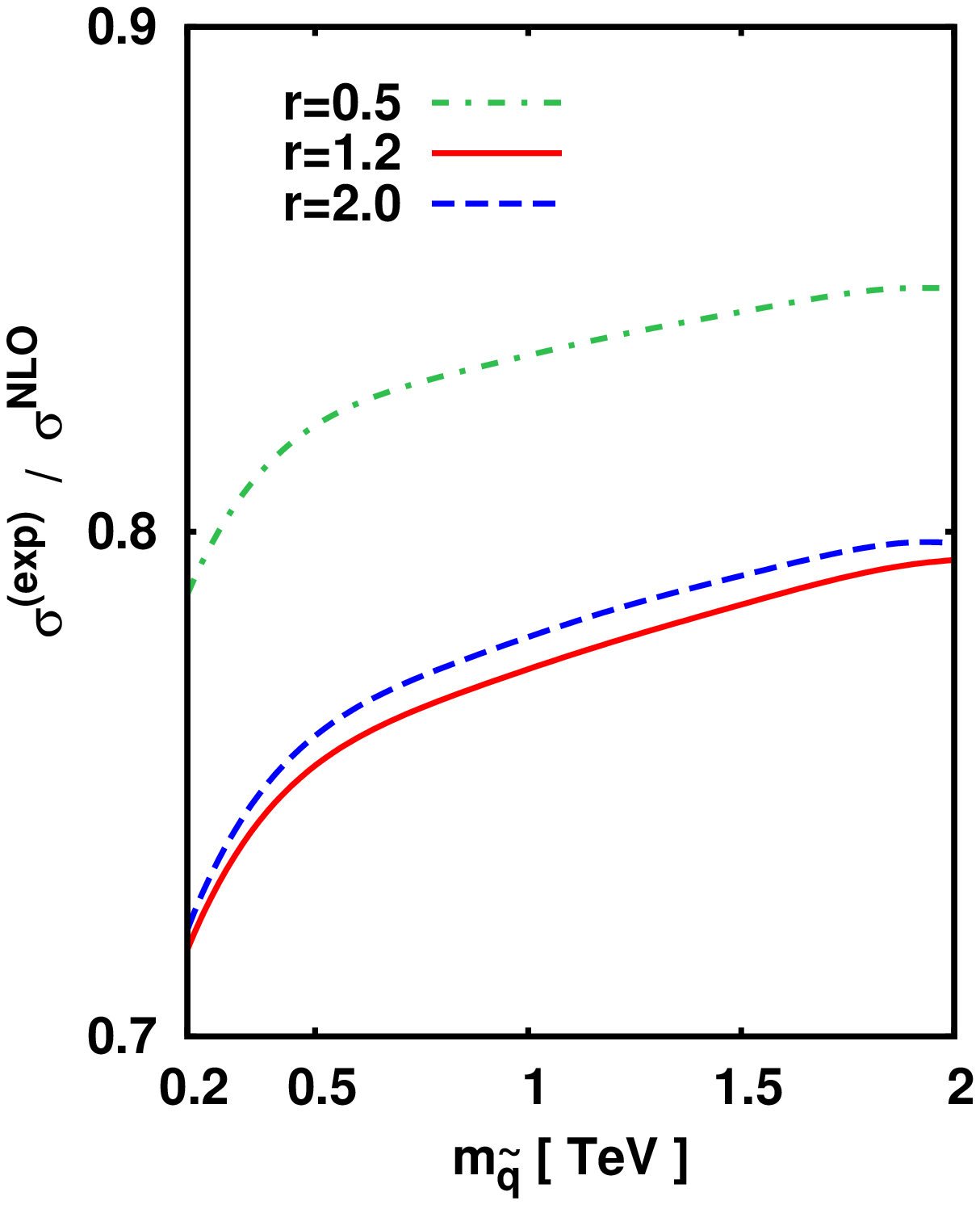,width=0.42\columnwidth}\\
{\large\bf a)} & {\large\bf b)} \end{tabular}
\end{center}
\caption{\it 
The fraction of the NLO correction exhausted by the soft gluon 
contribution, $\sigma^{\mathrm{(exp)}} / \sigma_{\NLO}$, for
(a) $q\bar q \to \sq\sqb$ and (b)  $gg \to \sq\sqb$; $r=m_{\gl} / m_{\sq}$.
\label{fig:qfrac}
} 
\end{figure}

It is interesting to check what fraction of the NLO corrections
is generated by the soft gluon contributions.
In order to answer this question we truncate at the NLO the expansion
of the resummed hadronic cross sections $\sigma_{pp \to ij \to kl}
^{\mathrm{(res)}}$ mediated by the partonic subchannel $ij$ and
define: $\sigma^{\mathrm{(exp)}} _{pp \to ij \to kl} =
\left. \si^{\mathrm{(res)}} _{pp \to ij \to
    kl}\right|_{(\NLO)}\,$. In Fig.\ \ref{fig:gfrac} and in  Fig.\
\ref{fig:qfrac} we plot the obtained values of the ratio,
$\sigma_{pp \to ij \to kl} ^{\mathrm{(exp)}}\, / \, \sigma_{pp \to ij
  \to kl} ^{\mathrm{(\NLO)}}$, for the $\gl\gl$ production
($kl=\gl\gl$) and the $\sq\sqb$ ($kl=\sq\sqb$) production,
respectively. It follows from the figures that the soft gluon
correction provides the dominant part (more than 2/3 in the studied
cases) of the NLO correction in both the $q \bar q$ and $gg$
subchannels, and both for the $\gl\gl$ and $\sq\sqb$ production.
The cusps visible in the curves in  Fig.\ \ref{fig:gfrac} originate from the 
NLO supersymmetric QCD correction to the $pp\to \gl\gl$ cross section. 
In more detail, the stop-top loop contribution to the gluino self-energy
exhibits the singular behaviour in vicinity of threshold for the  
decay of gluino into the top quark and its superpartner, the stop.

\begin{figure}[p]
\begin{center}
\begin{tabular}{ll}
\epsfig{file=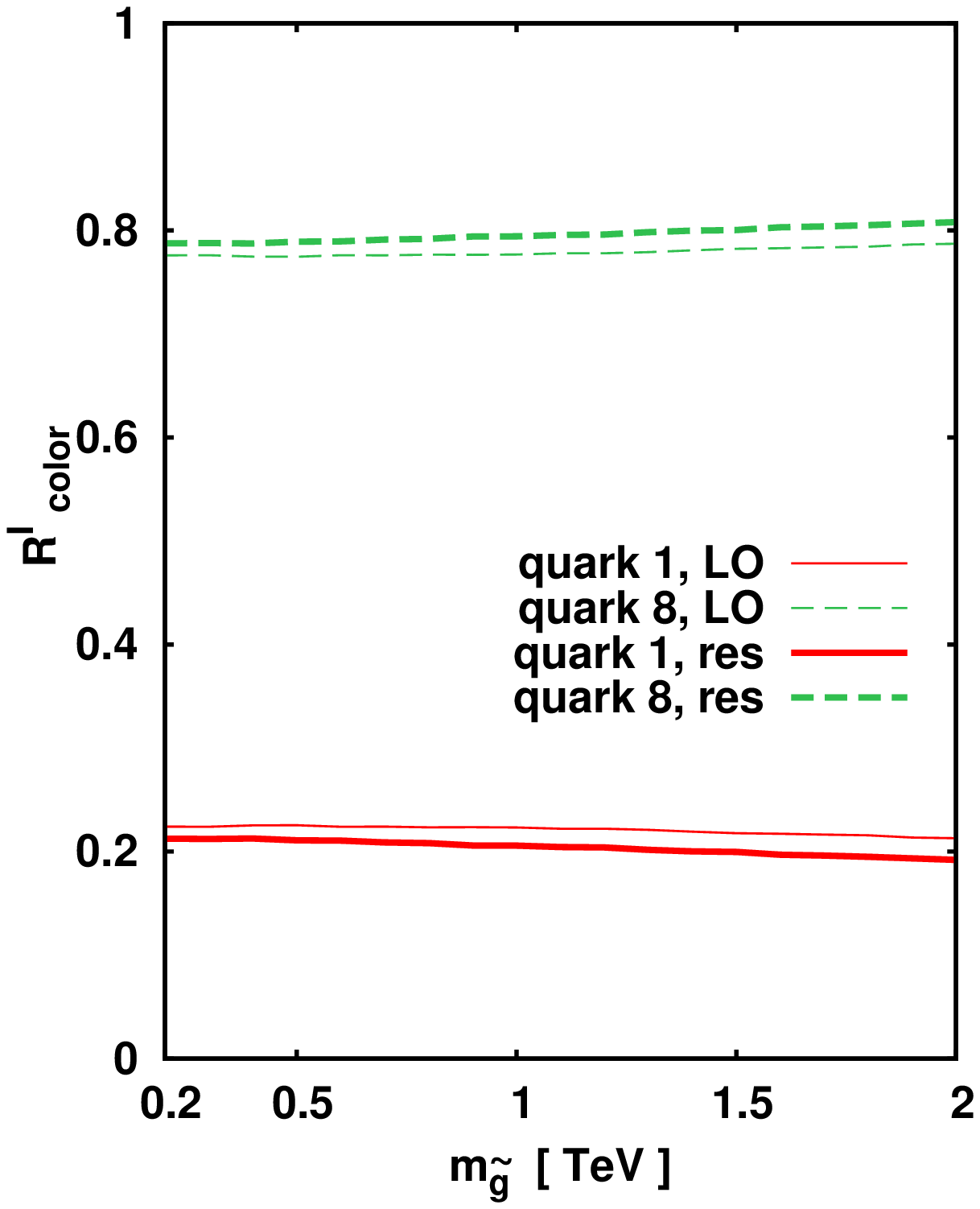,width=0.41\columnwidth} \hspace{0.05\columnwidth} &
\epsfig{file=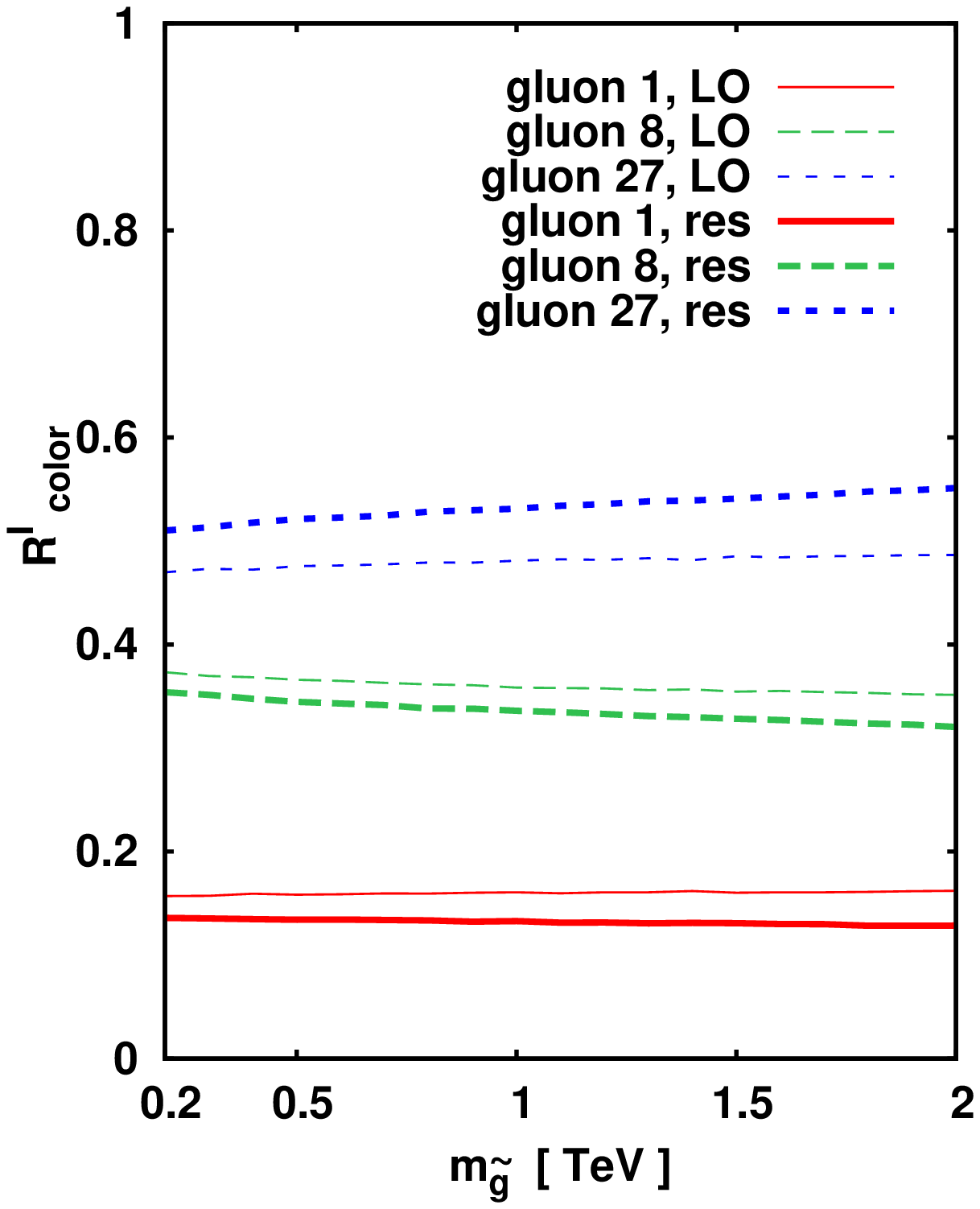,width=0.41\columnwidth}\\
{\large\bf a)} & {\large\bf b)} \end{tabular}
\end{center}
\caption{\it 
The decomposition of the cross section for the $\gl\gl$ hadroproduction into
the partonic colour channels, $R^{\mai} _{\mathrm{color}}$, for: (a) $q\bar q \to \gl\gl$ and (b) $gg \to \gl\gl$; $r=1.2$. The thick lines correspond to the resummed cross sections and the thin 
ones to the LO cross sections.
\label{fig:colgr}
} 
\end{figure}
\begin{figure}[p]
\begin{center}
\begin{tabular}{ll}
\epsfig{file=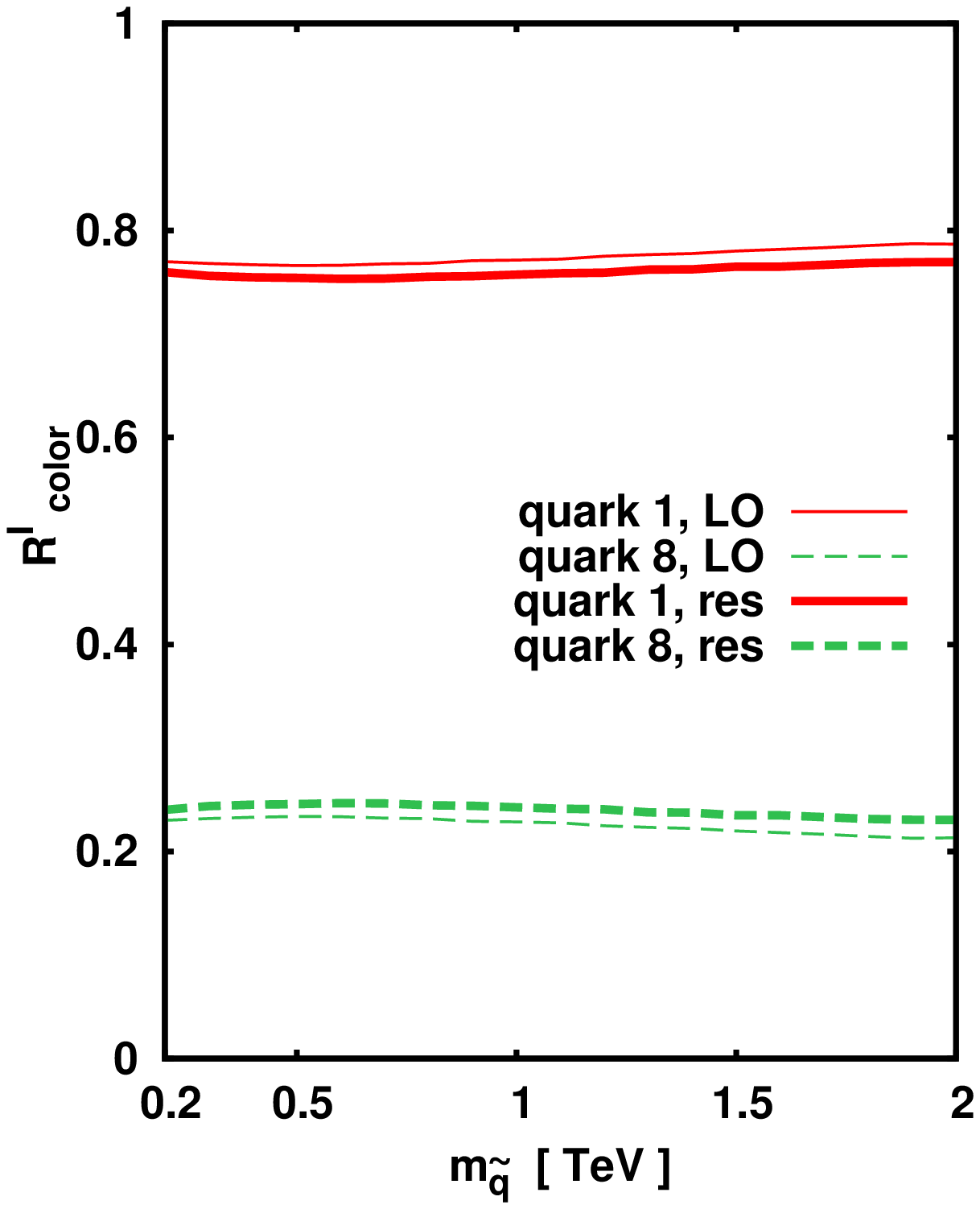,width=0.41\columnwidth} \hspace{0.05\columnwidth} &
\epsfig{file=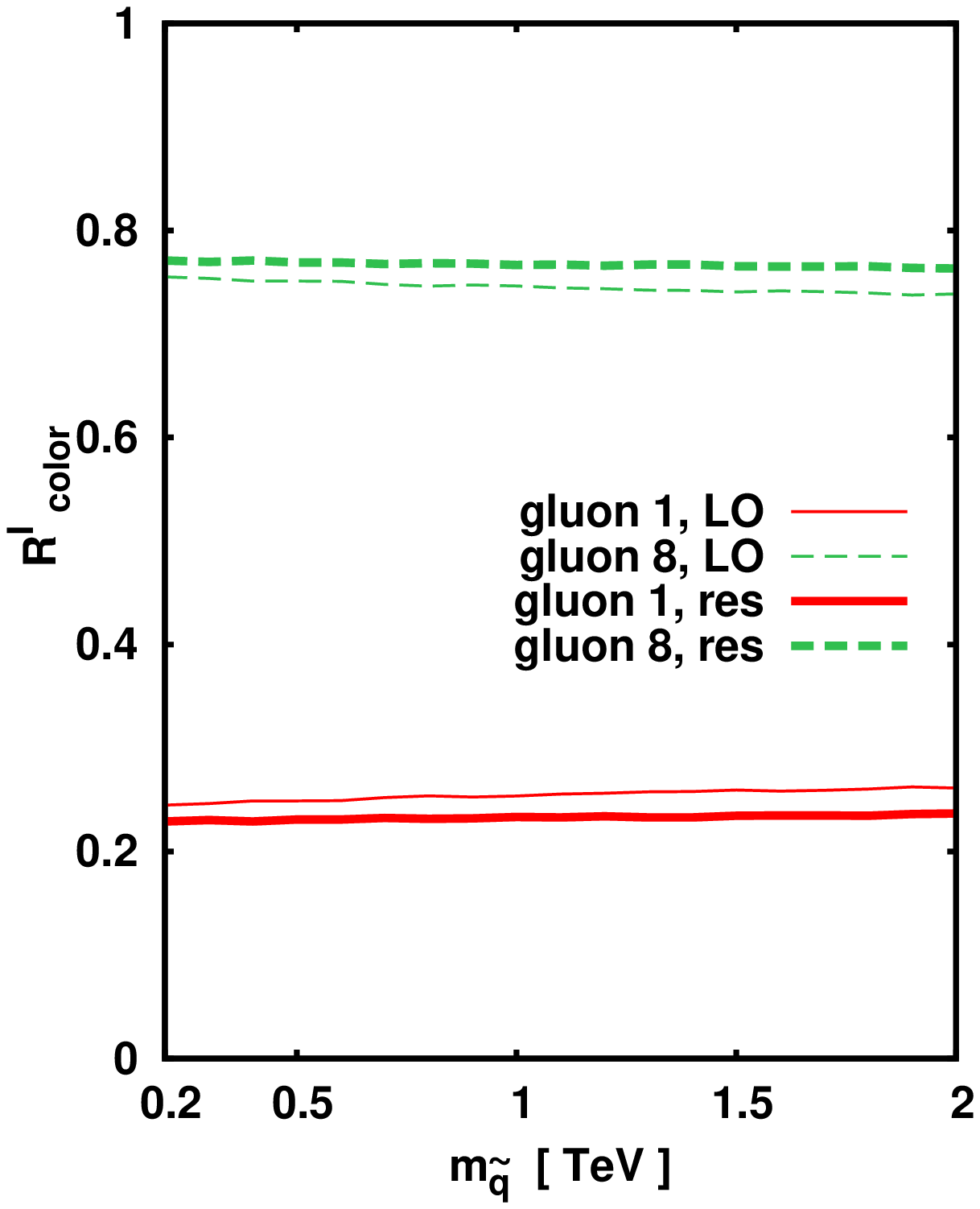,width=0.41\columnwidth}\\
{\large\bf a)} & {\large\bf b)}  \end{tabular}
\end{center}
\caption{\it 
The decomposition of the cross section for the $\sq\sqb$ hadroproduction into
the partonic colour channels, $R^{\mai} _{\mathrm{color}}$, for: (a) $q\bar q \to \sq\sqb$ and (b)  $gg \to \sq\sqb$; $r=1.2$.
The thick lines correspond to the NLL cross sections and the thin ones 
to the LO cross sections.
\label{fig:colqr}
} 
\end{figure}

Finally, we analyse the effect of soft gluon resummation in the
partonic 
colour channels. Recall that the soft gluon corrections at threshold do not lead 
to mixing of the $s$-channel colour representations. 
Let us denote by $\sigma^{\mathrm{(\LO)}} _{pp \to ij \to kl;\; \mai}$
the contribution to the $pp \to kl$ cross section at the LO coming from the 
partonic channels $ij$ in the $s$-channel colour representation 
$\mai$, and by $\sigma^{\mathrm{\mathrm{(res)}}} _{pp \to ij \to kl;\;
  \mai}$ the analogous contribution to the resummed hadronic
cross section (not matched to the NLO). Furthermore, we define ratios 
\be
R^{\mai} _{\mathrm{color}} \, = \,   
{\sigma_{pp \to ij \to kl;\; \mai} \over \sum_{\mai} 
\sigma_{pp \to ij \to kl;\; \mai}}
\ee
that correspond to the relative contribution of the colour representation
$\mai$ to the partonic channel $ij$ of the hadronic cross section (we
suppress here the indices of the initial, intermediate and final state
particles in   $R^{\mai} _{\mathrm{color}}$).  
In Fig.\ \ref{fig:colgr} and in Fig.\ \ref{fig:colqr} we show the ratios
$R^{\mai} _{\mathrm{color}}$ for $\gl\gl $ and $\sq\sqb$ production, respectively, both for the LO cross sections and the resummed cross sections with $r=1.2$. 
In general we find that the dependence of $R^{\mai} _{\mathrm{color}}$ on the mass of produced particle is weak. We find that the colour singlet $s$-channel representation dominates in the $q\bar q$ channel of $pp\to \sq\sqb$, and the colour octet  $s$-channel representation dominates in the $q\bar q$ channel
of  $pp\to \gl\gl$ and in the $gg$ channel of $pp\to \sq\sqb$. In all these channels the soft gluon effects only weakly affect $R^{\mai} _{\mathrm{color}}$. In the case of the $gg$ channel of $pp\to \gl\gl$, the pattern is more interesting.  We observe the following hierarchy of the contributions from the three colour channels,  $R^{\mathbf{27}} _{\mathrm{color}} > R^{\mathbf{8}} _{\mathrm{color}} > R^{\mathbf{1}} _{\mathrm{color}}$.
The largest contribution coming from the $\mathbf{27}$ $s$-channel
representation is most strongly enhanced due to soft gluon effects.

\begin{figure}[p]
\begin{center}
\begin{tabular}{ll}
\epsfig{file=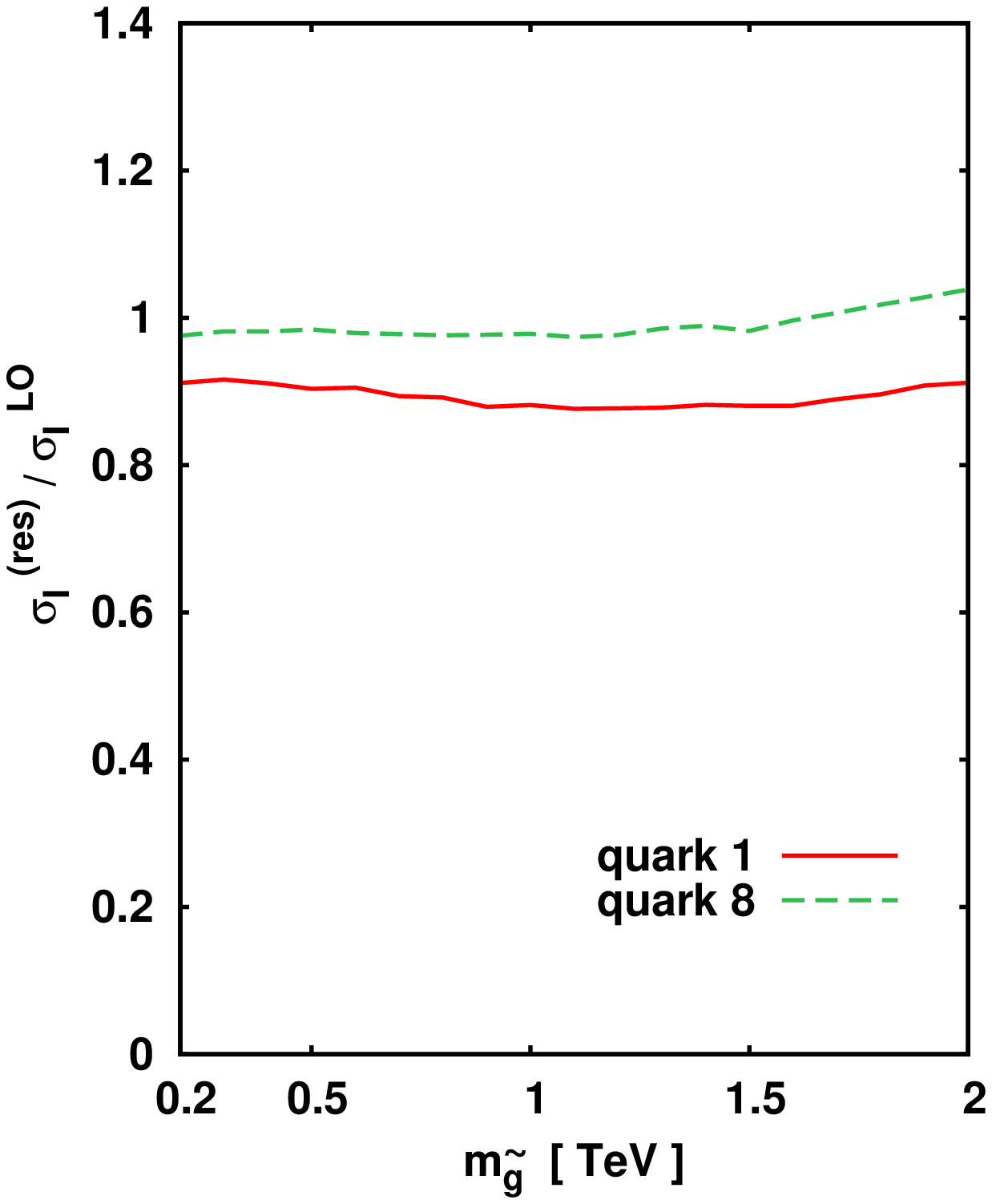,width=0.41\columnwidth} \hspace{0.05\columnwidth} &
\epsfig{file=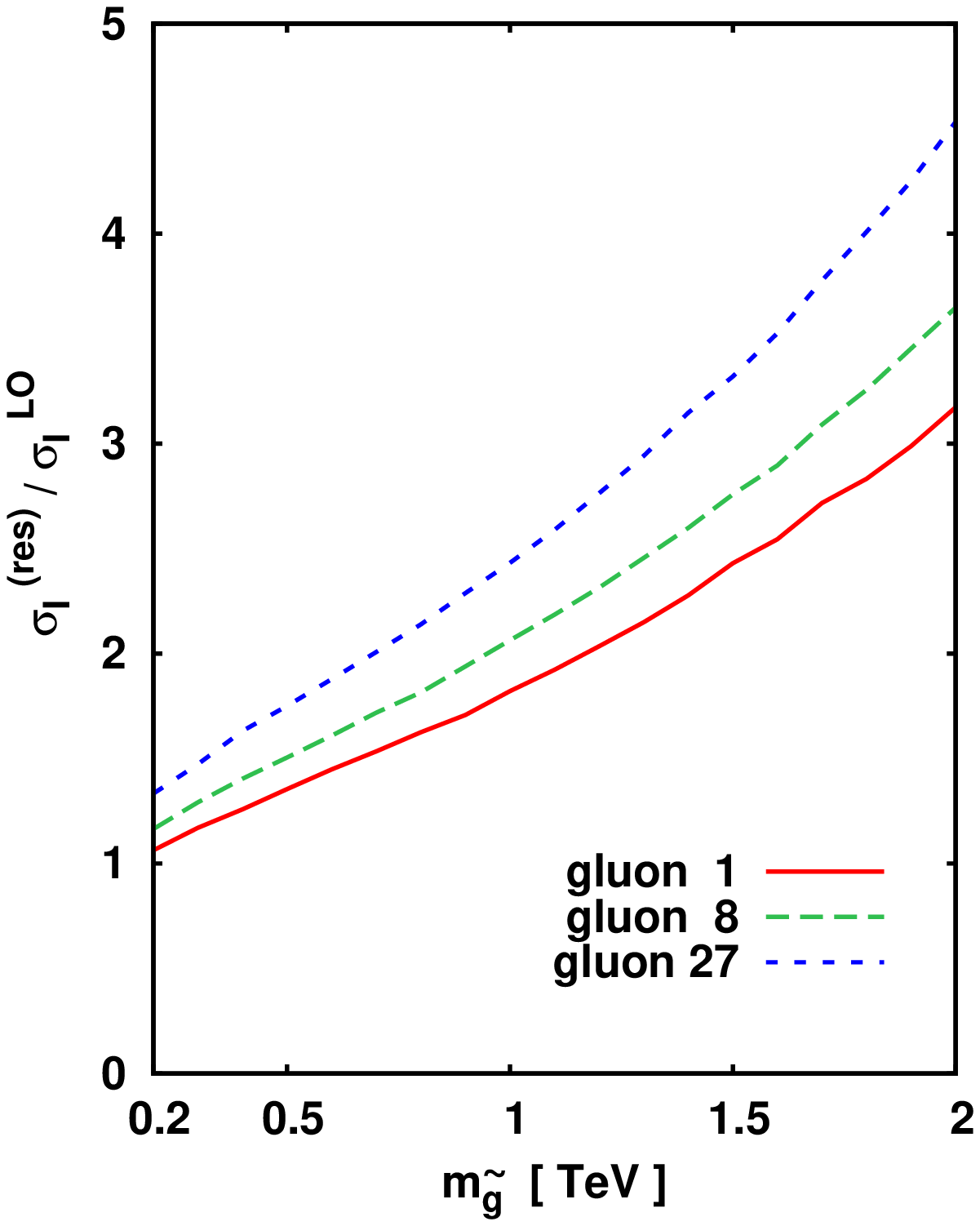,width=0.41\columnwidth}\\
{\large\bf a)} & {\large\bf b)}  \end{tabular}
\end{center}
\caption{\it 
The $K$-factors, $K=\sigma_{\mai}^{\mathrm{(res)}} / \sigma^{(\LO)}$, for partonic 
colour channels: (a) $q\bar q \to \gl\gl$ and (b) $gg \to \gl\gl$; $r=1.2$. 
The dimensions of the $s$-channel colour representations are indicated in the legend.
\label{fig:colgk}
} 
\end{figure}
\begin{figure}[p]
\begin{center}
\begin{tabular}{ll}
\epsfig{file=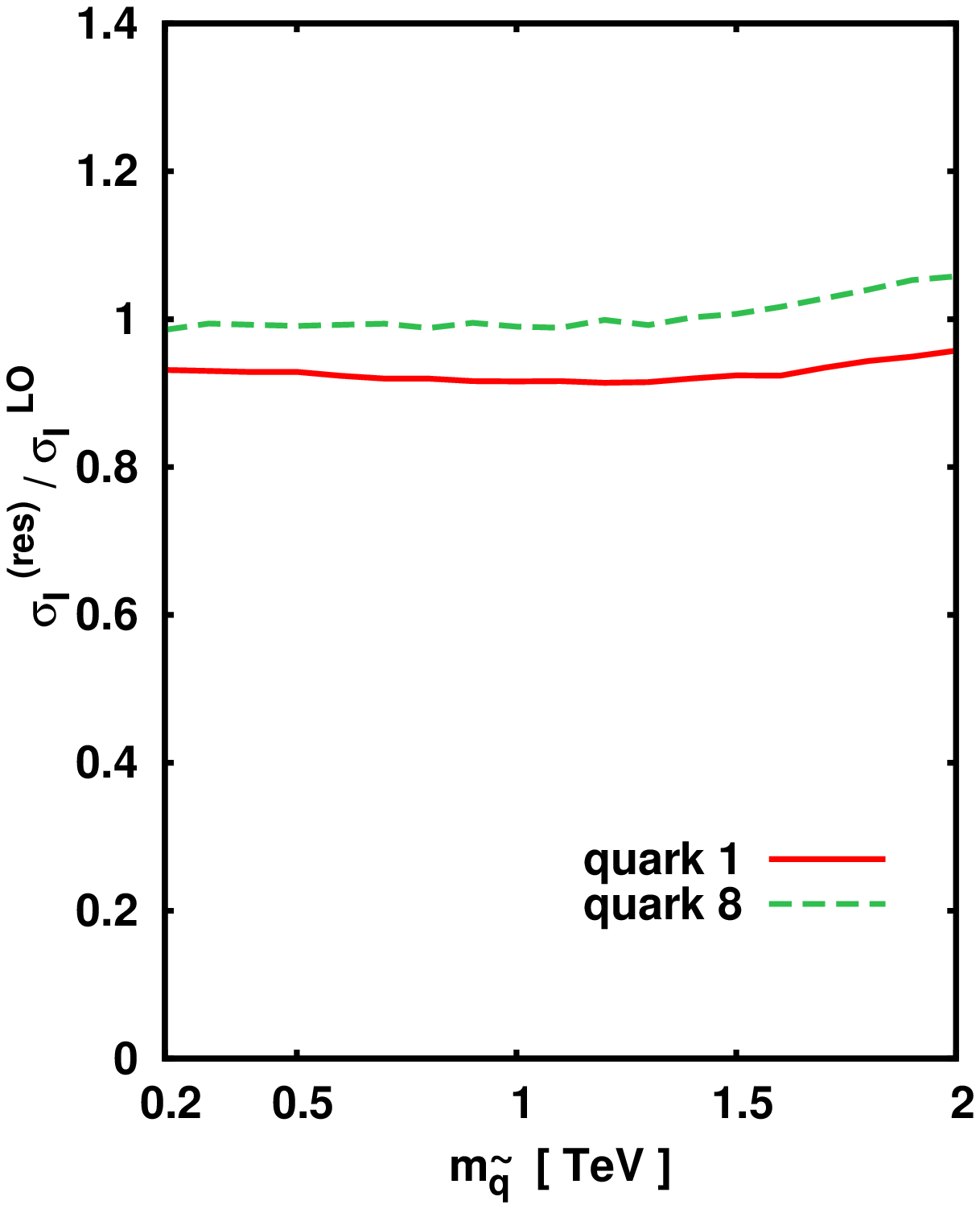,width=0.41\columnwidth} \hspace{0.05\columnwidth} &
\epsfig{file=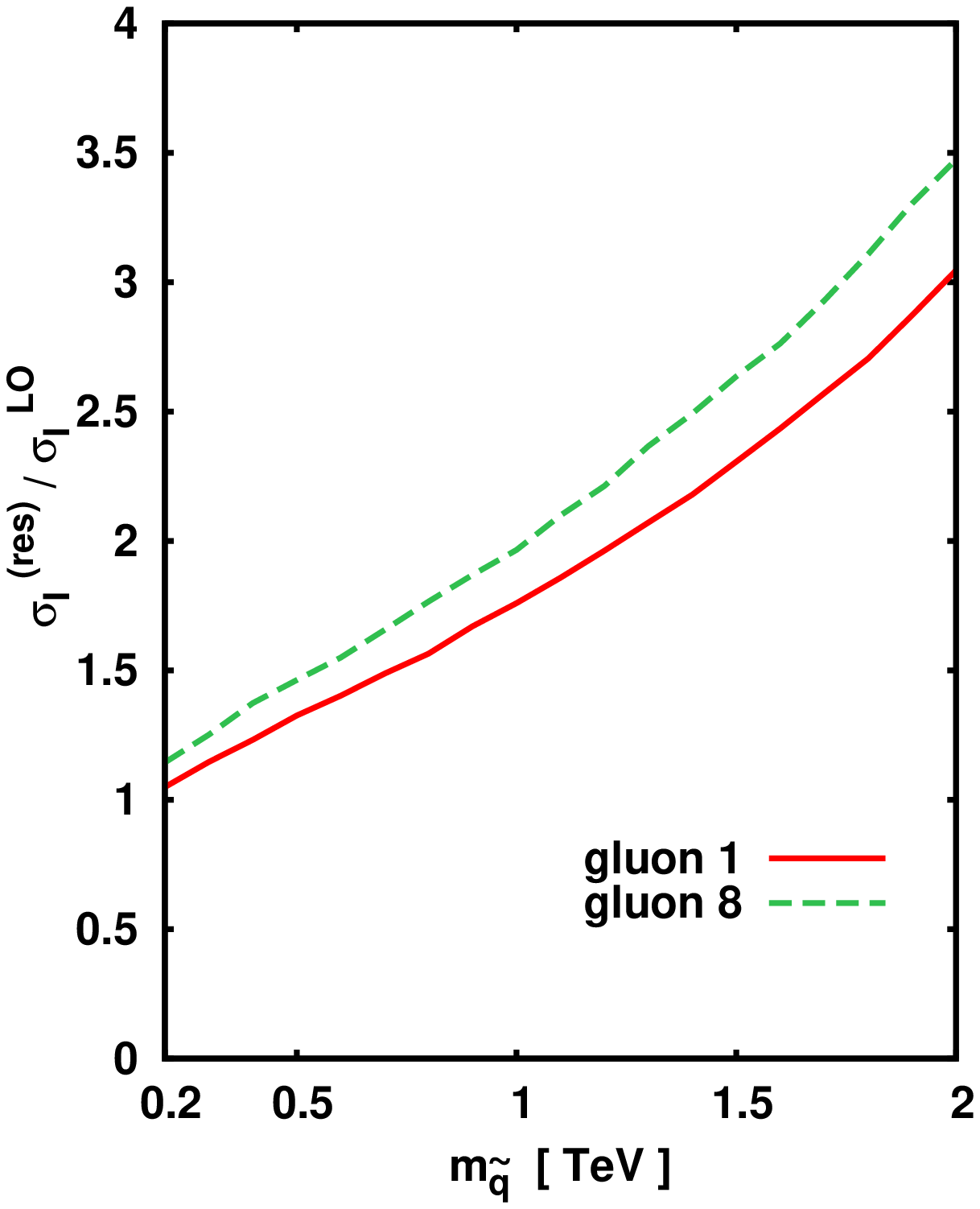,width=0.41\columnwidth}\\
{\large\bf a)} & {\large\bf b)} \end{tabular}
\end{center}
\caption{\it 
The $K$-factors, $K=\sigma_{\mai}^{\mathrm{(res)}} / \sigma^{(\LO)}$, for partonic colour channels: $q\bar q \to \sq\sqb$ and (b)  $gg \to \sq\sqb$; $r=1.2$. The dimensions of the $s$-channel colour representations are indicated in the legend.
\label{fig:colqk}
} 
\end{figure}

In Fig.\ \ref{fig:colgk} and Fig.\ \ref{fig:colqk} we show the effect
of soft gluon corrections on the cross sections in partonic-colour
channels for the $\gl\gl$ and $\sq\sqb$ production, respectively.  We plot the $K$-factors, $K=\sigma^{\mathrm{\mathrm{(res)}}} _{\mai} /
\,\sigma^{\mathrm{\mathrm{(\LO)}}} _{\mai}\equiv \sigma_{ij \to
  kl,I}^{\mathrm{\mathrm{(res)}}} /\sigma_{ij \to
  kl,I}^{(0)} $ in these channels. The pattern is quite clear: the enhancement from soft gluon resummation for the $gg$ partonic channels is strong and steeply rising with mass of the produced particles. This is not the case for the 
$q\bar q$ channel, where the soft gluons effects are small and weakly depending
on the mass of the produced particles. Note that the $K$-factors in
the $q\bar q$ channels are smaller than 1 for most values of the masses. In these channels, the size of the enhancement coming from soft gluon emission does not sufficiently compensate the suppresion due to changing from the LO to the NLO approximation of the pdfs and running $\as$.
For all considered channels we observe larger soft gluon enhancement for higher $s$-channel colour representations.

\section{Conclusions}

In this work we have studied the effect of soft gluon emission on the
production of gluino-gluino and squark-antiquark pairs in the proton--proton
collisions at the LHC.  
A detailed description of the derivation of the soft anomalous dimension matrices
for the $q\bar q \to \gl\gl$ and $gg\to \gl\gl$ scattering processes~\cite{KM} has been
given. It should be stressed that these matrices
govern the soft non-collinear gluon radiation in the hadronic pair production of any
heavy particles in the colour octet representation, like for instance
the supersymmetric heavy colour scalar particles considered in
\cite{CDKKPZ}. In the threshold limit the obtained soft anomalous
dimension matrices become
diagonal in the $s$-channel colour basis.  The diagonal
elements of the soft matrices have been found to correspond to values of the
quadratic Casimir operators of the SU(3) group for the outgoing pair
of heavy particles. This observation
confirms simple physical picture of the direct dependence of the soft gluon radiation at threshold on the total colour charge of the final state.

The NLL resummation of the soft gluon corrections to the
hadroproduction processes, $pp\to \sq\sqb$ and $pp\to \gl\gl$ at the
LHC has been explicitly performed for the values of masses of the
produced particles between 0.2~TeV and 2~TeV and for the gluino-to-squark
mass ratio $0.5 < r <2$.  The obtained  results
have been matched to the corresponding cross sections computed at the
NLO accuracy. We have determined the NLO $K$-factors,  $K_{\NLL}$,
for the squark--antisquark and gluino pair-production for these
processes and studied their scale dependence. Futhermore, we have 
investigated the dependence of the NLL $K$-factors on the choice of
the parton distribution functions. Additionally, the effect of the leading Coulomb corrections 
on the considered hadronic cross sections has been discussed. 
We have also analysed the partonic channel decomposition of the
$pp \to \gl\gl$ and $pp\to \sq\sqb$ total cross sections. In
particular, we have studied the effect of the NLO and NLL corrections on the
total cross sections in the partonic channels. Finally, we
have decomposed the LO and NLL hadronic cross sections for the
partonic subchannels in the colour basis and observed that the soft gluon enhancement grows with the 
the total colour charge of the pair of the produced particles.

\paragraph{Acknowledgments} 
A.~K.\ thanks W.~Beenakker, S.~Brensing, M.~Kr\"amer, E.~Laenen and
I.~Niessen for many discussions. L.M.\ is grateful to J.~Bartels for 
stimulating conversations. The work of A.K.\ is supported by the Initiative and
Networking Fund of the
Helmholtz Association, contract HA-101 ("Physics at the Terascale").
L.~M.\ acknowledges the DFG grant No.\ SFB 676 and of the Polish Ministry of Education grant No.\ N202 249235.

\begin{appendix}

\section{Mellin-moment transforms of the leading-order cross sections}

We define 
\bear
J_N &\equiv& \int_0^1 d z \frac{z^{N+1}}{1+ \left(\frac{1-r^2}{2r^2}\right)z}
\log
  \left( \frac{2r^2(1+\sqrt{1-z})+(1-r^2)z}{2r^2(1-\sqrt{1-z})+(1-r^2)z}
  \right) \;, \nn \\
B_{N} &\equiv& \beta (N+1, 1/2) \;, \nn \\
G_{N}^{(1)} &\equiv& _2 F_1 \left( 1,1/2,N+5/2,
  \left(\frac{r^2-1}{r^2+1}\right)^2 \right) \;, \nn
\\
G_{N}^{(2)} &\equiv& _2 F_1 \left( 1,N+1,N+5/2,
  -\frac{\left(r^2-1\right)^2}{4 r^2} \right) \;, \nn
\eear
where $r=\mgl/\msq$. With these definitions, the LO cross sections in $N$-space read

\bear
\tilde \si_{q_i \qbar_j \to \sq \sqb, \onebf}^{(0)} &=& \frac{ \text{\as}^2
  \pi B_{N}}{\text{\msq}^2 }
\frac{ 8}{81 \left(2 N^3+9 N^2+13 N+6\right) (r^2+1)}\\
&\times& \left[ -(r^2+1) N (2 N+3) + 2  (r^2 N+N+2)(N+1) G_{N}^{(1)} 
+ (r^2+1)(N^2+3N+2)G_{N}^{(2)} \right] \;, \nn \\
\tilde \si_{q_i \qbar_j \to \sq \sqb, \eigbf}^{(0)} &=& \frac{\text{\as}^2  \pi \text{B_{N}}}{\text{\msq}^2}
\left[
\frac{ \text{\delta_{ij}} \text{n_f}  }{9  \left(4 N^2+16 N+15\right)}
\right. \nn \\
&+&
\left. \frac{2 \text{\delta_{ij}}   \left((r^2+1) (3 r^2+N \
(2 r^2-1)-2)-2  r^4 (N+1)\text{G_N^{(1)}}\right)}{27  \left(2 N^3+13 \
N^2+27 N+18\right) \left(r^4-1\right)} \right]
+
\frac{1}{8} \sih_{q \qbar \to \sq \sqb, \onebf}^{(0)} \;, \\
\tilde \si_{gg \to \sq \sqb, \onebf}^{(0)} &=&\frac{\text{\as}^2 \pi
  \text{B_{N}}}{\text{\msq}^2}\frac{ \left(N^2+3 N+4\right) \text{n_f}
}{96 \ \left(2 N^3+13 N^2+27 N+18\right)} \;, \nn \\
\tilde \si_{gg \to \sq \sqb, \eigbf}^{(0)} &=&
\frac{\text{\as}^2  \pi\text{B_{N}}}{ \text{\msq}^2} \frac{\left(5 N^3+32 N^2+71 N+68\right) \
\text{n_f}  }{96  \left(4 N^4+36 N^3+119 N^2+171 N+90\right)} \;, \\
\tilde \si_{q \qbar \to \gl \gl, \onebf}^{(0)} &=&
-\frac{\text{\as}^2 \pi \text{B_{N}}}{\text{\mgl}^2}\frac{1}{54
  \left(2 N^2+7 N+6\right) (r^2+1)} \left[(r^2+1) (-8 N -12) \right. \nn \\
&+& \left. (8N+8)  \text{G_N^{(1)}} + 2 (r^2+1)(N+2)\text{G_N^{(2)}} +
(r^2+1) \left(2 N^2+7 N+6\right)\frac{\text{J_N}}{B_N} \right] \;, \\
\tilde \si_{q \qbar \to \gl \gl, \eigbf}^{(0)} &=&
\frac{\text{\as}^2 \pi\text{B_{N}}}{\text{\mgl}^2}
\frac{1}{27  \
\left(4 N^4+36 N^3+119 N^2+171 N+90\right) \left(r^4-1\right)}\nn \\
&\times&
 \left[ 
2 (r^2+1)( (r^2-1)N^3 - (43-25 r^2) N^2  - (162-99 r^2) N +108 r^2-153)
\right. \nn \\
&+& 
\left.
4 (2 N (r^2-1)+6 r^2+3)  \left(2 N^2+7 N+5\right)
\text{G_N^{(1)}}   \right. \nn \\
&-& \left.  7  (r^4 -1) \left(2 N^3+15 N^2+37 \
N+30\right) \text{G_N^{(2)}}\right.\nn \\
&+&
 \left.  \left(r^4-1\right)\left(4 N^4+36 N^3+119 N^2+171 \
N+90\right) \frac{\text{J_N}}{B_N}\right] \;, \\
\tilde \si_{{\rm sym}} &=&
\frac{\text{\as}^2 \pi \text{B_{N}}}{\text{\mgl}^2 }\frac{9 \left(N^3+9 N^2+20 N+14\right)  }{64 \
\left(2 N^4+15 N^3+40 N^2+45 N+18\right)} \;, \\
\tilde \si_{{\rm asym}} &=&
\frac{\text{\as}^2 \pi \text{B_{N}}}{\text{\mgl}^2 }\frac{9\left(N^3+11 N^2+30 N+26\right)  \
}{64 \left(4 N^5+40 N^4+155 N^3+290 N^2+261 N+90\right)} \;,
\eear
where we put $\hat \as =\as$.
For the numerical evaluation of $J_N$  we separately use two forms of
its expansion 
\be
J_N  =  \frac{2 r^2}{1+r^2} \sum_{m=0} ^{\infty} \left({ 1-r^2 \over 1+r^2}\right)^m \, 
\sum_{k=0} ^{\infty} \, 
{1 -\left({1-r^2\over 1+r^2}\right)^{2k+1} \over k+1/2}\, \beta (N+2,k+m+3/2)\, , \nn
\ee
and
\bear
&&J_N  =  \frac{2 r^2}{1+r^2} \sum_{m=0} ^{\infty} \left({ 1-r^2 \over 1+r^2}\right)^m \, 
\frac{1}{1+m} \sum_{k=0} ^{m} \, 
\frac{\left(-1\right)^k}{\beta\left(k+1,m-k+1\right)} \\
&&\times \left[ \frac{\beta \left(k+N+2, 1/2 \right)}{k+N+2} -2  \left({ 1-r^2
      \over 1+r^2}\right)
\beta \left(k+N+2, 3/2 \right)\; _2F_1 \left(1,1/2,k+N+7/2,\left({ 1-r^2 \over
    1+r^2}\right)^2\right)\right] \,. \nn
\eear

\section{LL and NLL functions }

The expressions for the resummed factors, expanded up to NLL, are 
\bear
&&\log \Delta_i (N,4m^2,\mu^2) \stackrel{\NLL}{=}  
g_i ^{(1)}\left(b_0\,\alpha_s(\mu^2)\log N\right)\, \log N\;
+ \; g_i ^{(2)}\left(b_0\,\alpha_s(\mu^2)\log N, 4 m^2, \mu^2\right)\,,\qquad \\
&&\log\Delta^{(s)} _{ij \to kl,\, \mai}(N, 4m^2, \mu^2) \stackrel{\NLL}{=} \;
h^{(2)}_{ij \to kl,\, \mai}\left(b_0\,\alpha_s(\mu^2)\log N\right)
\eear
with
\bear
g_i ^{(1)} (\lambda) & \;\;=\;\; & 
{A_i^{(1)} \over 2\pi b_0 \lambda} \, 
\left[ 2\lambda +(1-2\lambda)\, \log(1-2\lambda)\right]\, , \\
g_i ^{(2)} (\lambda, 4 m^2, \mu^2) & = &  -{ A_i^{(1)}\gamma_E \over \pi b_0}\, \log(1-2\lambda) 
\, + \, {A_i^{(1)} b_1 \over 2 \pi b_0^3}\, 
\left[2\lambda + \log(1-2\lambda) + {1\over 2}\log^2(1-2\lambda)\right]\qquad \nn\\
& & \; - \;{A_i^{(2)} \over 2\pi^2 b_0^2}\,\left[\,2\lambda + \log(1-2\lambda)\,\right]
\, - \, {A_i^{(1)} \over 2 \pi b_0}\, \log(1-2\lambda)\, 
\log\left( {\mu^2 \over 4 m^2} \right)\, , \\
h^{(2)}_{ij \to kl,\, \mai} (\lambda) & \;\;=\;\; &
{\log(1-2\lambda)\over 2\pi b_0} \,
D^{(1)} _{ij \to kl,\,{\mai}} \,,
\eear
where we took $\mu=\mu_F=\mu_R$ and  $b_0$ and $b_1$ are the first two 
coefficients of the QCD $\beta$-function, 
\be
b_0 = {11C_A - 4T_R\nf \over 12\pi}\,, \qquad 
b_1 =  {17C^2_A - 10C_AT_R\nf - 6C_FT_R\nf \over 24\pi^2}\;.
\ee
The values of the coefficients
$A_i^{(1)},\ A_i^{(2)}$, $D^{(1)} _{ij \to kl,\,{\mai}}$ are
defined in Section 3.1. 

\end{appendix}

\end{document}